%% file: Barmak_instability_pipe.tex
\documentclass[
amsmath,amssymb,
aps,
]{revtex4-2}
\usepackage[]{graphicx,color}
\usepackage{amssymb}                    
\usepackage{amsmath}

\usepackage{bm}							
\usepackage[position=t,caption=false,justification=centering]{subfig}
\usepackage{hyperref}
\usepackage{mathtools}
\usepackage[makeroom]{cancel}
\usepackage{stmaryrd} 
\usepackage{tabularx}
\usepackage{multirow}
\renewcommand{\vec}[1]{{\mbox{\boldmath $ #1 $}}}
\allowdisplaybreaks
\newcommand{\Fr}{\mbox{Fr}}	 
\newcommand{\Rey}{\mbox{Re}} 
\newcommand{\We}{\mbox{We}}  
\newcommand{\Eo}{\mbox{Eo}}  
\graphicspath{{./figures/}}
\begin{document}

\preprint{APS/123-QED}

\title{Instability of stratified air--water flows in circular pipes}

\author{Ilya Barmak}
\email{ilyab@tauex.tau.ac.il}
\affiliation{School of Mechanical Engineering, Tel Aviv University, Tel Aviv 6997801, Israel}
\affiliation{Soreq NRC, Yavne 8180000, Israel 
}%

\author{Alexander Gelfgat}
\email{gelfgat@tauex.tau.ac.il}
\affiliation{School of Mechanical Engineering, Tel Aviv University, Tel Aviv 6997801, Israel 
}%

\author{Neima Brauner}
\email{brauner@tauex.tau.ac.il}
\affiliation{School of Mechanical Engineering, Tel Aviv University, Tel Aviv 6997801, Israel 
}%


\maketitle

\section*{Abstract}

This work deals with stability of two-phase stratified air-water flows in horizontal circular pipes. For this purpose, we performed a linear stability analysis, which considers all possible three-dimensional infinitesimal disturbances and takes into account deformations of the air-water interface. The main results are presented in form of stability maps, which compare well with the available experimental data. The neutral stability curves are accompanied by the corresponding wavenumbers and wave speeds of the critical perturbations, as well as by spatial patterns of their velocity components. Accordingly, several modes of the critical perturbation are revealed. Long waves are found to be the critical perturbation over part of the stability boundary, and they are affected by the surface tension due to the confinement effect of the lateral direction. Exploring the effect of pipe diameter on the stability boundary and critical perturbations shows that for small water holdups (i.e., thin water film) the scaling of the critical gas velocity by the gas Froude number is valid for pipe diameters larger than about 0.1m, where the surface tension effects due to the lateral confinement become negligible. Comparing results obtained in pipe, square duct, and two-plate geometries, we show that there are cases where the simplified geometry of two parallel plates can be employed to model the realistic geometry reasonably well.

\section{Introduction}

In this paper, we extend our previous studies on stability of two-phase stratified flows between two infinite plates \cite{Barmak16a,Barmak16b,Barmak19} and ducts with rectangular cross-section \cite{Gelfgat20a,Gelfgat21} to flows in circular pipes. The main motivation is to find the stability boundary, beyond which the stratified flow with a smooth interface is unstable and transition to other flow patterns, e.g., stratified-wavy, annular, intermittent flow, can take place. The stability diagrams are commonly depicted in coordinates of superficial velocities of each phase \cite[e.g.,][]{Taitel15,Brauner03}. Although circular pipes are of main practical interest and they are utilized in many experimental studies \cite[e.g., ][]{Mandhane74,Barnea80,Barnea82}, the only attempt of rigorous theoretical study to treat instability of two-phase stratified pipe flows was done recently by \citet{Barmak23}, where a general numerical framework for performing such an analysis was presented and verified. In this study we apply this methodology to perform a detailed parametric study of stability of stratified air-water flows in circular pipes.

In theoretical studies, a widely used simplified approach to stability of the two-phase stratified flow is the so-called Two-Fluid (TF) mechanistic model \cite[e.g., ][]{Andritsos89,Brauner91,Barnea93,Ullmann06,Kushnir17}. This approach is critically dependent on the closure relations required to model the interaction of the base flow with the interfacial disturbance (e.g., steady and wave induced wall and interfacial shear stresses, velocity profile shape factors, see details in \cite{Kushnir14,Kushnir17}). Moreover, an inherent assumption of the stability analyses based on the TF model is long-wave perturbations, which are not always the critical perturbation that trigger instability. Another approach is to refer to the simpler two-plate (TP) geometry that allows to account for perturbations of all possible wavelengths (see \cite{Barmak16a,Barmak16b} and references therein). This approach, nevertheless, cannot reveal how stability characteristics are affected by the real geometry of a circular pipe, in particular by the presence of walls in the lateral direction.

The effect of the lateral walls was recently studied both numerically and experimentally for ducts of a rectangular cross-section \cite{Gelfgat20a,Gelfgat21,Nezihovski22}. The studies showed a good agreement between the numerical results and experimental findings. Nevertheless, stratified flows in circular pipes still remain unexplored.

Therefore, in this study, we adopt the numerical approach of \cite{Barmak23} to explore the linear stability of stratified air--water pipe flows with respect to arbitrary wavenumber disturbances. The stability boundaries are plotted on the flow pattern maps. Along with the neutral stability curves, we report the critical perturbation wavenumber and wave velocity. We also present cross-sectional patterns of the critical perturbations that are responsible for the onset of the instability (i.e., the most unstable perturbation modes). Furthermore, the effects of the pipe diameter and the conduit geometry on the stability boundary and the critical perturbation are explored and discussed.

\section{Problem formulation} \label{Sec: Formulation}

\begin{figure}[h!]
	\centering
	\def\svgwidth{0.7\textwidth}
	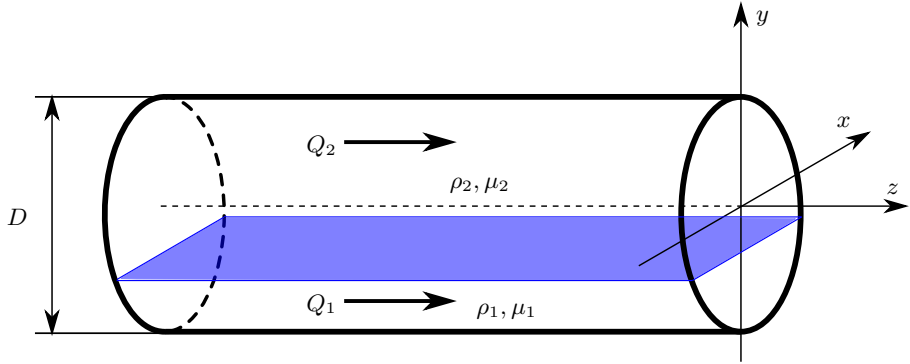	
	\caption{Schematics of stratified two-phase pipe flow (plane interface).}
	\label{Fig: flow_geometry}
\end{figure}

We consider a two-phase stratified flow in a horizontal circular pipe of diameter $D$. The flow is assumed to be isothermal. It is driven by a pressure gradient imposed in the direction of the pipe axis. The flow configuration is sketched in Fig.\ \ref{Fig: flow_geometry}. The volumetric flow rates of the heavy and light fluids are $Q_1$ and $Q_2$, respectively. The superficial velocities of each phase are defined as the corresponding flow rate divided by the pipe cross section, or $\displaystyle U_{1S} = Q_1 / A$ and $\displaystyle U_{2S} = Q_2 / A$, where $\displaystyle A = \pi D^2/4$.

The two-phase stratified flow is described by the velocity $\vec{u}^{(k)}$ and pressure $p^{(k)}$ fields that satisfy the dimensionless continuity and momentum equations defined in each layer $k=1,2$ (1 - heavy fluid, 2 - light fluid):
\begin{equation} \label{Eq: Continuity_vector}
	\nabla \cdot \vec{u}^{(k)} = 0,
\end{equation}
\begin{equation} \label{Eq: N-S_vector_form}
	\frac{\partial \vec{u}^{(k)}}{\partial t} + \left(\vec{u}^{(k)}\cdot\nabla\right)\vec{u}^{(k)}
	= -\frac{1}{\rho^{(k)}}\nabla p^{(k)} + \frac{1}{\Rey^{(k)}} \Delta \vec{u}^{(k)} + \frac{1}{\Fr} \vec{e_g},
\end{equation}
where the velocity is scaled by the mixture velocity $\displaystyle U_m = U_{1S} + U_{2S}$, while the time and pressure are scaled by $\displaystyle D/U_m$ and $\rho_2 U_m^2$, respectively. The dimensionless governing parameters are the density and viscosity ratios, $\displaystyle\rho_{12} = \rho_1/\rho_2$ and $\displaystyle\mu_{12} = \mu_1/\mu_2$, respectively, the Reynolds number of each phase $\displaystyle\Rey^{(k)}=\rho_k D U_m/\mu_k$ and the Froude number $\displaystyle\Fr = U_m^2/(gD)$, where $g$ is the gravitational acceleration. $\displaystyle\vec{e_g}$ is a unit vector in the direction of gravity. $\rho^{(k)}=\rho_{1 2}\rho_k/\rho_1$ is introduced for brevity of the notation.

\begin{figure}[h!]
	\setcounter{subfigure}{0}
	\subfloat[Cartesian coordinates]
	{\def\svgwidth{0.35\textwidth}
		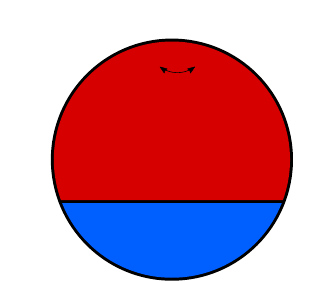}
	\qquad
	\subfloat[Bipolar coordinates]
	{\def\svgwidth{0.6\textwidth}
		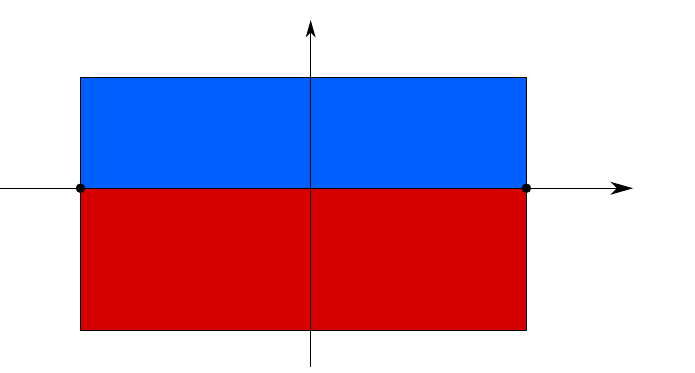}	
	\caption{Stratified two-phase pipe flow with a plane interface ($\displaystyle\phi^*=\pi$). Cross-section of a circular pipe in (a) Cartesian and (b) bipolar coordinates.}
	\label{Fig: Geometry_cross_section}
\end{figure}

 Due to big difference in densities between air and water, air--water flows are gravity-dominated with large E\"otv\"os number, $\displaystyle\Eo = (\rho_1 - \rho_2) g D^2/\sigma \gg 1$, where $\sigma$ is the surface tension coefficient. The smooth stratified flow in such systems has a plane interface \cite{Gorelik99}. As discussed in detail in \cite{Barmak23}, the convenient coordinate system for this flow geometry is the bipolar cylindrical coordinates $(\xi,\phi,z)$ \cite[e.g.,][]{Korn00,Goldstein15}. In these coordinates, the pipe wall corresponds to an isoline of the coordinate $\phi$: $\phi_0$ -- for the upper section of the pipe wall and $\phi_0+\pi$ -- for the bottom of the pipe. The interface separating two fluids is a surface defined as $\phi_\eta = \phi^* + \eta\left(\xi,z\right)$, where $\phi^*=\pi$ for the plane interface. The cross-sectional geometry is shown in Fig.\ \ref{Fig: Geometry_cross_section}. The heavy fluid (shown in blue) occupies an infinite strip of $\displaystyle\bigl(-\infty<\xi<+\infty,\pi<\phi<\phi_0+\pi\bigr)$ and the corresponding cross-sectional area is $\displaystyle A_1 = 0.25 D^2 \big(\phi_0 - 0.5\sin2\phi_0\big)$. The light fluid (shown in red) is located in an infinite strip of $\displaystyle\bigl(-\infty<\xi<+\infty,\phi_0<\phi<\pi\bigr)$ and occupies the area of $\displaystyle A_2 = A - A_1$. The heavy phase holdup is then defined as $\displaystyle h = A_1/A=(\phi_0 - 0.5\sin2\phi_0)/\pi$, so that $h=0$ corresponds to single-phase air flow, while $h=1$ -- to single-phase water flow. On the other hand, the relative height of the lower water layer differs, in general, from the holdup and defined as $\displaystyle h_1 = H_1 / D = 0.5(1 - \cos\phi_0)$, where $H_1$ is the height of the water layer. The dimensionless interface length also varies with the holdup and equals to $l_{int}=\sin\phi_0$.
 
The governing equations for the velocity field are coupled with the no-slip conditions at the pipe wall, including the triple points ($\xi\to\pm\infty$)
\begin{equation} \label{Eq: BC_no-slip}
	\vec{u}^{(1)}(\phi=\phi_0 + \pi)
	= \vec{u}^{(2)}(\phi=\phi_0) 
	= 0,
\end{equation}
and the boundary conditions at the interface.

The absence of mass transfer across the interface is accounted by the kinematic boundary condition
\begin{equation}	\label{Eq: BC_kinematic}	
	u_\phi^{(k)}(\phi=\phi_\eta) = H_\phi \frac{\partial \eta}{\partial t}
	+ u_\xi^{(k)}(\phi=\phi_\eta) \frac{H_\phi}{H_\xi} \frac{\partial \eta}{\partial \xi}
	+ u_z^{(k)}(\phi=\phi_\eta) H_\phi \frac{\partial \eta}{\partial z},
\end{equation}
where the Lam\'e coefficients (scale factors) for the bipolar coordinates are defined as
\begin{equation} \label{Eq: Lame_bipolar}
	H_\xi = H_\phi 
	= \frac{1}{2} \frac{\sin\phi_0}{\cosh \xi - \cos \phi}.
\end{equation}
The velocity field is continuous across the interface, therefore
\begin{equation}
	\vec{u}^{(1)}(\phi=\phi_\eta) 
	= \vec{u}^{(2)}(\phi=\phi_\eta).
\end{equation} 

The interfacial boundary conditions require continuity of the tangential components of the viscous stress tensor, i.e., $\displaystyle\tau_{\xi \phi}$ in the plane $\displaystyle(\xi,\phi)$ and $\displaystyle\tau_{\phi z}$ in the plane $\displaystyle(\phi,z)$, while the discontinuity in the normal component, which is the sum of the normal viscous stress $\displaystyle\sigma_n$ and the pressure, is balanced by the surface tension
\begin{subequations}
	\begin{align}
		\tau_{\xi \phi}^{(1)}(\phi=\phi_\eta) 
		&= \tau_{\xi \phi}^{(2)}(\phi=\phi_\eta),
		\\
		\tau_{\phi z}^{(1)}(\phi=\phi_\eta) 
		&= \tau_{\phi z}^{(2)}(\phi=\phi_\eta),
		\\		
		\biggl\llbracket -p + \sigma_n \biggr\rrbracket_{\phi=\phi_\eta} 
		&= - \frac{1}{\We}\nabla \cdot \vec{n},	
	\end{align}
\end{subequations}
where $\vec{n}$ is a unit normal vector to the interface and double square brackets denote a jump in the corresponding quantity across the interface. The Weber number is defined as $\displaystyle\We=\rho_2 D U_m^2/\sigma$. Detailed description of the governing equations and boundary conditions can be found in \cite{Barmak23}.

\section{Base flow} \label{Sec: Base_flow}

In the stability analysis, the base flow is assumed to be steady, laminar, and fully-developed. The axial velocity is the only non zero component, and it depends only on the cross-sectional coordinates, $\displaystyle\vec{U^{(k)}}=\left(0,0,U_z^{(k)}(\xi,\phi)\right)$. Then base flow is defined by the steady-state momentum equation in the $z$-direction that reads

\begin{subequations} \label{Eq: Base_flow_eqs}
	\begin{align}
		\frac{1}{H_\phi H_\xi}
		\left[\frac{\partial}{\partial \xi}
		\left(\frac{H_\phi}{H_\xi}
		\frac{\partial U_z^{(k)}}{\partial \xi}\right)
		+\frac{\partial}{\partial \phi}
		\left(\frac{H_\xi}{H_\phi}
		\frac{\partial U_z^{(k)}}{\partial \phi}\right)\right]
		&= \frac{\mu_1}{\mu_{12} \mu_k}
		\frac{\partial P^{(k)}}{\partial z},						
		\\
		\frac{\partial P^{(1)}}{\partial z}
		&= \frac{\partial P^{(2)}}{\partial z}.
	\end{align}
\end{subequations}

The base flow velocity is subject to the no-slip conditions at the pipe wall
\begin{equation}
	U_z^{(1)} \left(\phi=\phi_0+\pi\right) = 0, \qquad
	U_z^{(2)} \left(\phi=\phi_0\right) = 0.
\end{equation}
The base flow velocity and shear stress are continuous at the interface
\begin{equation}
	U_z^{(1)}(\phi=\pi) = U_z^{(2)}(\phi=\pi),
\end{equation}
\begin{equation}
	\mu_{12} \bigg(\frac{\partial U_z^{(1)}}{\partial \phi}\bigg)_{\phi=\pi} 
	= \bigg(\frac{\partial U_z^{(2)}}{\partial \phi}\bigg)_{\phi=\pi}. 	
\end{equation}
For a plane unperturbed interface, continuity of the normal stress results in continuity of the pressure across the interface
\begin{equation}
	P^{(1)}(\phi=\pi) = P^{(2)}(\phi=\pi).
\end{equation}
Integration of the base-flow velocity profiles over the lower- and upper-layer cross-sectional areas yields a relative contribution of each fluid into the prescribed total volumetric flow rate, which can be written as a ratio of the corresponding superficial velocity to the mixture velocity
\begin{subequations} \label{Eq: BF_flow_rates}
	\begin{align}
		\int_{\pi}^{\phi_0 + \pi}\int_{-\infty}^{+\infty} U_z^{(1)} H_\xi H_\phi d\xi d\phi 
		&= \frac{U_{1S}}{U_m},
		\\
		\int_{\phi_0}^{\pi}\int_{-\infty}^{+\infty} U_z^{(2)} H_\xi H_\phi d\xi d\phi 
		&= \frac{U_{2S}}{U_m}.
	\end{align}
\end{subequations}

For a specific two-phase system with particular physical properties of fluids and pipe diameter, the holdup and base flow velocity profile are uniquely defined by the volumetric flow rate ratio, $\displaystyle q_{12} = Q_1/Q_2=U_{1S}/U_{2S}$ \cite[see, e.g.,][]{Barmak23}.

\section{Linear stability}

Linear stability of the base flow is studied with respect to infinitesimal perturbations. The velocity and pressure of the perturbed flow are
\begin{align}
	\vec{u^{(k)}} &= \vec{U^{(k)}}\bigl(\xi,\phi\bigr)
	+ \vec{u'^{(k)}}\bigl(\xi,\phi,z\bigr),
	\\
	p^{(k)} &= P^{(k)}\bigl(\xi,\phi\bigr) 
	+ p'^{(k)}\bigl(\xi,\phi,z\bigr),
\end{align}
while the perturbed interface is described as a surface $\displaystyle \phi = \pi + \eta\left(\xi,z,t\right)$, where $\eta$ denotes the deviation of the interface from the undisturbed plane state. The infinitesimal perturbations are presented as
\begin{equation} \label{Eq: Perturbation}
	\begin{pmatrix}
		\vec{u'^{(k)}} \\
		p'^{(k)} \\
		\eta
	\end{pmatrix}
	=
	\begin{pmatrix}
		\vec{\tilde{u}^{(k)}}(\xi,\phi)
		= \left[\tilde{u}_\xi^{(k)}(\xi,\phi),\tilde{u}_\phi^{(k)}(\xi,\phi),\tilde{u}_z^{(k)}(\xi,\phi)\right] 
		\\
		\tilde{p}^{(k)}(\xi,\phi)
		\\
		\tilde{\eta}(\xi)
	\end{pmatrix}
	e^{\left(i\alpha z + \lambda t\right)},
\end{equation}
where $\alpha$ is the real wavenumber and $\lambda=\lambda_{R} + i\lambda_{I}$ is the complex time increment of the perturbation. The dimensionless wave speed is then defined as $c=-\lambda_{I}/\alpha$.

The linearized equations for the perturbation amplitude (Eq.\ \ref{Eq: Perturbation}) are
\begin{equation} \label{Eq: Stability_continuity}
	\frac{1}{H_\xi}\frac{\partial \tilde{u}_\xi^{(k)}}{\partial \xi}
	+ \frac{1}{H_\phi} \frac{\partial \tilde{u}_\phi^{(k)}}{\partial \phi}
	- 2 \frac{\sinh\xi}{\sin\phi_0} \tilde{u}_\xi^{(k)}
	- 2 \frac{\sin\phi}{\sin\phi_0} \tilde{u}_\phi^{(k)}
	+ i \alpha \tilde{u}_z = 0,
\end{equation}
\begin{subequations}
	\begin{align} \label{Eq: Stability_xi}
		&\begin{aligned}
			\lambda \tilde{u}_\xi^{(k)}
			&+ i \alpha U_z^{(k)} \tilde{u}_\xi^{(k)}
			= - \frac{\rho_1}{\rho_{12} \rho_k} 
			\frac{1}{H_\xi}
			\frac{\partial \tilde{p}^{(k)}}{\partial \xi}
			\\
			&+ \frac{1}{\Rey}\frac{\rho_1}{\rho_{12} \rho_k} \frac{\mu_{12} \mu_k}{\mu_1} 	\Biggl[
			\frac{1}{H_\xi H_\phi}
			\frac{\partial}{\partial \xi} 
			\biggl(\frac{H_\phi}{H_\xi}\frac{\partial \tilde{u}_\xi^{(k)}}{\partial \xi}\biggr)
			+ \frac{1}{H_\xi H_\phi}
			\frac{\partial}{\partial \phi}
			\biggl(\frac{H_\xi}{H_\phi} 
			\frac{\partial \tilde{u}_\xi^{(k)}}{\partial \phi}\biggr)
			- \alpha ^2 \tilde{u}_\xi^{(k)}
			\\
			&- 4 \frac{\sin\phi}{\sin\phi_0}
			\frac{1}{H_\xi}\frac{\partial \tilde{u}_\phi^{(k)}}{\partial \xi} 	
			+ 4 \frac{\sinh\xi}{\sin\phi_0}
			\frac{1}{H_\phi} \frac{\partial \tilde{u}_\phi^{(k)}}{\partial \phi}
			+ 4 \frac{\cos^2\phi - \cosh^2\xi}{\sin^2\phi_0} 
			\tilde{u}_\xi^{(k)}
			\Biggr],
		\end{aligned}
	\end{align}
	\begin{align} \label{Eq: Stability_phi}
		&\begin{aligned}
			\lambda \tilde{u}_\phi^{(k)}
			&+ i \alpha U_z^{(k)} \tilde{u}_\phi^{(k)}			
			=
			- \frac{\rho_1}{\rho_{12} \rho_k} 
			\frac{1}{H_\xi}
			\frac{\partial \tilde{p}^{(k)}}{\partial \phi}
			\\
			&+ \frac{1}{\Rey}\frac{\rho_1}{\rho_{12} \rho_k} \frac{\mu_{12} \mu_k}{\mu_1} 
			\Biggl[
			\frac{1}{H_\xi H_\phi}
			\frac{\partial}{\partial \xi}
			\biggl(\frac{H_\phi}{H_\xi}\frac{\partial \tilde{u}_\phi^{(k)}}{\partial \xi}\biggr)
			+ \frac{1}{H_\xi H_\phi}
			\frac{\partial}{\partial \phi}	\biggl(
			\frac{H_\xi}{H_\phi}
			\frac{\partial \tilde{u}_\phi^{(k)}}{\partial \phi}\biggr)	
			- \alpha ^2 \tilde{u}_\phi^{(k)}
			\\
			&+ 4 \frac{\sin\phi}{\sin\phi_0}
			\frac{1}{H_\xi} 
			\frac{\partial \tilde{u}_\xi^{(k)}}{\partial \xi}
			- 4 \frac{\sinh\xi}{\sin\phi_0}
			\frac{1}{H_\phi}
			\frac{\partial \tilde{u}_\xi^{(k)}}{\partial \phi}
			+ 4 \frac{\cos^2\phi - \cosh^2\xi}{\sin^2\phi_0}
			\tilde{u}_\phi^{(k)}
			\Biggr],
		\end{aligned}
	\end{align}
	\begin{align} \label{Eq: Stability_z}
		&\begin{aligned}
			\lambda \tilde{u}_z^{(k)}
			&+ \frac{\tilde{u}_\xi^{(k)}}{H_\xi}\frac{\partial U_z^{(k)}}{\partial \xi}
			+ \frac{\tilde{u}_\phi^{(k)}}{H_\phi}\frac{\partial U_z^{(k)}}{\partial \phi}			
			+ i \alpha U_z^{(k)} \tilde{u}_z^{(k)}
			= - i \alpha \frac{\rho_1}{\rho_{12} \rho_k} \tilde{p}^{(k)}
			\\
			&+ \frac{1}{\Rey} \frac{\rho_1}{\rho_{12} \rho_k} \frac{\mu_{12} \mu_k}{\mu_1} \Biggl[
			\frac{1}{H_\xi H_\phi}
			\frac{\partial}{\partial \xi}
			\left(\frac{H_\phi}{H_\xi} \frac{\partial \tilde{u}_z^{(k)}}{\partial \xi}\right)
			+ \frac{1}{H_\xi H_\phi} 
			\frac{\partial}{\partial \phi}
			\left(\frac{H_\xi}{H_\phi} \frac{\partial \tilde{u}_z^{(k)}}{\partial \phi}\right)
			\\
			&- \alpha ^2 \tilde{u}_z^{(k)}\Biggr].
		\end{aligned}
	\end{align}
\end{subequations}

The no-slip condition at the pipe wall is formulated for the perturbation amplitude (Eq.\ \ref{Eq: Perturbation}) as
\begin{equation} \label{Eq: BC_no-slip_lin}
	\vec{\tilde{u}}^{(1)}(\phi=\phi_0 + \pi) 
	= \vec{\tilde{u}}^{(2)}(\phi=\phi_0)
	= 0.
\end{equation}

To derive the linearized form of the interfacial boundary conditions, we use the Taylor expansion around the unperturbed interface $\phi=\pi$. The kinematic boundary condition (Eq.\ \ref{Eq: BC_kinematic}) for the perturbed flow reads
\begin{equation} \label{Eq: BC_kinematic_lin}
	\lambda H_\phi \tilde{\eta}
	= \tilde{u}_\phi^{(1)}
	- i \alpha U_z^{(1)} (\phi=\pi) H_\phi \tilde{\eta}
	= \tilde{u}_\phi^{(2)}
	- i \alpha U_z^{(2)} (\phi=\pi) H_\phi \tilde{\eta}.
\end{equation}
The continuity of velocity components is
\begin{subequations} \label{Eq: BC_continuity_velocity}
	\begin{align} 
		\tilde{u}_\xi^{(1)}(\phi=\pi) 
		&= \tilde{u}_\xi^{(2)}(\phi=\pi),
		\\
		\tilde{u}_\phi^{(1)}(\phi=\pi) 
		&= \tilde{u}_\phi^{(2)}(\phi=\pi),
		\\
		\tilde{u}_z^{(1)}(\phi=\pi)
		&= \tilde{u}_z^{(2)}(\phi=\pi) 
		+ \bigg(\frac{1}{H_\phi} \frac{\partial U_z^{(2)}}{\partial \phi} 
		\biggr|_{\phi=\pi}
		- \frac{1}{H_\phi} \frac{\partial U_z^{(1)}}{\partial \phi} 
		\biggr|_{\phi=\pi}\bigg) 
		H_\phi \tilde{\eta}.
	\end{align}
\end{subequations}
The interfacial boundary conditions require also continuity of the tangential stresses in planes $\bigl(\xi,\phi\bigr)$ and $\bigl(\phi,z\bigr)$ and a jump in the normal shear stress. After linearization, they read
\begin{subequations} \label{Eq: BC_shear_stress_lin}
	\begin{align} 
		&\begin{aligned}
			&\mu_{1 2}\frac{1}{H_\phi}\frac{\partial \tilde{u}_\xi^{(1)}}{\partial \phi}\Biggr|_{\phi = \pi}
			- \frac{1}{H_\phi}\frac{\partial \tilde{u}_\xi^{(2)}}{\partial \phi}\Biggr|_{\phi = \pi}
			+ (\mu_{1 2}-1) \frac{1}{H_\xi}\frac{\partial \tilde{u}_\phi}{\partial \xi}\Biggr|_{\phi = \pi}
			\\
			&- i \alpha (\mu_{1 2}-1) \frac{1}{H_\xi} \frac{\partial U_z}{\partial \xi} H_\phi \tilde{\eta} \Biggr|_{\phi = \pi}
			+ 2 (\mu_{1 2}-1)  \frac{\sinh\xi}{\sin\phi_0} \tilde{u}_\phi \left(\phi = \pi\right)
			= 0,
		\end{aligned}
		\\
		&\begin{aligned}
			&i \alpha \left(\mu_{1 2}-1\right) \tilde{u}_\phi (\phi = \pi)
			- (\mu_{1 2}-1) \frac{1}{H_\xi}
			\frac{\partial U_z}{\partial \xi} \frac{1}{H_\xi} \frac{\partial \big(H_\phi \tilde{\eta}\big)}{\partial \xi} \Biggr|_{\phi = \pi}
			\\
			&+ \Biggl[\mu_{1 2} \frac{1}{H_\phi} \frac{\partial}{\partial \phi} \biggl(\frac{1}{H_\phi}\frac{\partial U_z^{(1)}}{\partial \phi}\biggr)
			- \mu_{1 2} \frac{1}{H_\xi} \frac{\partial U_z^{(1)}}{\partial \xi} \frac{2 \sinh\xi}{\sin\phi_0} 
			\\
			&- \frac{1}{H_\phi} \frac{\partial}{\partial \phi} \biggl(\frac{1}{H_\phi}\frac{\partial U_z^{(2)}}{\partial \phi}\biggr)
			+ \frac{1}{H_\xi}
			\frac{\partial U_z^{(2)}}{\partial \xi} \frac{2 \sinh\xi}{\sin\phi_0} \Biggr]_{\phi = \pi} H_\phi \tilde{\eta}
			+ \biggl(\mu_{1 2} \frac{1}{H_\phi} \frac{\partial \tilde{u}_z^{(1)}}{\partial \phi} 
			- \frac{1}{H_\phi} \frac{\partial \tilde{u}_z^{(2)}}{\partial \phi}\biggr)\Biggr|_{\phi = \pi}
			= 0,
		\end{aligned}
		\\
		&\begin{aligned}
			&\left(\tilde{p}^{(2)} - \tilde{p}^{(1)}\right)\biggr|_{\phi = \pi}
			+ \frac{1}{\Fr} \left(\rho_{1 2}-1\right)
			\left(-\cosh\xi + \frac{\sinh^2\xi}{\cosh\xi -\cos\phi}\right) 
			H_\phi \tilde{\eta} \Biggr|_{\phi = \pi}
			\\
			&+ \frac{2}{\Rey} \biggl(
			\frac{\mu_{1 2}}{H_\phi}\frac{\partial \tilde{u}_\phi^{(1)}}{\partial \phi}	
			- \frac{1}{H_\phi}\frac{\partial \tilde{u}_\phi^{(2)}}{\partial \phi}\biggr) \Biggr|_{\phi = \pi}
			- \frac{4}{\Rey} \left(\mu_{1 2}-1\right) \frac{\sinh\xi}{\sin\phi_0} \tilde{u}_\xi \Biggr|_{\phi = \pi}
			\\
			&= - \frac{1}{\We} \biggl[\frac{1}{H_\xi}
			\frac{\partial}{\partial \xi}\left(\frac{H_\phi}{H_\xi} \frac{\partial \tilde{\eta}}{\partial \xi}\right)
			- 2 \frac{H_\phi}{H_\xi} \frac{\partial \tilde{\eta}}{\partial \xi}
			\frac{\sinh\xi}{\sin\phi_0}
			- \alpha ^2 H_\phi \tilde{\eta}\biggr].
		\end{aligned}
	\end{align}
\end{subequations}

\section{Numerical method} \label{Sec: Numerics}

The stability problem formulated in the previous section is solved using the finite--volume method. We use a staggered quadrilateral grid, whose faces coincide with the coordinate lines in the bipolar coordinates, so that the scalar values of pressure $p$, the base flow velocity $U_z^{(k)}$, and the axial velocity of the perturbation $u_z^{(k)}$ are calculated in the cell centers defined by the integer indices $\big[\xi_i,\phi_j\big]$, where $i=0,1,...,N_\xi$ and $j=0,1,...,N_\phi$. In the following, the tilde marks  are omitted above the perturbation amplitude. The two other components of the perturbation velocity, $u_\xi^{(k)}$ and $u_\phi^{(k)}$, are calculated on the cell faces $\displaystyle\big[\xi_{i+1/2},\phi_j\big]$ and $\displaystyle\big[\xi_i,\phi_{j+1/2}\big]$, respectively, where $\displaystyle\xi_{i+1/2} = \big(\xi_i + \xi_{i+1}\big)/2$ and $\displaystyle\phi_{j+1/2} = \big(\phi_j + \phi_{j+1}\big)/2$. The holdup is found as a function of the specified input flow rates, so that the mesh distribution in the $\phi$-direction is changing in the process of the base flow solution. The number of cells in each sublayer (phase) is then proportional to its relative height ($N_\text{lower}/N_\phi = (\pi-\phi_0)/\pi$). 

The hyperbolic tangent stretching is used in the $\xi$-direction
\begin{equation} \label{Eq: tanh_stretching}
	\xi \to \xi_\text{max}\bigg(\frac{\tanh\big[a_\xi\big(\xi - 1\big)\big]}{\tanh(a_\xi)}
	+ \frac{\tanh(a_\xi \xi)}{\tanh(a_\xi)}\bigg),
\end{equation}
where $\xi$ is originally uniformly distributed grid cells and $a_\xi=5$ is the stretching parameter that determines the degree of clustering, so that the cells are redistributed to be more dense near the midplane (around $\xi=0$). In the $\phi$-direction, the grid is stretched using the $sin$-function near the pipe walls and at both sides of the interface as following  
\begin{equation} \label{Eq: sin_stretching}
	\phi \to \phi - a_\phi \sin \big(2 \pi \phi\big),
\end{equation} 
where $a_\phi=0.12$. In \cite{Barmak23}, it was demonstrated that the number of computational cells in each direction sufficient to obtain solution converged at least in three decimal digits is equal to $\displaystyle N_\xi=N_\phi=200$ for the considered problem.

The discretization of the linearized governing equations (Eqs.\ \ref{Eq: Stability_continuity} - \ref{Eq: Stability_z}) and boundary conditions(Eqs.\ \ref{Eq: BC_no-slip_lin} - \ref{Eq: BC_shear_stress_lin}) reduces the linear stability problem to a generalized eigenvalue problem
\begin{equation} \label{Eq: Eigenvalue_problem}
	\lambda \cdot \vec{B} 
	\cdot \vec{v}
	= \vec{J} 
	\cdot \vec{v},
\end{equation}
where $\lambda = \lambda_{R} + i \lambda_{I}$ is the eigenvalue and the corresponding eigenvector is the perturbation amplitude, $\displaystyle\vec{v}= \big(u_\xi,u_\phi,u_z, p,\Delta\eta\big)^T$. $\vec{B}$ is a diagonal matrix with the elements corresponding to time derivatives of $\tilde{u}$ and $\eta$ equal to one. On the other hand, the diagonal elements corresponding to $p$ and the boundary conditions without time derivatives (all except for the kinematic one) are zeros, so that $\det \vec{B} = 0$. $\vec{J}$ is the Jacobian matrix defined by the right-hand-side of the linearized equations. The real part of the eigenvalue, $\lambda_{R}$ determines the growth rate of the perturbation, so that the flow is considered unstable when there exists an eigenvalue with a positive real part. Considering all possible wavenumbers, the eigenvalue with the largest $\lambda_{R}$ and the corresponding eigenvector are referred as leading or the most unstable one.

The generalized eigenvalue problem \ref{Eq: Eigenvalue_problem} is solved by the Arnoldi iteration in the shift-and-inverse mode using the ARPACK package of \cite{Lechouq98} in FORTRAN
\begin{equation}
	(\vec{J} - \lambda_0 \cdot \vec{B})^{-1} 
	\cdot \vec{B} 
	\cdot \vec{v}
	= \vartheta 
	\cdot \vec{v},
\end{equation}
where $\displaystyle\vartheta = 1 /\bigl(\lambda - \lambda_0\bigr)$ and $\lambda_0$ is a complex shift. More details on the present numerical approach and its validation can be found in \cite{Barmak23}.

\section{Results and discussion}

There are six dimensionless parameters that govern the stability problem of horizontal two-phase stratified flow in a circular pipe. These parameters include the flow rate ratio, $q_{12} = U_{1S}/U_{2S}$, and the viscosity ratio, $\mu_{12}$, that uniquely define the base flow solution, i.e. the heavy phase holdup and dimensionless velocity profile. Four additional dimensionless parameters are required to determine the flow stability, such as the density ratio, $\rho_{12}$, the Reynolds number of one of the phases, e.g., $\Rey^{(2)}$, the Froude number, $\Fr$, and the Weber number, $\We$. For such a large parameter space, an overall parametric analysis seems unfeasible. Therefore, we focus on air--water flows, thereby fixing $\mu_{12}$ and $\rho_{12}$. 

In the following, we study the flow stability with respect to all possible three-dimensional infinitesimal disturbances, using the numerical approach presented and verified in \cite{Barmak23}. We vary the superficial velocity of air and water to cover the whole range of holdups starting from very thin film of the heavy phase (low holdup) and up to very thin film of the light phase (high holdup).  For the sake of easier comparison with experimental data and further physical interpretations, the obtained results are summarized in form of the stability boundaries on a flow pattern map with the axes of superficial velocities of each phase.

\subsection{Case study}

Air--water flows are associated with large density and viscosity ratios, e.g., $\displaystyle\rho_{1 2} = 1000$ and $\displaystyle\mu_{1 2} = 55$, respectively. The case study is a flow in a horizontal circular pipe of diameter $D=0.05$m. 

\begin{figure}[h!]
	\centering
	\subfloat[$h=0.1$]{\includegraphics[width=0.32\textwidth,clip]{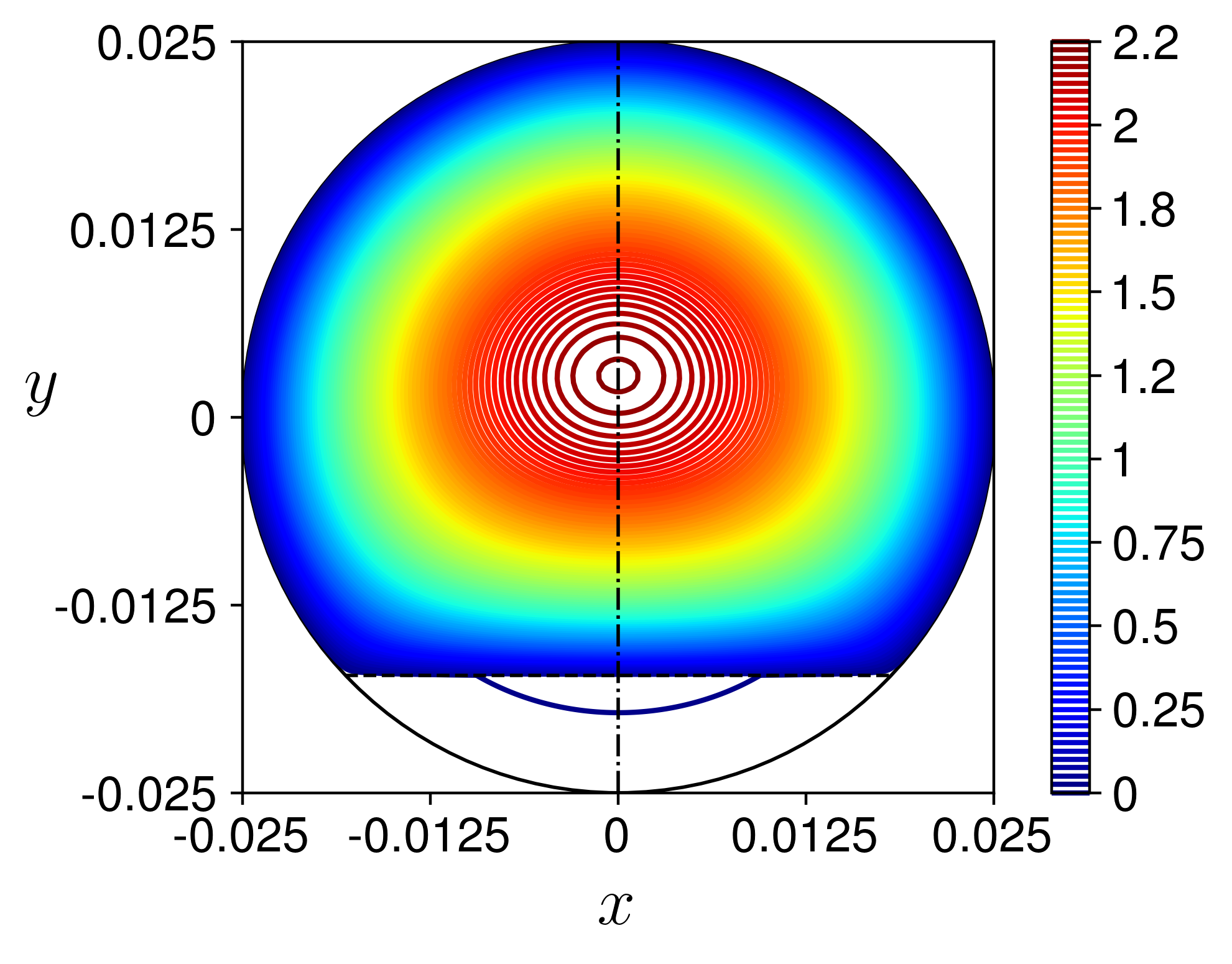}}
	\subfloat[$h=0.52$]{\includegraphics[width=0.32\textwidth,clip]{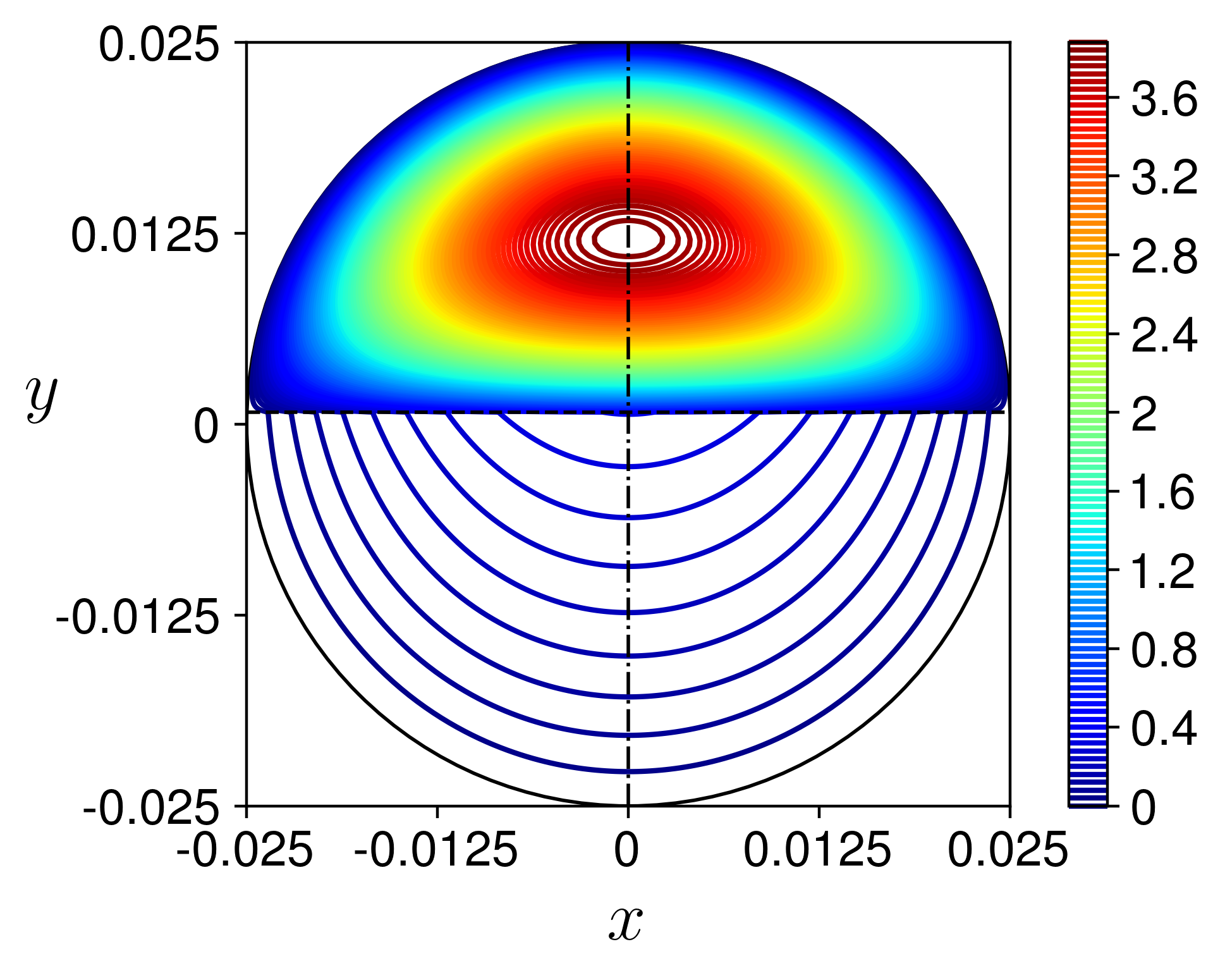}}
	\subfloat[$h=0.95$]{\includegraphics[width=0.32\textwidth,clip]{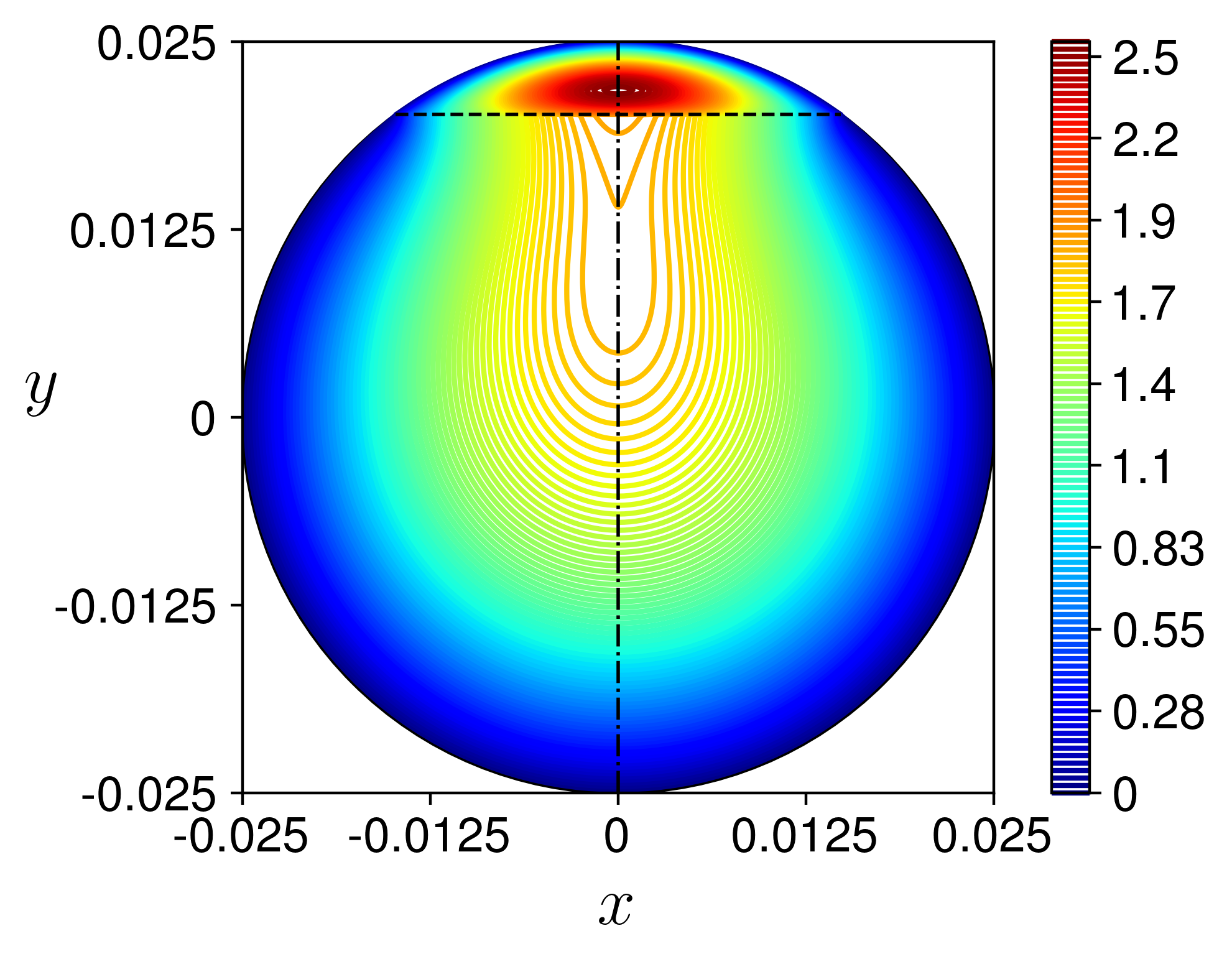}}
	\caption{\label{Fig: Base_flow}Base flow velocity, $U_z$. Air--water pipe flow, $D=0.05$m. (a) $h=0.1$; (b) $h=0.52$; (c) $h=0.95$. The unperturbed interface is denoted by horizontal dashed black line; the cross-section centerline -- vertical dash-dot black line.}	
\end{figure}

\begin{figure}[h!]
	\centering
	\subfloat[$U_z$, $x=0$]{\includegraphics[width=0.33\textwidth,clip]{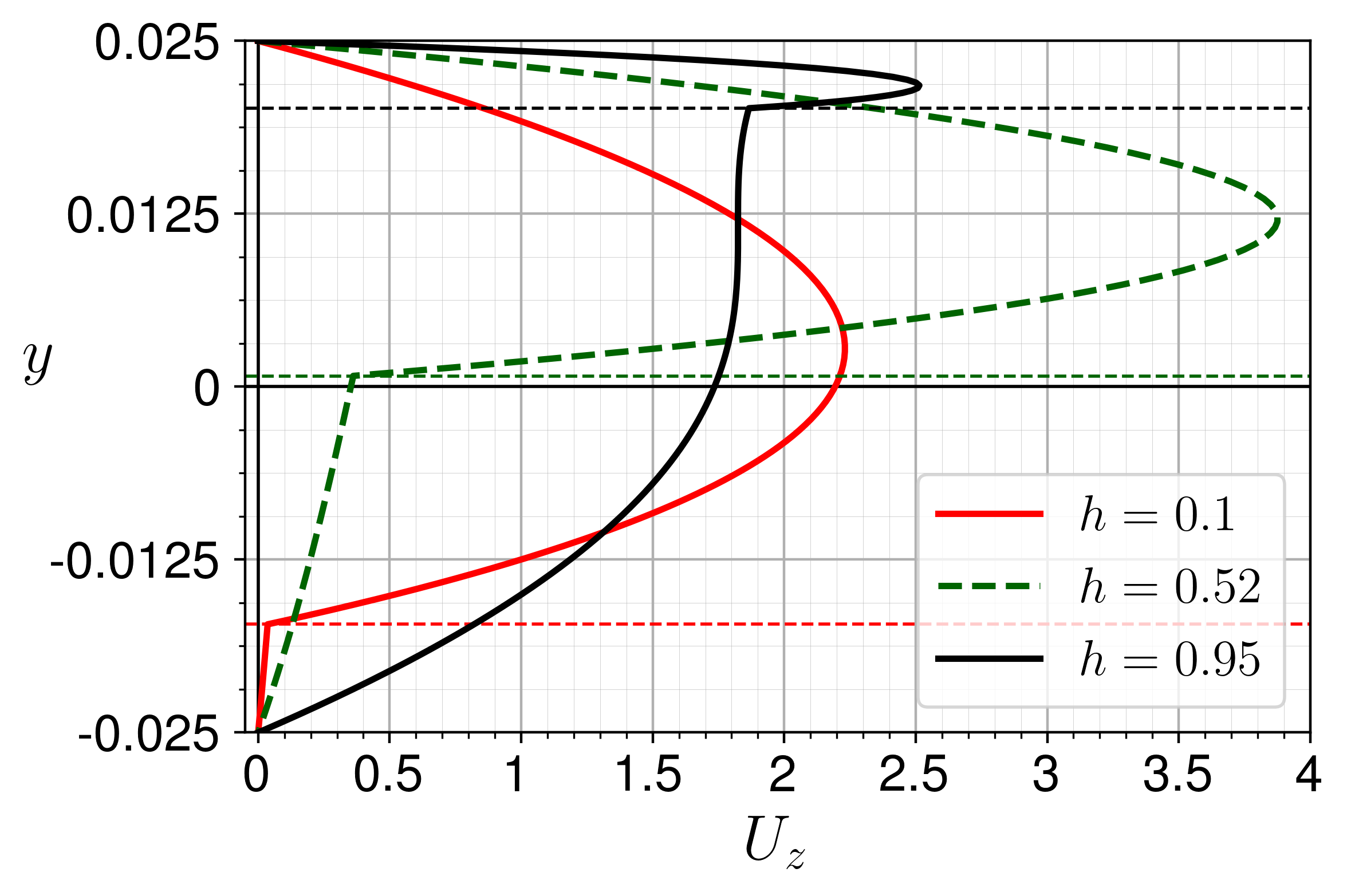}}
	\subfloat[$U_z$, air--water interface]{\includegraphics[width=0.33\textwidth,clip]{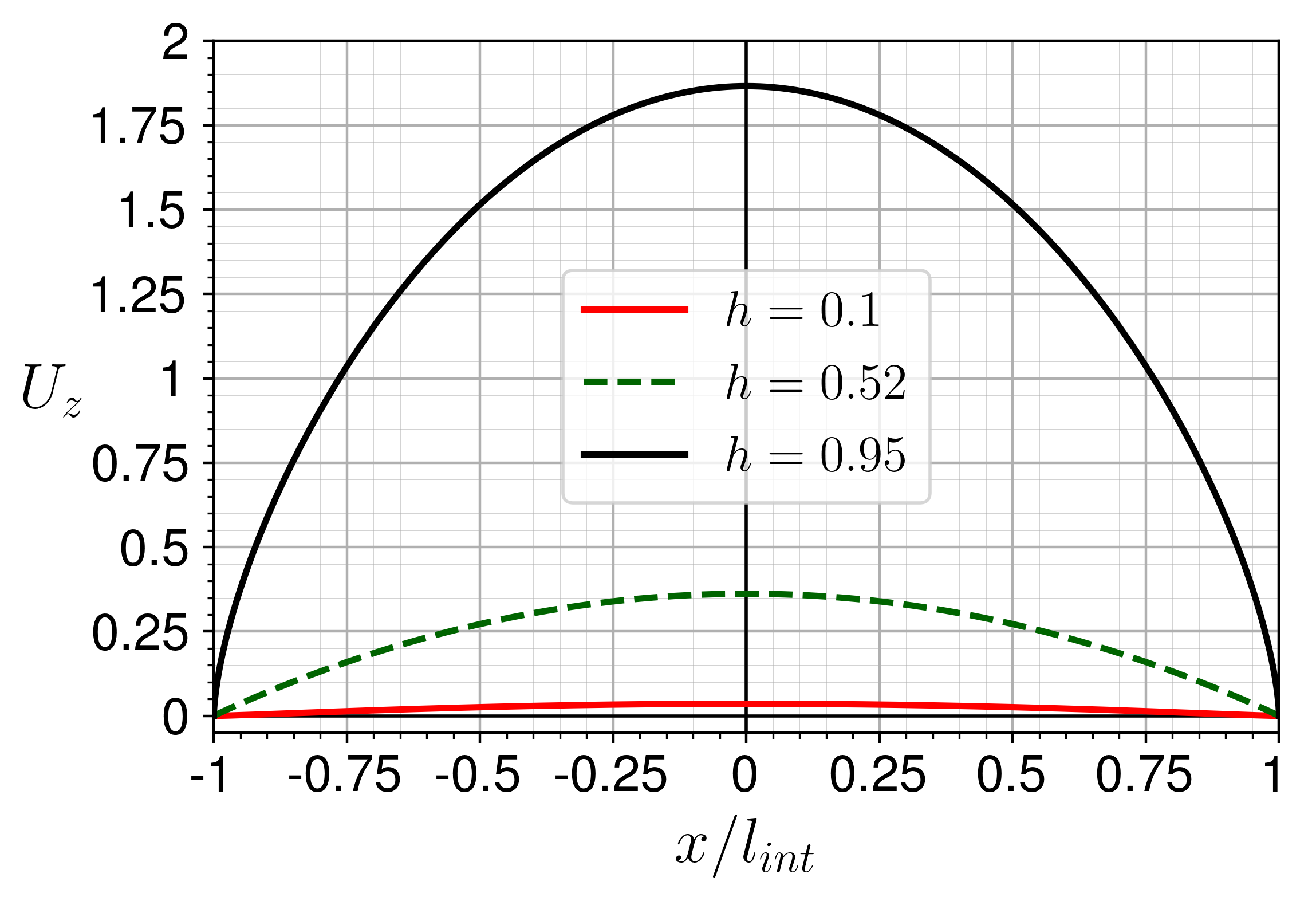}}
	\subfloat[Interfacial shear stress]{\includegraphics[width=0.33\textwidth,clip]{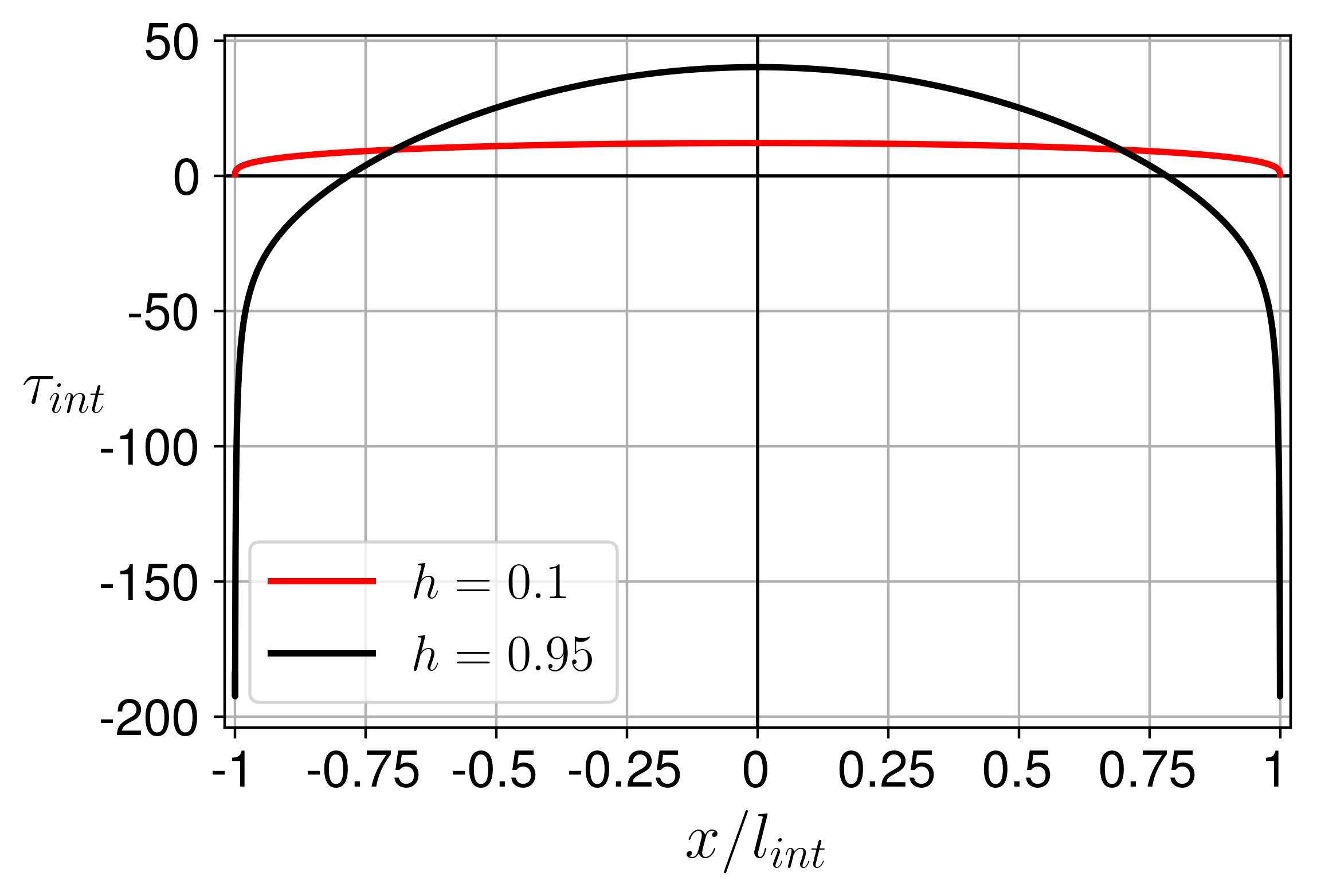}}
	\caption{\label{Fig: Base_flow_linegraphs}Base flow. Air--water pipe flow, $D=0.05$m. (a) Velocity profiles at $x=0$. The locations of the unperturbed interface is denoted by corresponding horizontal dashed lines. (b) Velocity profiles at the interface. (c) Interfacial shear stress.}	
\end{figure}

For a given two-phase system, the base flow with a smooth interface is uniquely defined by the ratio of superficial velocities, $\displaystyle U_{1S}/U_{2S}$. In particular, the water holdup (i. e., relative cross-sectional area occupied by the lower (water) layer), can be found. The dimensionless base velocity (scaled by the mixture velocity, $U_m$) is also uniquely defined by the flow rate ratio. Its contours are shown in Fig.\ \ref{Fig: Base_flow} for holdups $0.1$, $0.52$, and $0.95$. For these holdups, the base flow velocity profiles are shown at the vertical centerline ($x=0$) and at the interface between air and water in Fig.\ \ref{Fig: Base_flow_linegraphs}a and b, respectively.  

The water flow rate is less than that of air ($U_{1S}<U_{2S}$) for a wide range of holdups, up to $\displaystyle h\approx0.81$. For all water holdups, even for very high, the maximum of the base flow velocity is reached at the air phase, on the vertical centerline of the cross section (Fig.\ \ref{Fig: Base_flow}). Since the base flow velocity distribution in horizontal pipe does not depend on densities of the phases (see sec.\ \ref{Sec: Base_flow}), this is a result of high viscosity ratio between the water and air. In case of a very thin air layer, $h\ge0.95$ ($\displaystyle U_{1S}/U_{2S}\ge12.885$), there is a secondary velocity maximum at the interface, and in its vicinity the water velocity gradients in the vertical direction are small (e.g., Fig.\ \ref{Fig: Base_flow}c and black line Fig.\ \ref{Fig: Base_flow_linegraphs}a for $h=0.95$). The maximum of the dimensionless base flow velocity, $U_z$, is reached for the holdup of about $0.75$ and is equal $U_{\max}\approx5.14$ (cf. $U_{\max}\approx3.61$ for $h=0.52$ in Fig.\ \ref{Fig: Base_flow}b and dashed green line in Fig.\ \ref{Fig: Base_flow_linegraphs}a). In a circular pipe, the interface length varies strongly with the holdup (see Sec.\ \ref{Sec: Formulation}) and equals to $0.72D$, $0.99D$, and $0.59D$ for the three considered holdups. Therefore, the profiles at the interface are shown with respect to coordinate $x$ scaled by $l_{int}$.

As shown in \cite{Goldstein15} and further investigated in \cite{Goldstein21a,Goldstein21b}, the solution for the base flow predicts very large and diverging wall and interfacial shear stresses upon approaching the triple point, when the contact angle between the interface and the pipe wall as seen from the less viscous phase is sharp (i.e., $\pi-\phi_0<\pi/2$). In case of the air-water flow, it happens when the water occupies more than half the pipe, i.e., $h>0.5$ and $\phi_0 >\pi/2$. This can be observed in Fig. \ref{Fig: Base_flow_linegraphs}c for $h=0.95$, when the interfacial shear stress exerted by the air on the water, $\tau_{int}$, achieves very large negative values (in fact, $\tau_{int}\to-\infty$) for $x\to\pm1$. For small water holdups, e.g., $h=0.1$ (red line in Fig.\ \ref{Fig: Base_flow_linegraphs}c), the faster air phase drags the water all along the interface, so that $\tau_{int}>=0$. Once $h>0.5$, there is a region near the triple point, where the air is pulled by the water phase ($\tau_{int}<0$). With higher holdup, this region expands to the pipe center. It was found that although the values of shear stress diverge (go to infinity), their integral around the triple point, i.e., the friction force, stays finite. Our numerical solution in bipolar coordinates allows us to reproduce the analytically obtained exponential growth of the interfacial shear stresses \cite{Goldstein21a}, when the cutoff length $\xi_{\max}>3$.

\begin{figure}[h!]
	\centering
	\includegraphics[width=0.6\textwidth,clip]{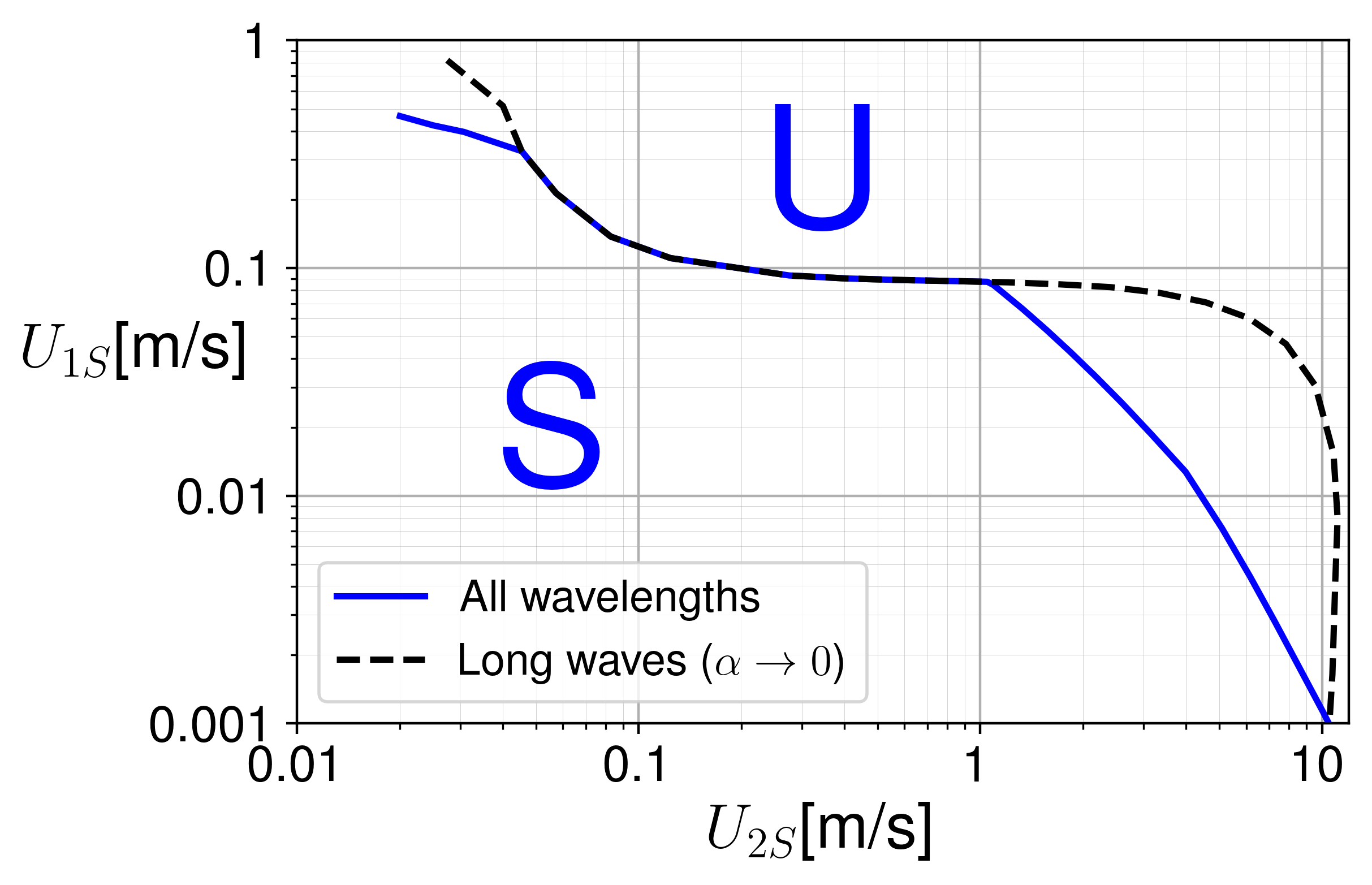}
	\caption{\label{Fig: Stability_map_D_0d05}Stability boundaries for all wavelengths (solid blue line) and long-wave (dashed black line) perturbations of the air-water flow in a pipe of diameter $D=0.05$m.}	
\end{figure}

\begin{figure}[h!]
	\centering
	\includegraphics[width=0.6\textwidth,clip]{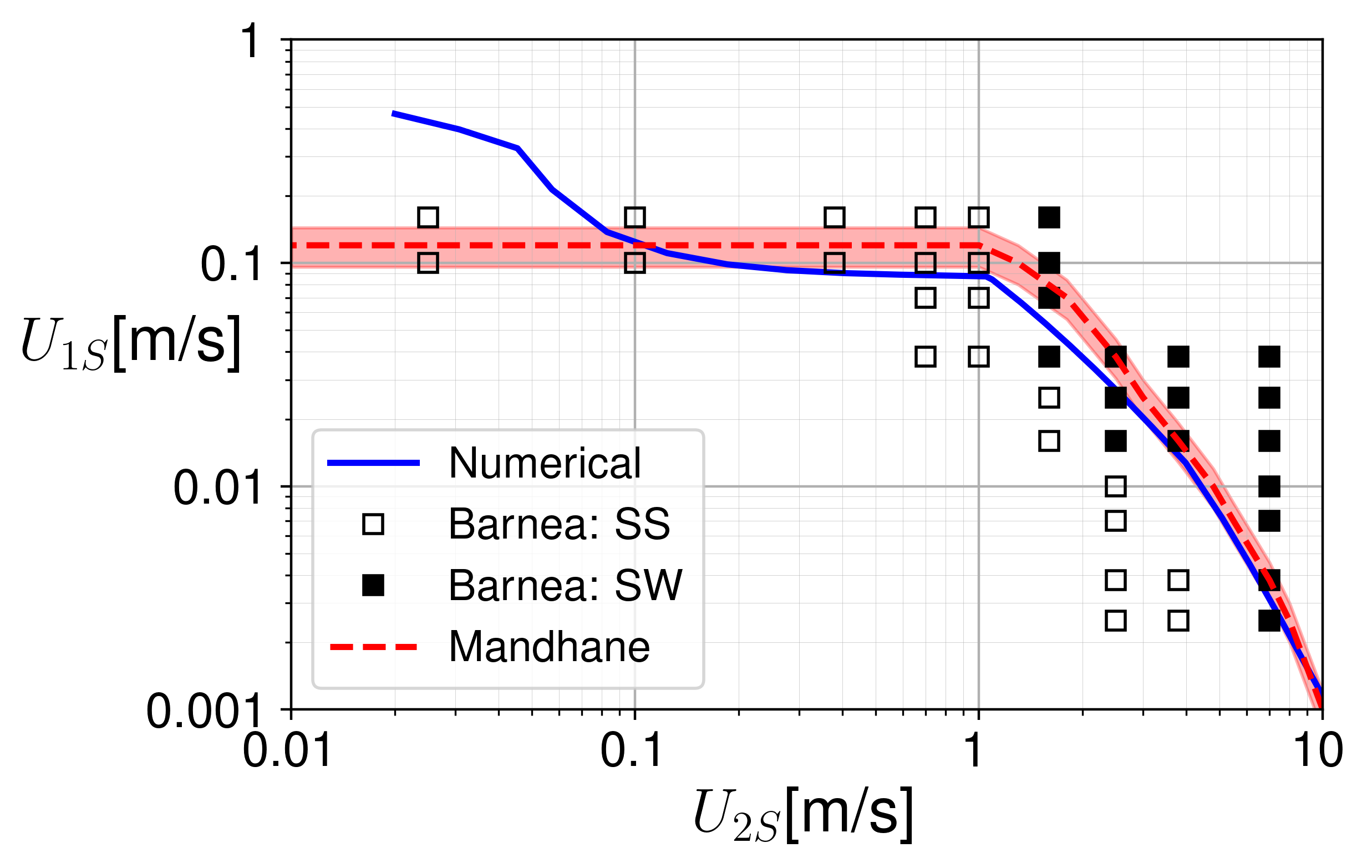}
	\caption{\label{Fig: Numerics_vs_experiment}Stability boundary predicted by the present numerical analysis (solid blue line, same as in Fig.\ \ref{Fig: Stability_map_D_0d05}) and the experimental data of \cite{Mandhane74} and \cite{Barnea82} for the air-water flow in a pipe of diameter $D=0.05$m.}	
\end{figure}

\begin{figure}[h!]
	\centering
	\subfloat[Critical water superficial velocity]{\includegraphics[width=0.49\textwidth,clip]{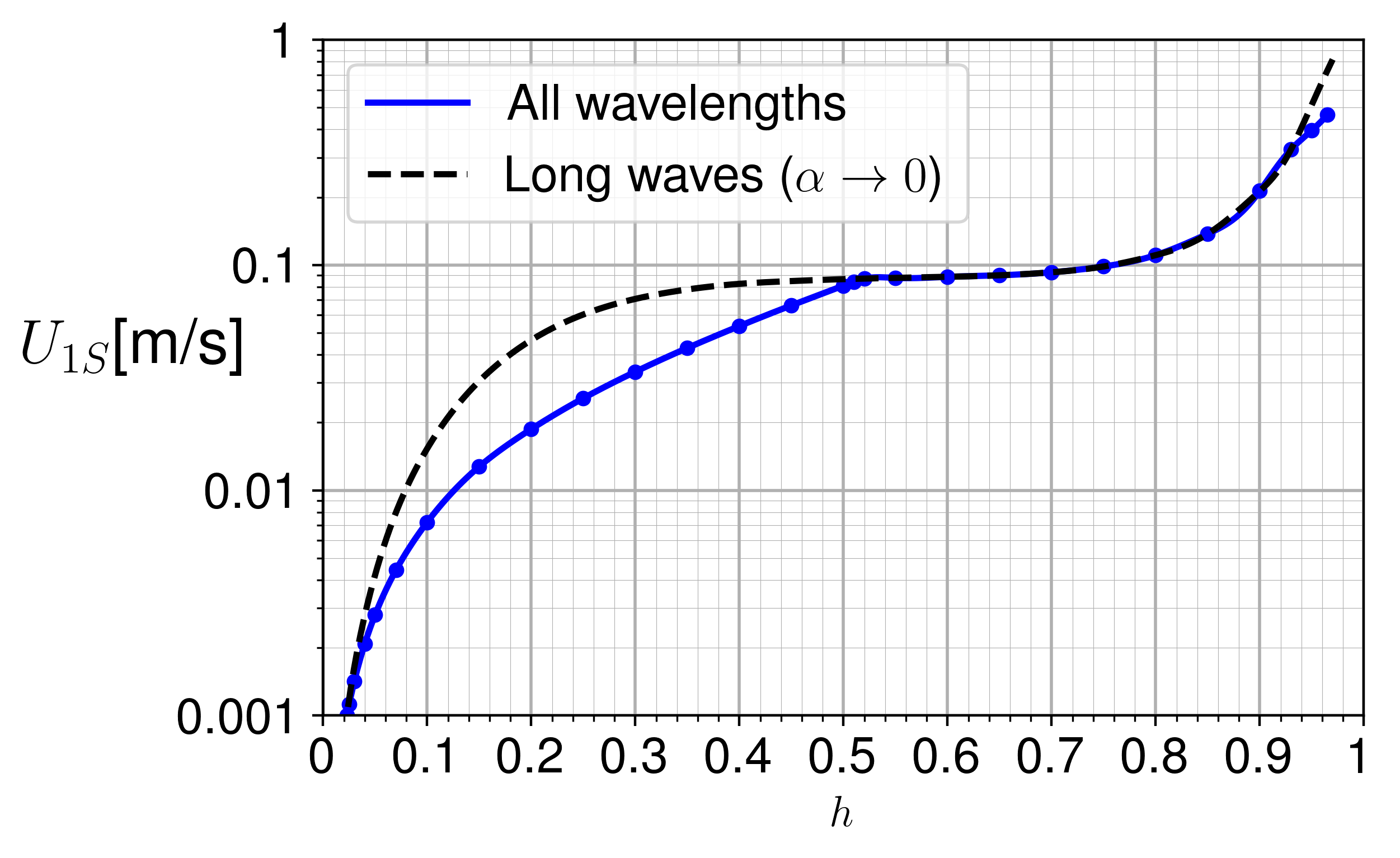}}
	\subfloat[Critical wave parameters]{\includegraphics[width=0.45\textwidth,clip]{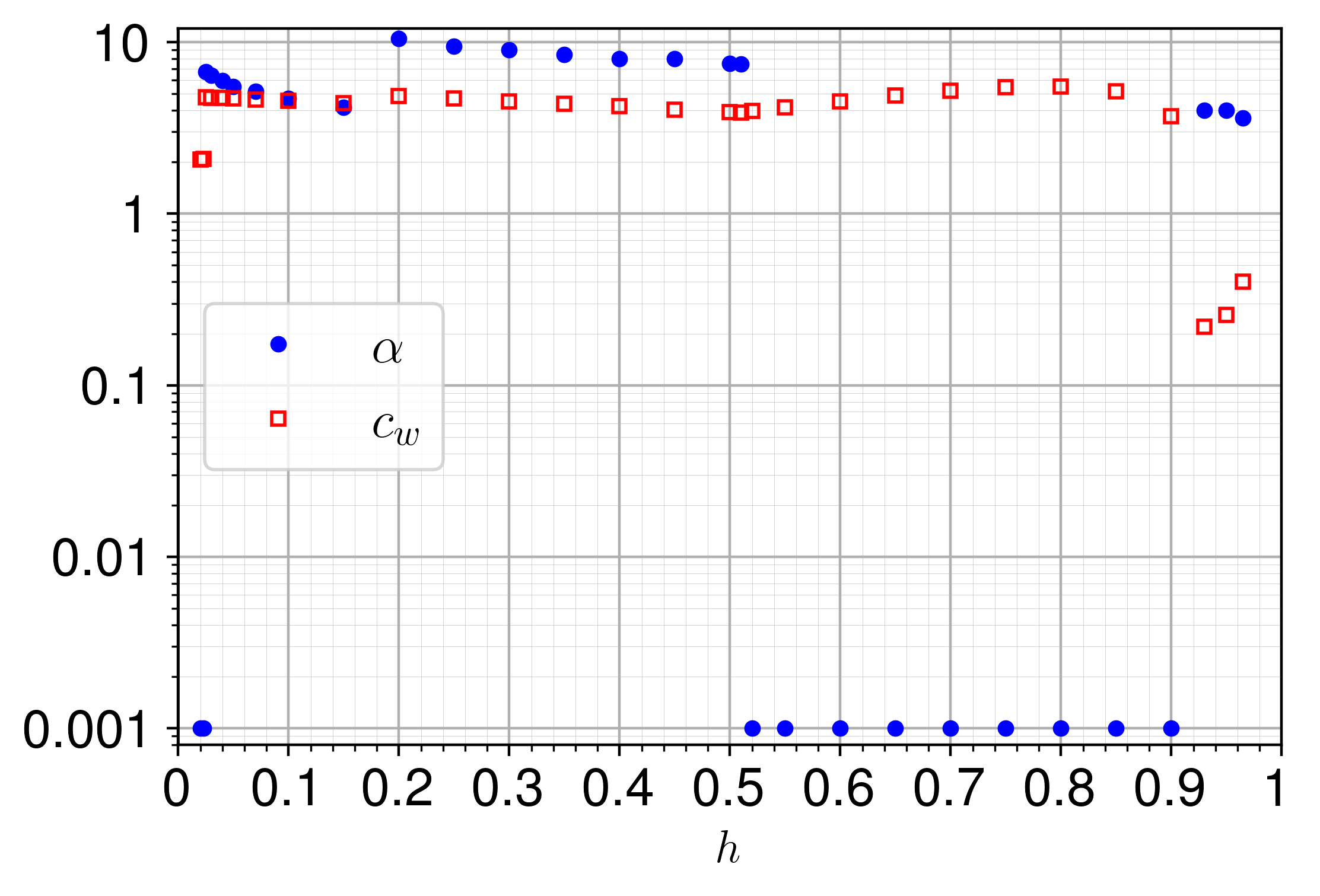}}
	\caption{\label{Fig: Stability_boundary_holdup}Critical parameters along the stability boundary (Fig.\ \ref{Fig: Stability_map_D_0d05}) as functions of holdup: (a) Critical superficial velocity of water, $U_{1S}$; (b) wavenumber (blue circles) and wave speed scaled by the average water base-flow velocity $c_w=c_\text{crit} U_m h/ U_{1S}$ (red squares) for some of the holdups.}	
\end{figure}

The critical point on the neutral stability boundary for a particular holdup is found by increasing both superficial velocities, starting from very small values, while keeping their ratio constant, until the real part of the leading eigenvalue for any wavenumber $\alpha$ changes sign from negative to positive, i.e., becomes unstable. At the critical point, the perturbations of all wavenumbers $\alpha$ are damped in time ($\lambda_{R} < 0$), except for the critical wavenumber $\alpha_\text{crit}$, for which $\max(\lambda_{R}) = 0$, meaning that the perturbation is neutrally stable. By identifying the critical superficial velocities for a wide range of water holdups ($h \in [0.02,0.97]$), the stability boundary of smooth stratified flow pattern is found with respect to perturbations of all wavenumbers. It is shown in Fig.\ \ref{Fig: Stability_map_D_0d05} as a solid blue curve on the flow pattern map with the air and water superficial velocities on the axes. The stable region for smooth-stratified flow obtained for relatively low superficial velocities is denoted by "S", while the unstable one for larger $U_{1S}$ and $U_{2S}$ is denoted by "U". Of particular interest is long-wave perturbation, which is the basic assumption in some mechanistic models, e.g., the Two-Fluid model. It is observed that the stability boundary with respect to long-wave perturbations, which is depicted by a dashed black line, overpredicts the stable region.

Experimental flow-pattern data are available for the pipe diameter chosen for the case study \cite[e. g., ][]{Mandhane74,Barnea82}. In \cite{Barnea82}, the pipe diameter was slightly different, i.e., $D=2''\approx5.1$cm. Moreover, the flow patterns in the experiments were determined by visual observations with relatively large steps in superficial velocities (see the experimental points depicted in Fig.\ \ref{Fig: Numerics_vs_experiment}). Therefore, only the qualitative comparison between theoretical and experimental results is possible. For the range of $U_{1S}\in[0.002,0.1]$, a good agreement between the critical $U_{2S|\text{crit}}$ obtained in the stability analysis and the experimentally found transition from the stratified-smooth to stratified-wavy flow patterns \cite[Fig. 2 in ][]{Barnea82}. The numerical results are also close to the average distribution of the large experimental datasets of \cite{Mandhane74}.

{\renewcommand{\arraystretch}{1.5}
	\begin{table}[h!]
		\caption{\label{Tab: Critical parameter}Critical parameters for air--water flow in a circular pipe $D=0.05$m. $U_{int}$ is the dimensionless interfacial velocity at $x=0$.}				
		\centering		
		\begin{tabular}{|c|c|c|c|c|c|}
			\hline\hline
			$U_{1S|\text{crit}}$ & $U_{2S|\text{crit}}$ & $h$ & $U_{int}$ & $\alpha_\text{crit}$ & $c_\text{crit}$ 
			\\
			$\bigg[\dfrac{m}{s}\bigg]$ & $\bigg[\dfrac{m}{s}\bigg]$ & 
			& & &
			\\
			\hline
			0.000781 & 10.410 & 0.02 & 0.009593 & $\to0$ & 0.007737 \\
			\hline
			0.007219 & 5.093 & 0.1 & 0.035995 & 4.683 & 0.064243 \\
			\hline
			0.033600 & 2.168 & 0.3 & 0.127212 & 9 & 0.230111 \\
			\hline
			0.084345 & 1.092 & 0.51 & 0.332493 & 7.454 & 0.545201 \\
			\hline
			0.087089 & 1.048 & 0.52 & 0.361668 & $\to0$ & 0.584886 \\
			\hline
			0.092859 & 0.276 & 0.7 & 0.859151 & $\to0$ & 1.875878 \\
			\hline
			0.362804 & 0.028 & 0.95 & 1.865712 & 4 & 0.251130 \\
			\hline\hline
		\end{tabular}
	\end{table}
}

The stability boundary can be alternatively presented in terms of the water holdup. The blue curve in Fig.\ \ref{Fig: Stability_boundary_holdup}a corresponds to the blue curve in Fig.\ \ref{Fig: Stability_map_D_0d05} and shows the critical superficial velocity of the heavy phase as a function of holdup. Fig.\ \ref{Fig: Stability_boundary_holdup}b shows the critical wavenumber, $\alpha_\text{crit}$, and the corresponding wave speed scaled by the average water velocity for points along the stability boundary. According to the critical wavenumbers, five critical modes can be distinguished. The critical wave speed is about four times the average water velocity and about twice the interfacial velocity (see Table\ \ref{Tab: Critical parameter}) for most of the holdups, except for the very small ($h<0.03$) and large ($h>0.9$) holdups. For these extreme holdups, the strong confinement effect due to the curved walls slows down the wave, and its velocity is lower than the  interfacial velocity at the pipe centerline.  For very low water holdups ($\displaystyle<0.03$) and when the water occupies more than half of the pipe, i.e., $h\in[0.52,0.9]$, the critical perturbation is long wave ($\displaystyle\alpha_\text{crit}\to0$), denoted as $\displaystyle\alpha_\text{crit}=0.001$ in Fig.\ \ref{Fig: Stability_boundary_holdup}b. This is due to the fact that convergence for small, but non-zero, values of the wavenumber $\alpha$ (long-wave limit) is reached, starting from $\displaystyle\alpha=0.001$, as was elaborated in \cite{Barmak23}. The long-wave stability boundary calculated for the whole range of holdups is also shown in Fig.\ \ref{Fig: Stability_boundary_holdup}a. 

\begin{figure}[h!]
	\centering
	\subfloat[]{\includegraphics[width=0.5\textwidth,clip]{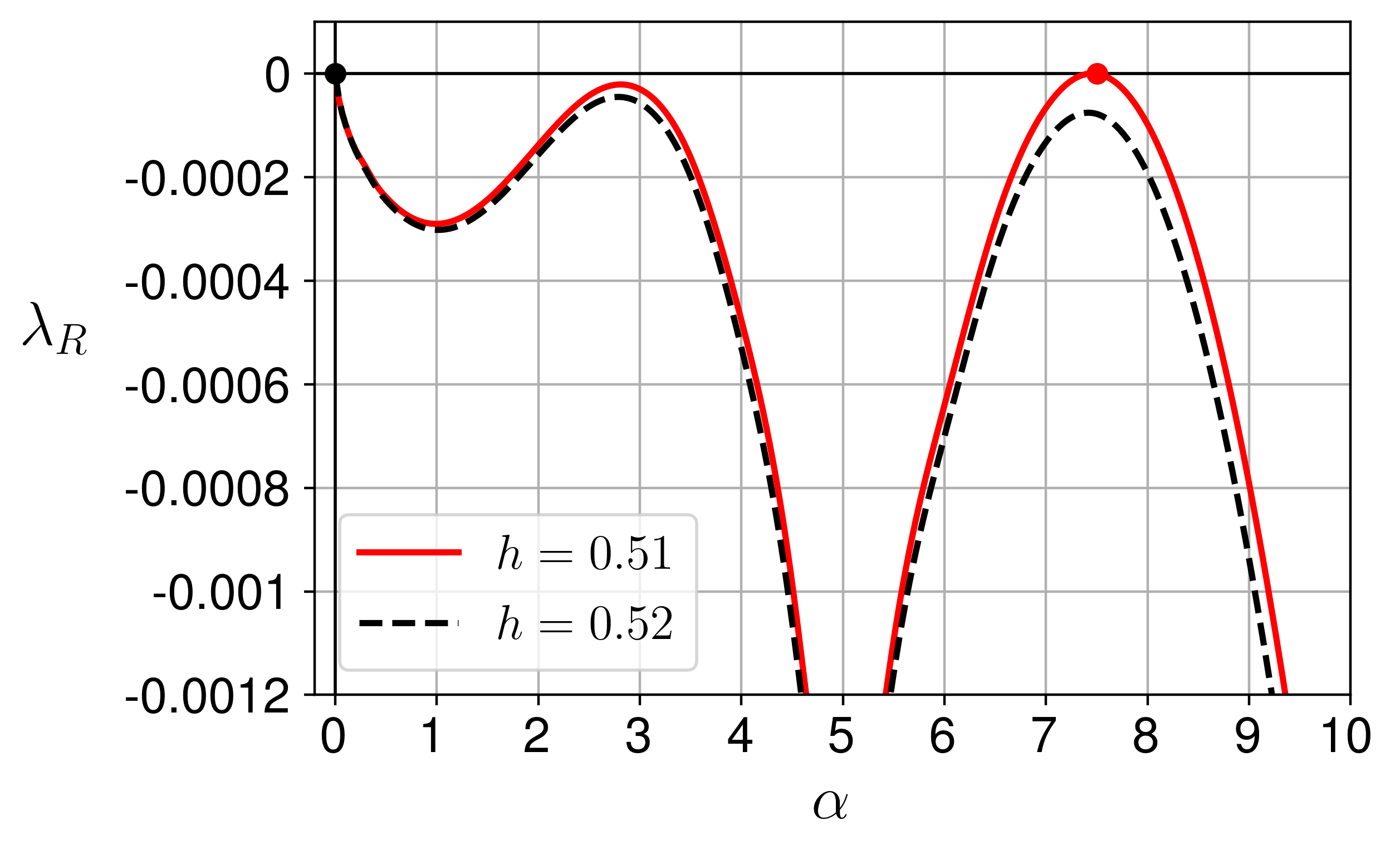}}
	\subfloat[]{\includegraphics[width=0.5\textwidth,clip]{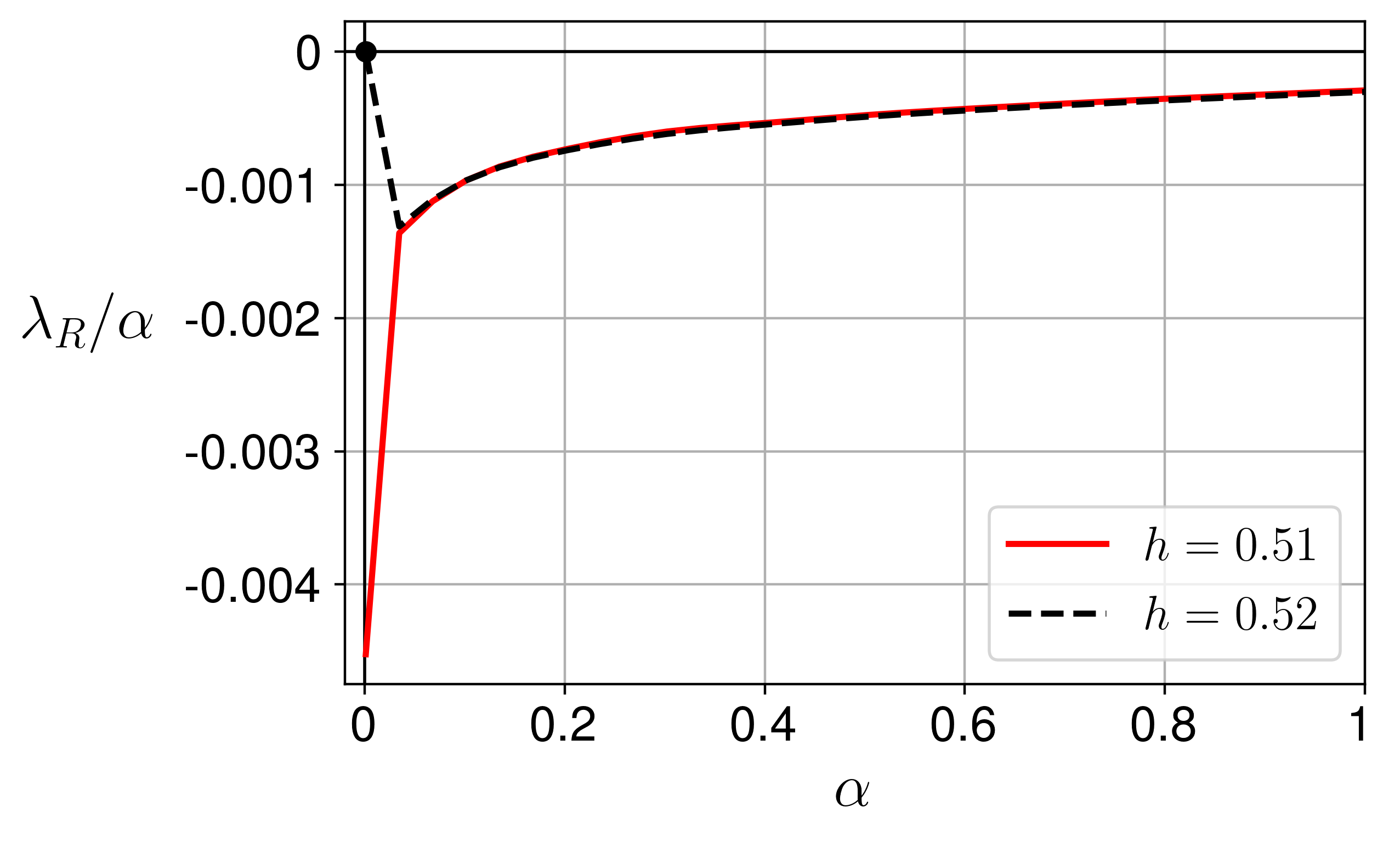}}
	\caption{\label{Fig: spectrum_a-w}(a) Growth rate of perturbation as a function of wavenumber for two different holdups along the stability boundary (Fig.\ \ref{Fig: Stability_map_D_0d05}). (b) Enlargement on the long-wave region (small wavenumbers $\alpha$). Growth rate scaled by the wavenumber.}	
\end{figure}

Contrarily to the two-plate geometry \cite{Barmak16a}, where for long waves there is no critical water superficial velocity for low air velocities, in pipe flow, the long-wave stability region is bounded for the whole range of holdups. However, long-wave perturbations, which is the basic assumption in widely-used Two Fluid model, are not critical for other holdups. Three distinct regions are observed along the stability boundary (Fig.\ \ref{Fig: Stability_boundary_holdup}b). In a range of holdups between $0.03$ and $0.2$, the flow stability is defined by short-wave perturbations, whose wavenumber decreases continuously from $\approx6.7$ to 4. When the holdup exceeds a value of 0.2 ($\displaystyle U_{1S} = 0.018$m/s, $\displaystyle U_{2S} = 3.159$m/s), an abrupt change in the critical mode takes place, from one with $\alpha_\text{crit}\approx4$ to a mode with a larger wavenumber, $\alpha_\text{crit}\approx10.5$ (smaller wavelength). Starting from the holdup of $0.2$ and up to $0.52$, $\alpha_\text{crit}$ decreases to $\approx7.5$, when another abrupt change in the critical wavenumber is observed and the long-wave perturbation becomes critical. To analyze this abrupt change in the critical wavenumber, one can analyze the growth rate of perturbations with respect to wavenumber at $h=0.51$ (red line in Fig.\ \ref{Fig: spectrum_a-w}). At this point, i.e., $\displaystyle U_{1S} = 0.084$m/s, $\displaystyle U_{2S} = 1.092$m/s, $\displaystyle\alpha_\text{crit} = 7.45$ (shown as red circle), while the secondary maximum is at $\displaystyle\alpha\approx3$. The values of the growth rate $\lambda_{R}$, gets close to $0$ also for $\displaystyle\alpha\to0$, since $\lambda_{R}$ is scaled with $\alpha^2$ for long waves, i.e., $\displaystyle\lambda_{R}/\alpha^2=\text{const}$. This is demonstrated by straight lines of $\lambda_{R}/\alpha$ for $\displaystyle\alpha\to0$ in Fig.\ \ref{Fig: spectrum_a-w}b, where an enlargement of the long-wave region is presented (the calculations for this figure were performed with resolution of $\Delta\alpha=0.01$). When increasing the holdup to $0.52$ (dashed black line in Fig.\ \ref{Fig: spectrum_a-w}), the critical mode switches to long wave (black circle), while the short waves are stable ($\displaystyle\lambda_{R}<0$) with local maxima at $\displaystyle\alpha\approx3$ and at $\displaystyle\alpha\approx7.45$.

\begin{figure}[h!]
	\centering
	\subfloat[$|u_x|/\max(|u_z|)$]{\includegraphics[width=0.33\textwidth,clip]{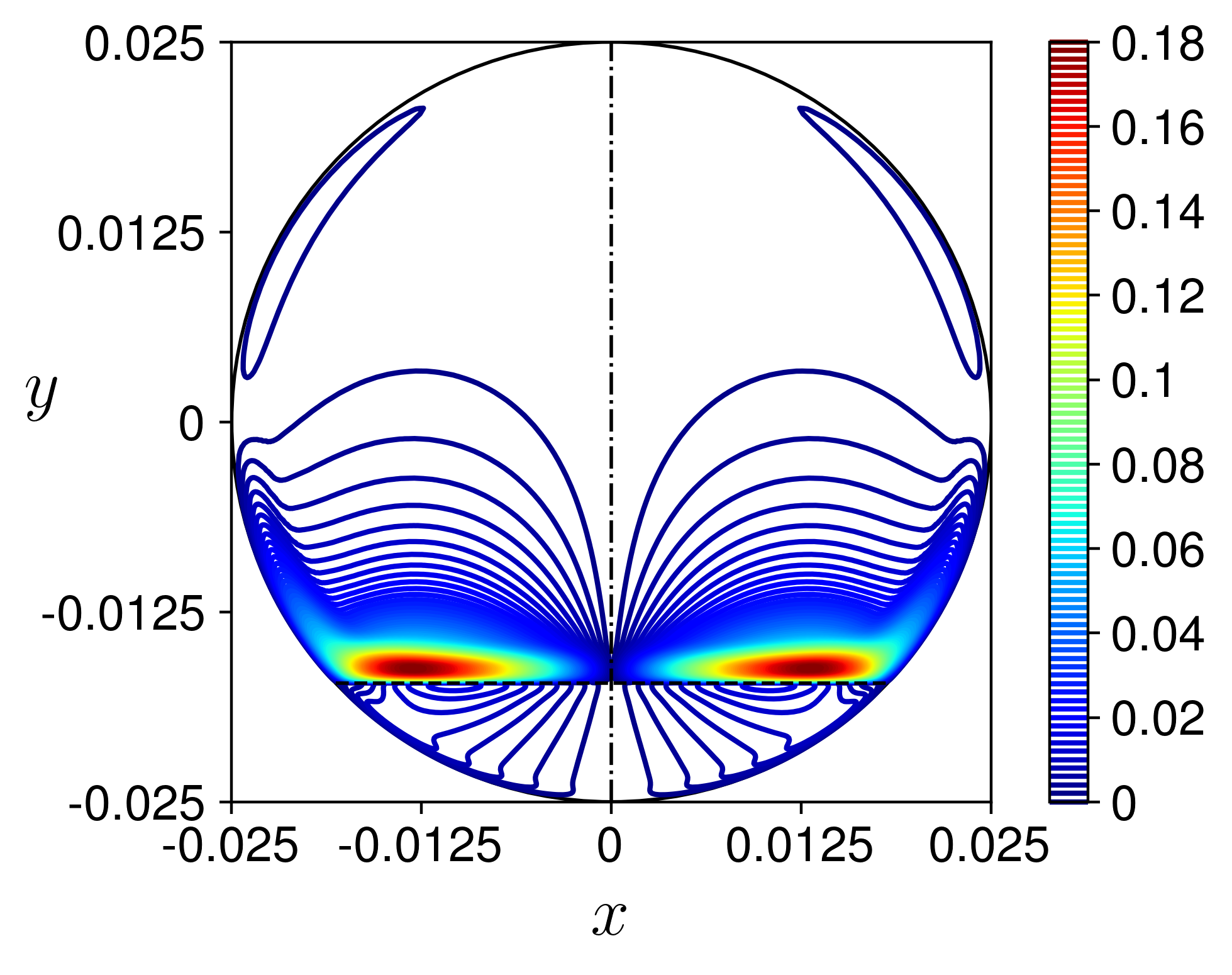}}
	\subfloat[$|u_y|/\max(|u_y|)$]{\includegraphics[width=0.33\textwidth,clip]{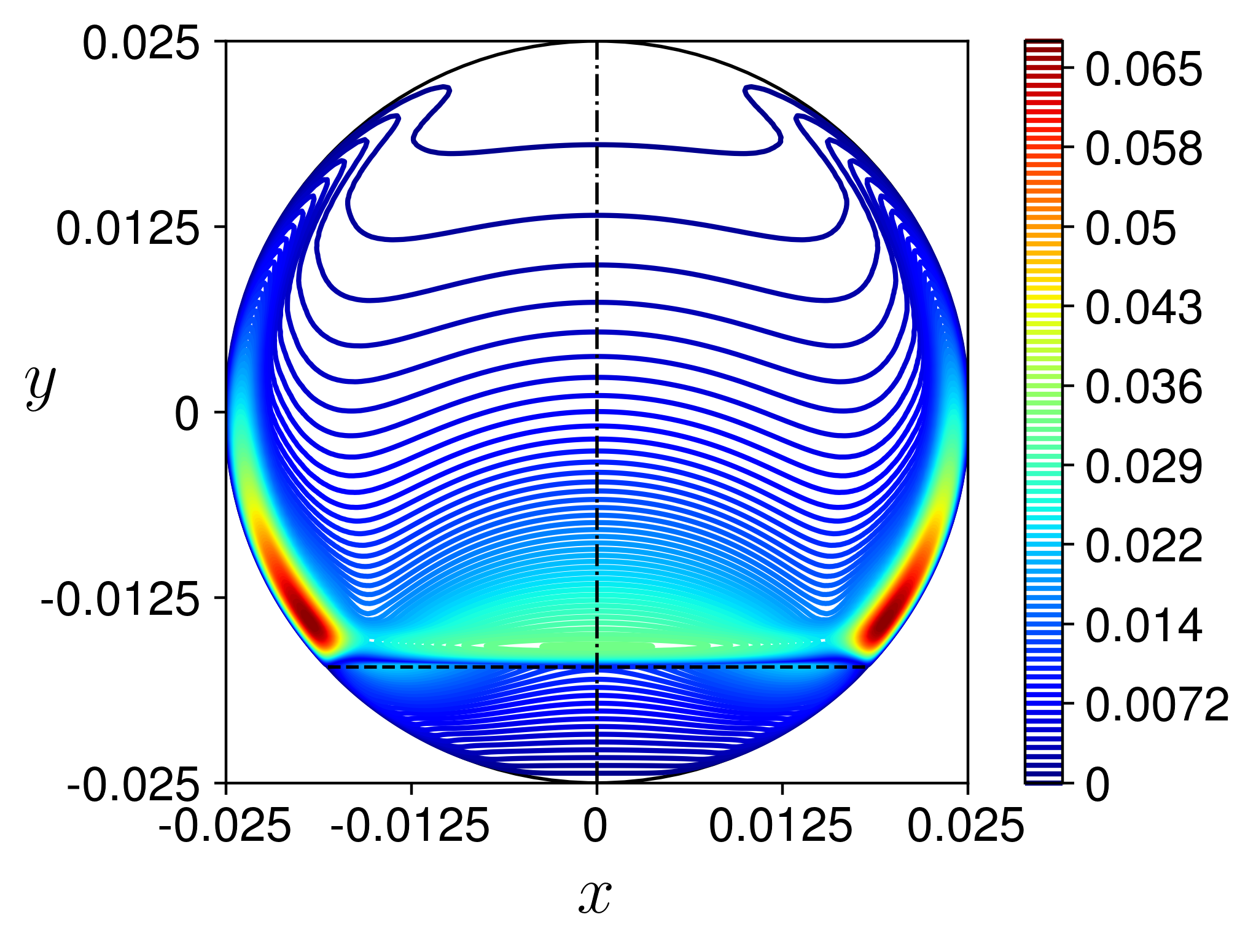}}
	\subfloat[$|u_z|/\max(|u_z|)$]{\includegraphics[width=0.33\textwidth,clip]{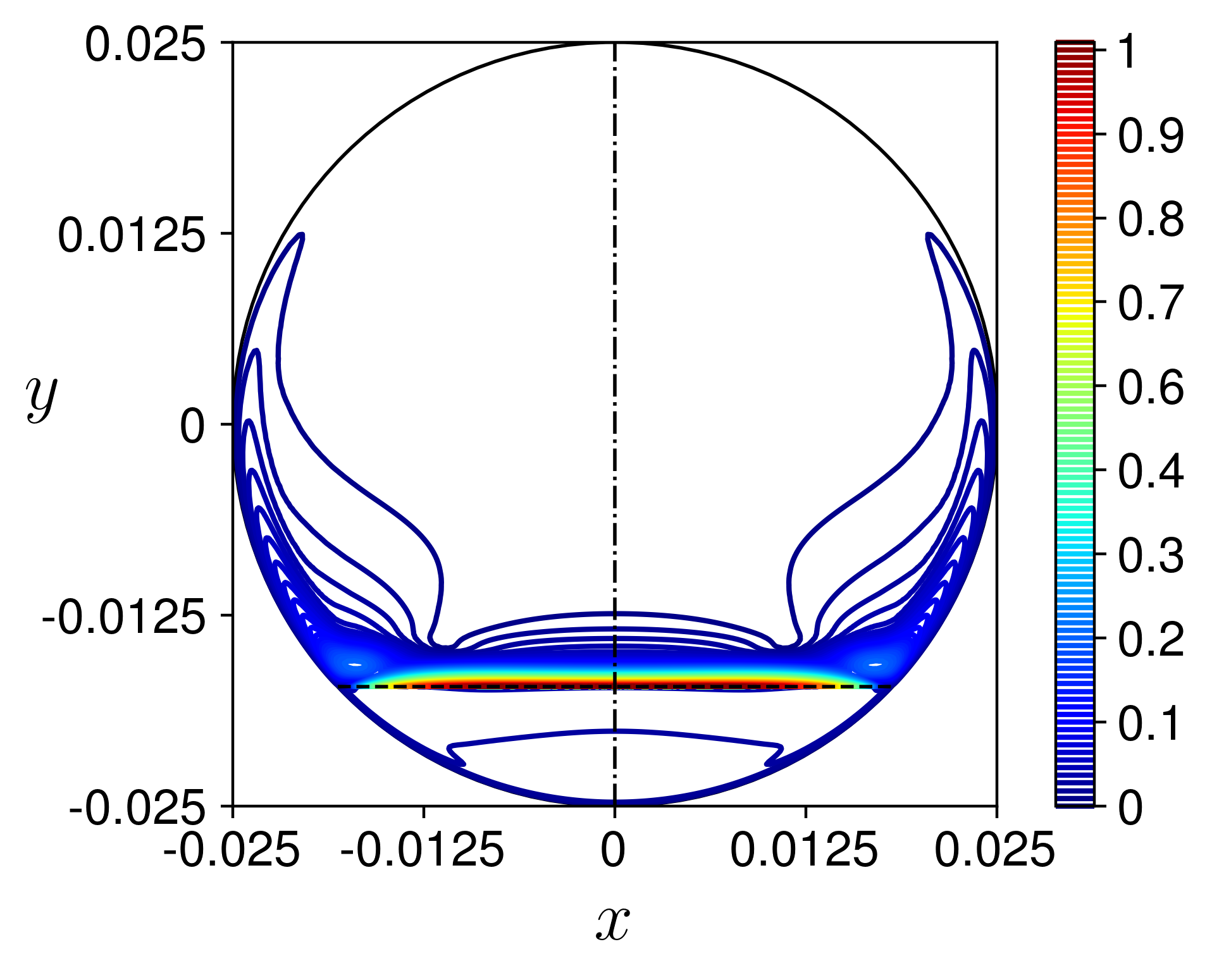}}
	\\
	\subfloat[$|u_x|/\max(|u_z|)$]{\includegraphics[width=0.33\textwidth,clip]{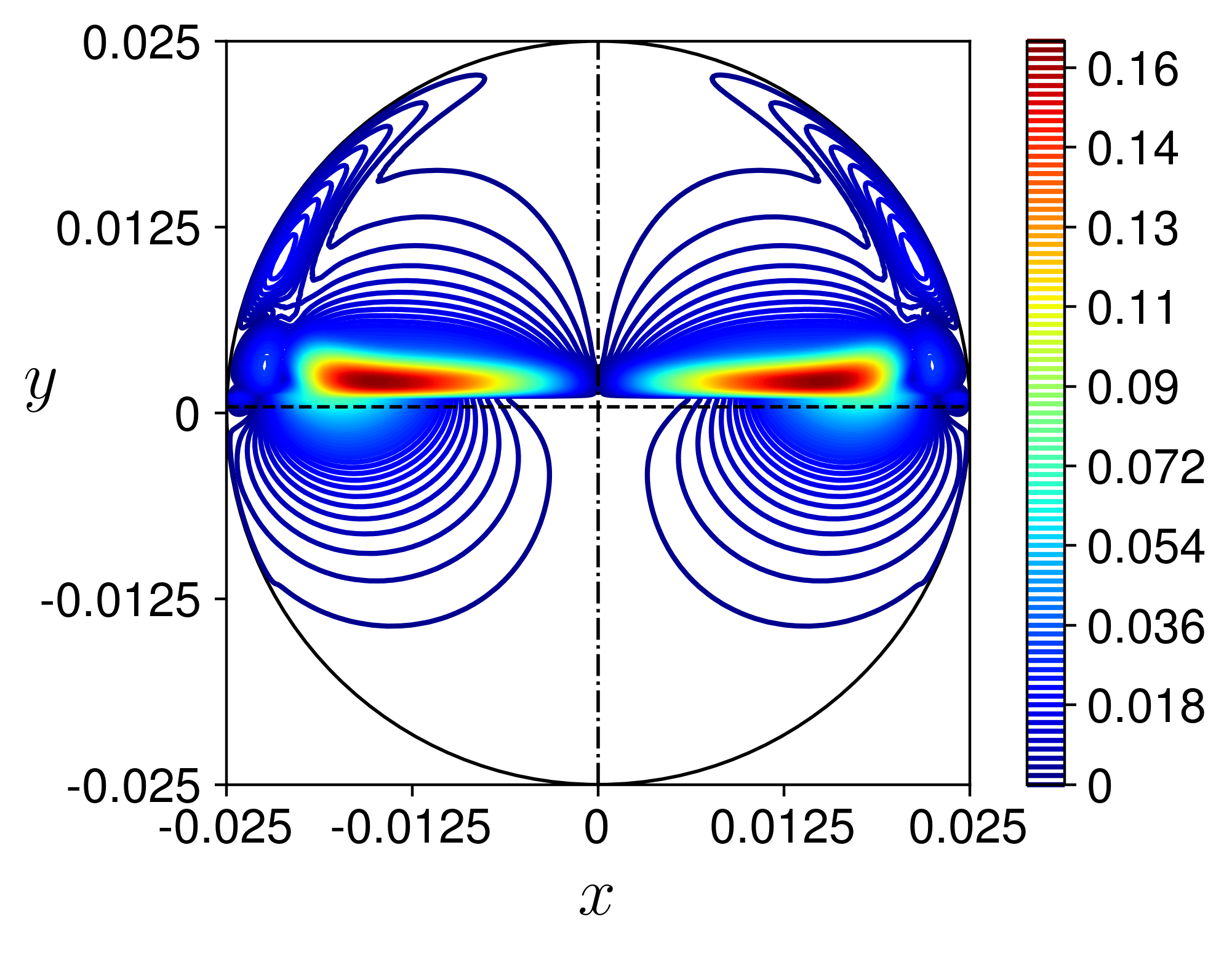}}
	\subfloat[$|u_y|/\max(|u_z|)$]{\includegraphics[width=0.33\textwidth,clip]{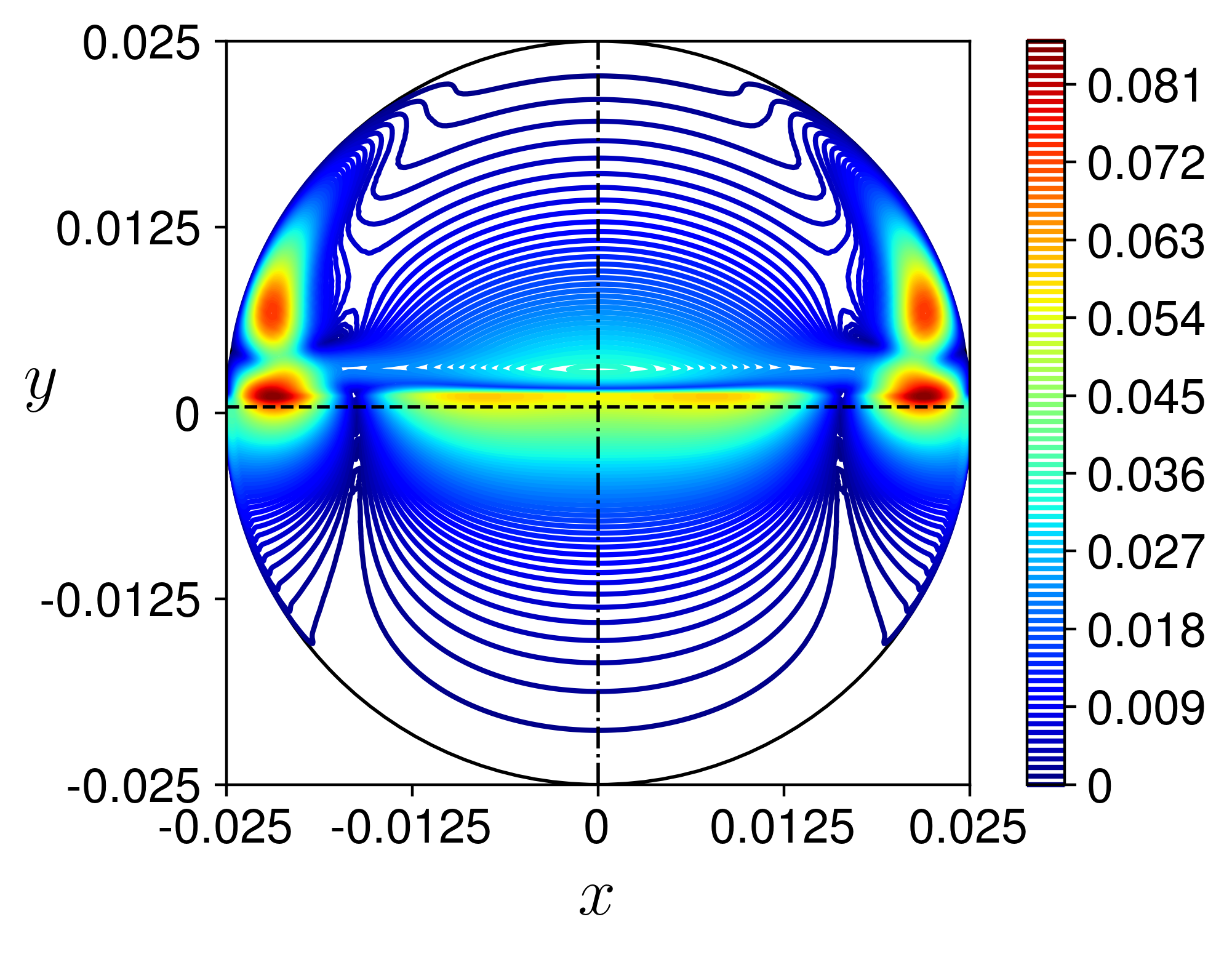}}
	\subfloat[$|u_z|/\max(|u_z|)$]{\includegraphics[width=0.33\textwidth,clip]{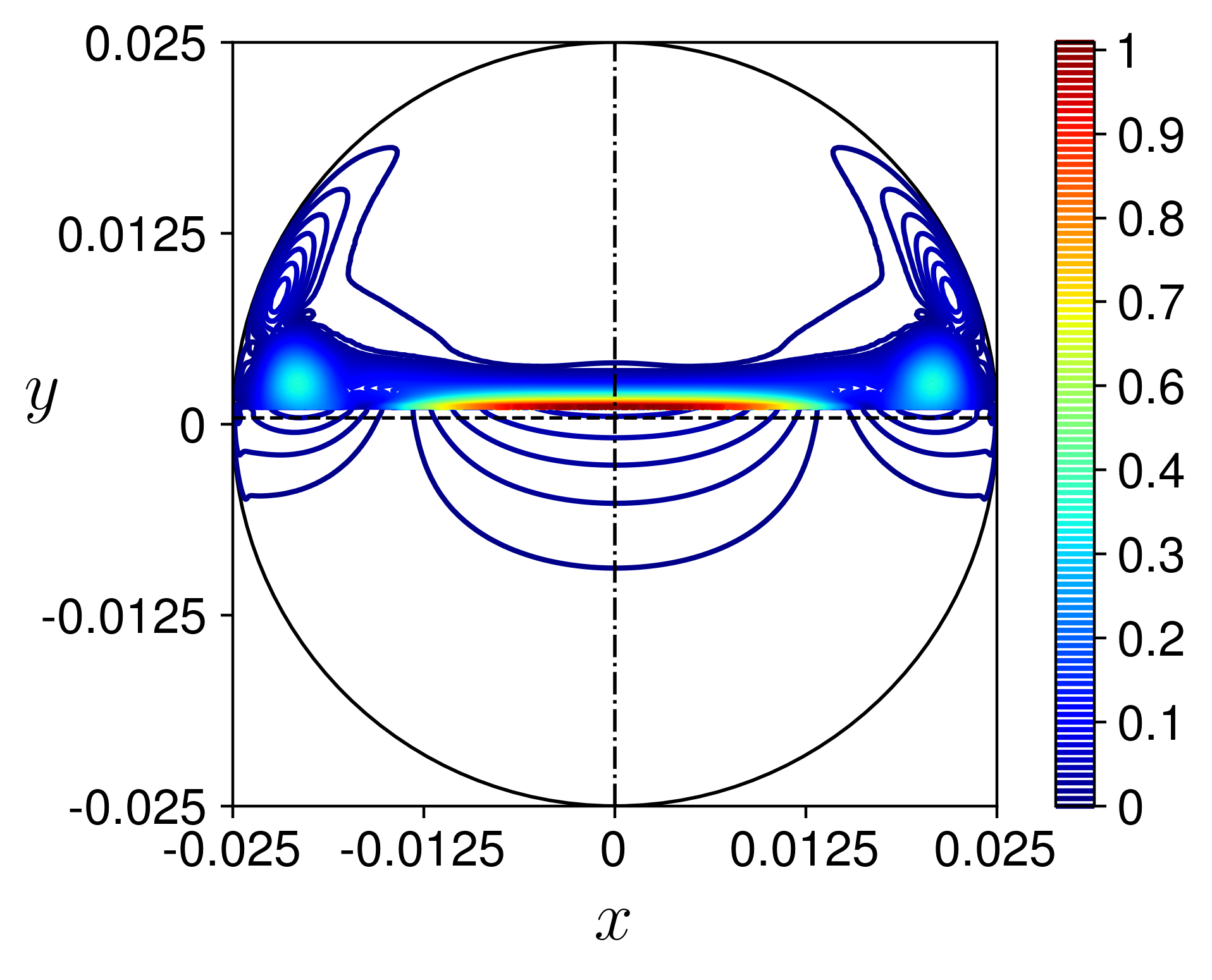}}
	\caption{\label{Fig: Holdup_0d1-0d51}Amplitude contours of the critical perturbation. Air--water pipe flow, $D=0.05$m. (a) - (c) $h=0.1$. (d) - (f) $h=0.51$. (a),(d) Lateral component of velocity; (b),(e) vertical component of velocity; (c),(f) axial velocity. The unperturbed interface is denoted by horizontal dashed black line, and the cross-section centerline -- by vertical dash-dot black line.}	
\end{figure}

The three components of the velocity of the critical perturbations are shown in Fig.\ \ref{Fig: Holdup_0d1-0d51} by their amplitude, which depends on two cross-sectional coordinates (Eq.\ \ref{Eq: Perturbation}). In linear stability analysis, the perturbation amplitude is defined to within multiplication by a constant. For the purpose of presentation, it is scaled by the maximal absolute value of the axial perturbation velocity, $\displaystyle\max(|u_z|)$. As pointed above, for very low holdups, the critical perturbation is long wave ($\alpha_\text{crit}\to0$). For $h=0.1$, the critical perturbation is short wave with $\alpha_\text{crit}=4.683$, and the amplitude of the velocity components of the critical perturbation is shown in Fig.\ \ref{Fig: Holdup_0d1-0d51}a-c. The maximal axial-velocity perturbation is located at the center of the interface on its upper (air) side. Note that there is a jump in the axial velocity of the perturbation across the interface because of its deformation (Eq.\ \ref{Eq: BC_continuity_velocity}c). The difference between the axial air and water velocities at the interface can be seen below in Fig.\ \ref{Fig: u_x_0}c for $\displaystyle h=0.1,0.52,0.95$. The axial velocities in the water phase are much smaller (less than 0.05) than the maximal value. Further from the interface, the axial velocity decreases drastically also in the air phase. However, once the air layer becomes thinner (i.e., $h>0.52$), the perturbation amplitude of the axial velocity reaches a local maximum of $\displaystyle\approx0.65\max(|u_z|)$ in the air bulk for $h=0.95$.

\begin{figure}[h]
	\centering
	\subfloat[$|u_x|/\max(|u_z|)$]{\includegraphics[width=0.33\textwidth,clip]{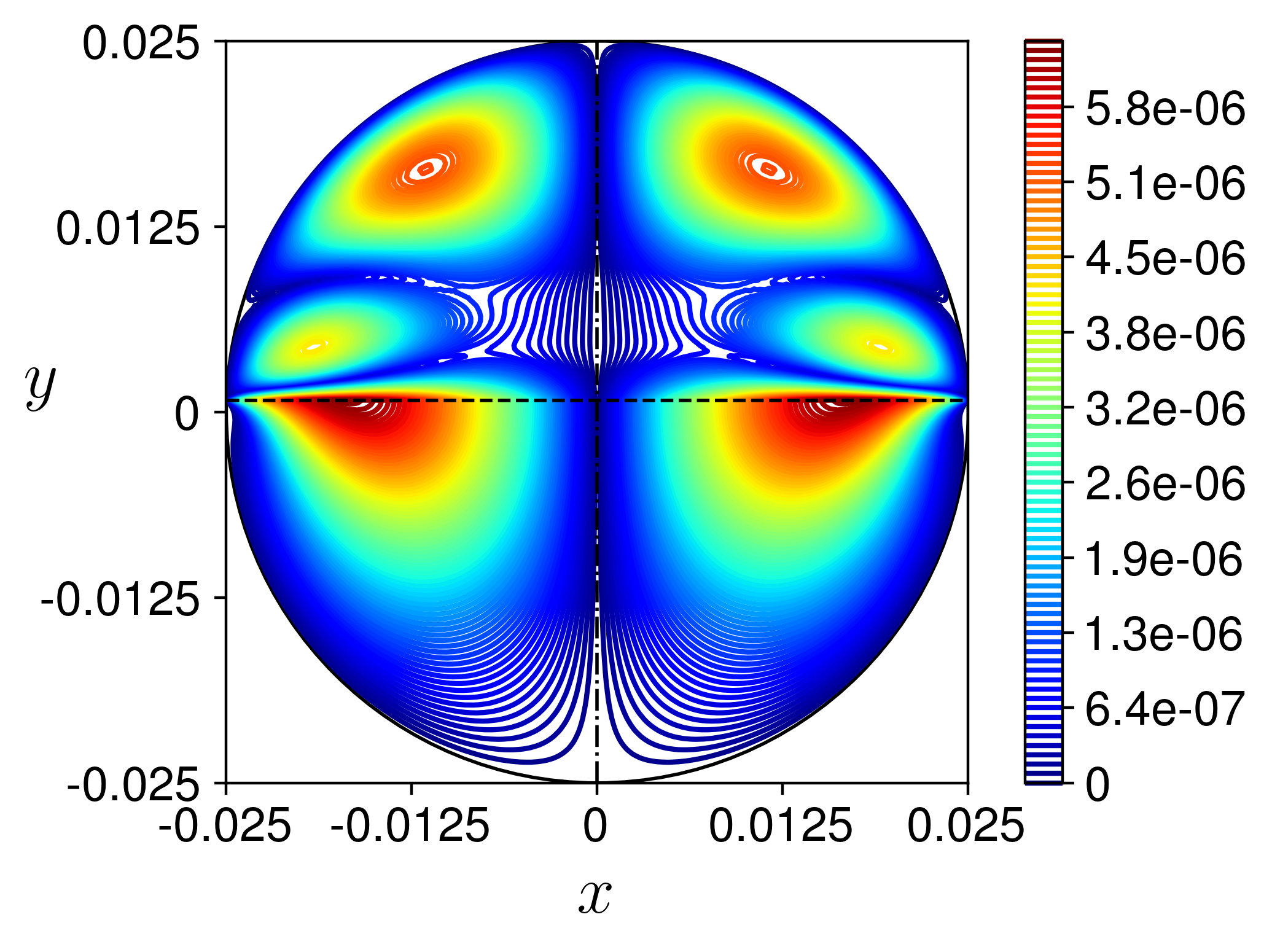}}
	\subfloat[$|u_y|/\max(|u_y|)$]{\includegraphics[width=0.335\textwidth,clip]{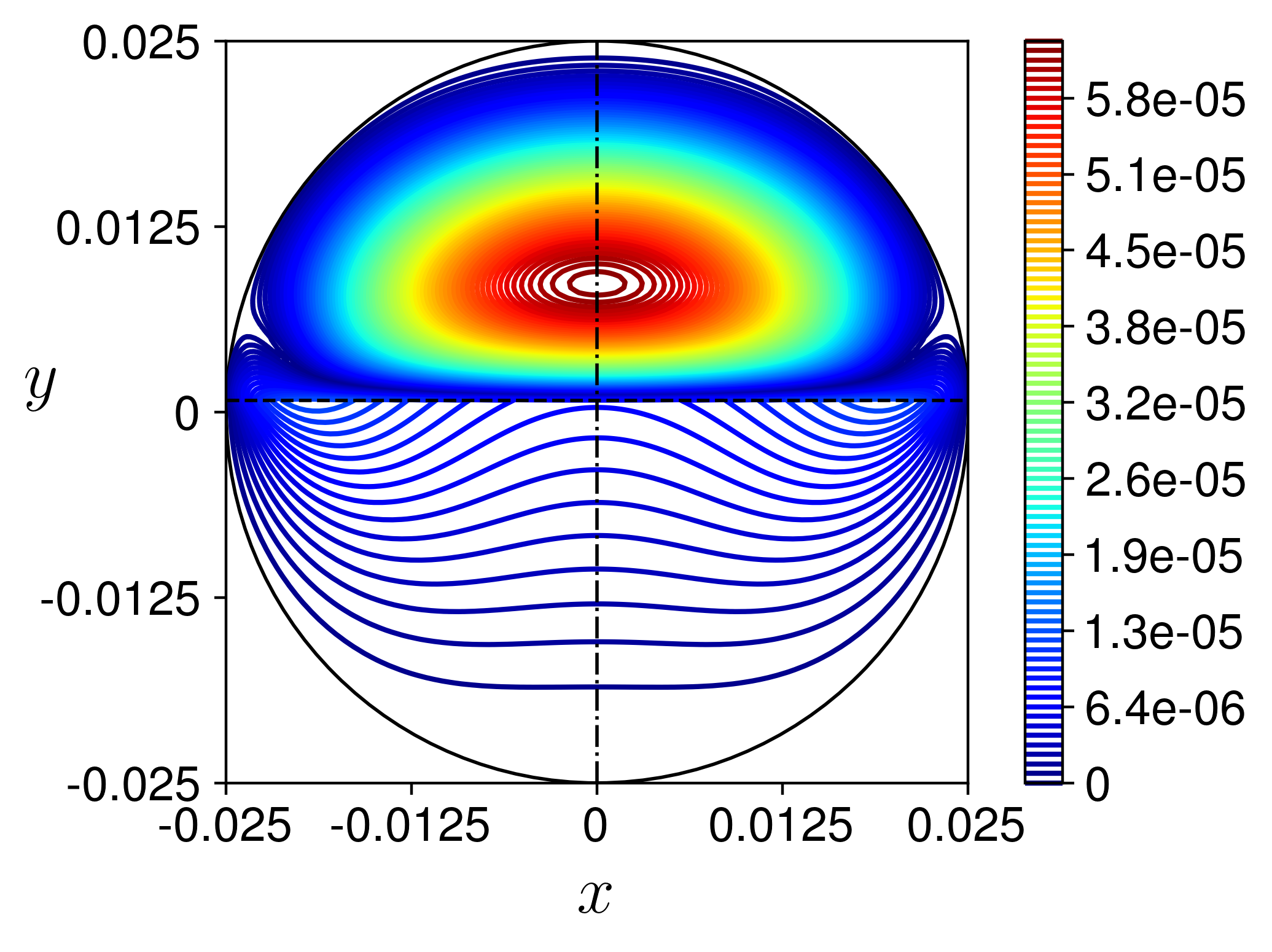}}
	\subfloat[$|u_z|/\max(|u_z|)$]{\includegraphics[width=0.31\textwidth,clip]{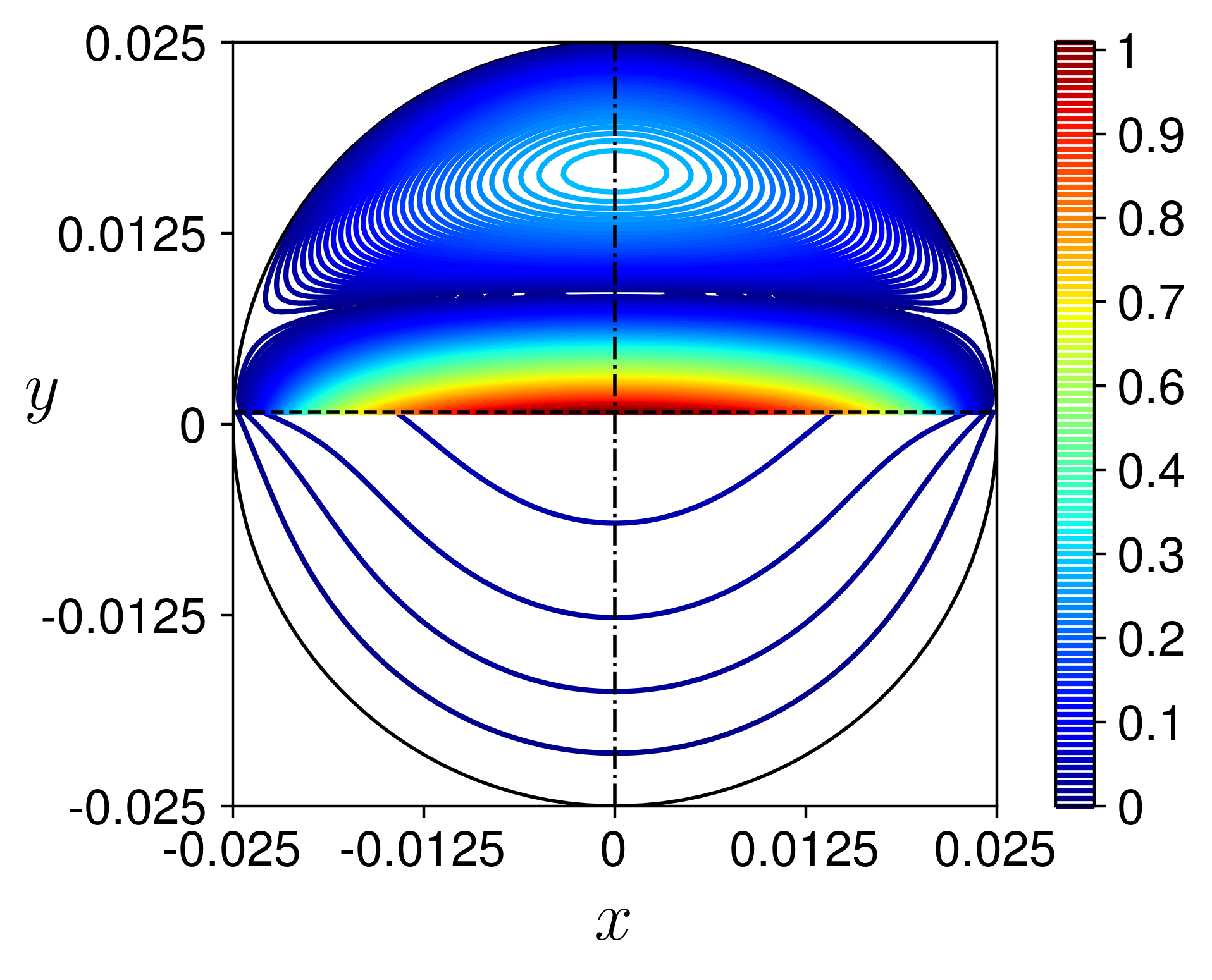}}
	\\
	\subfloat[$|u_x|/\max(|u_z|)$]{\includegraphics[width=0.33\textwidth,clip]{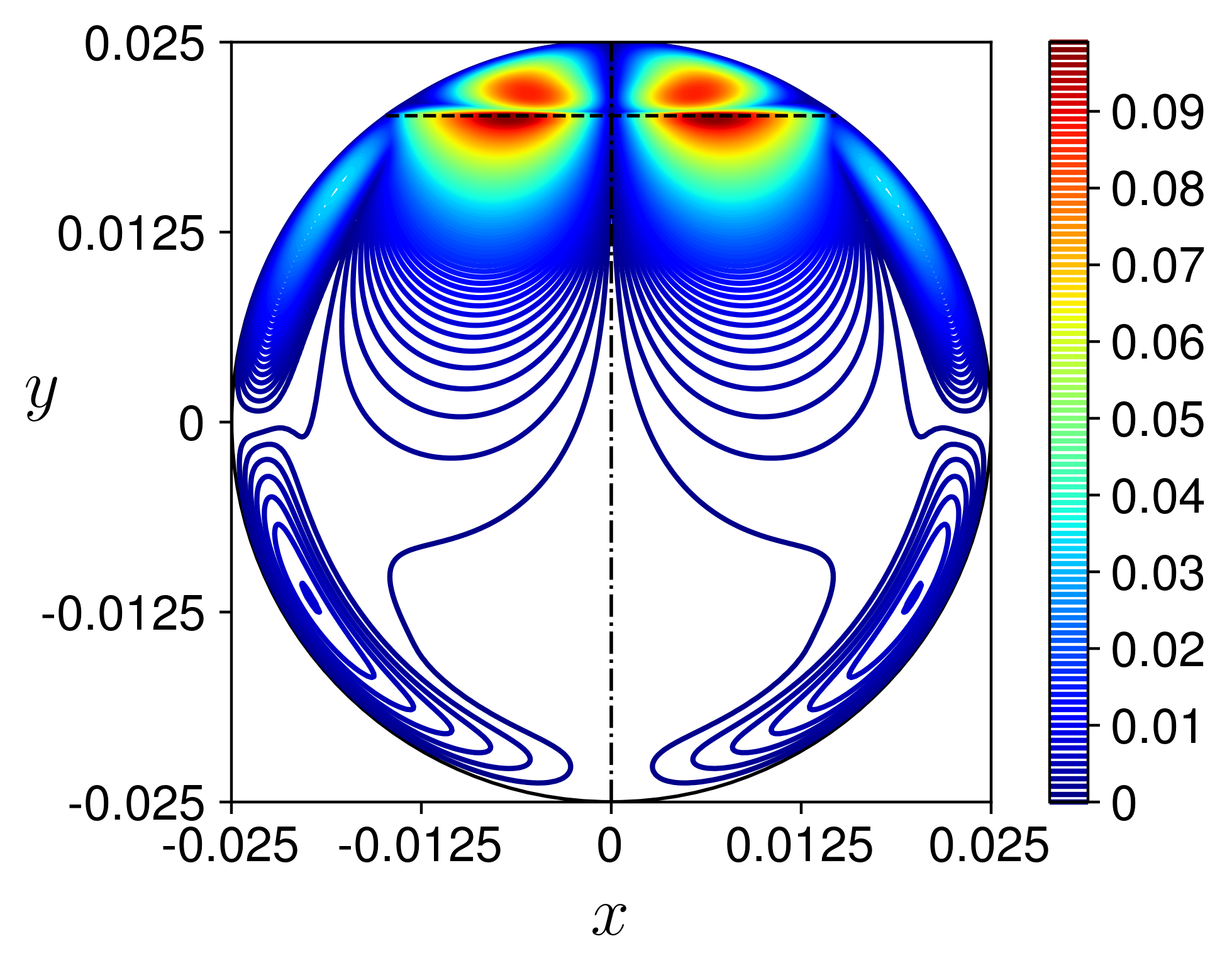}}
	\subfloat[$|u_y|/\max(|u_y|)$]{\includegraphics[width=0.33\textwidth,clip]{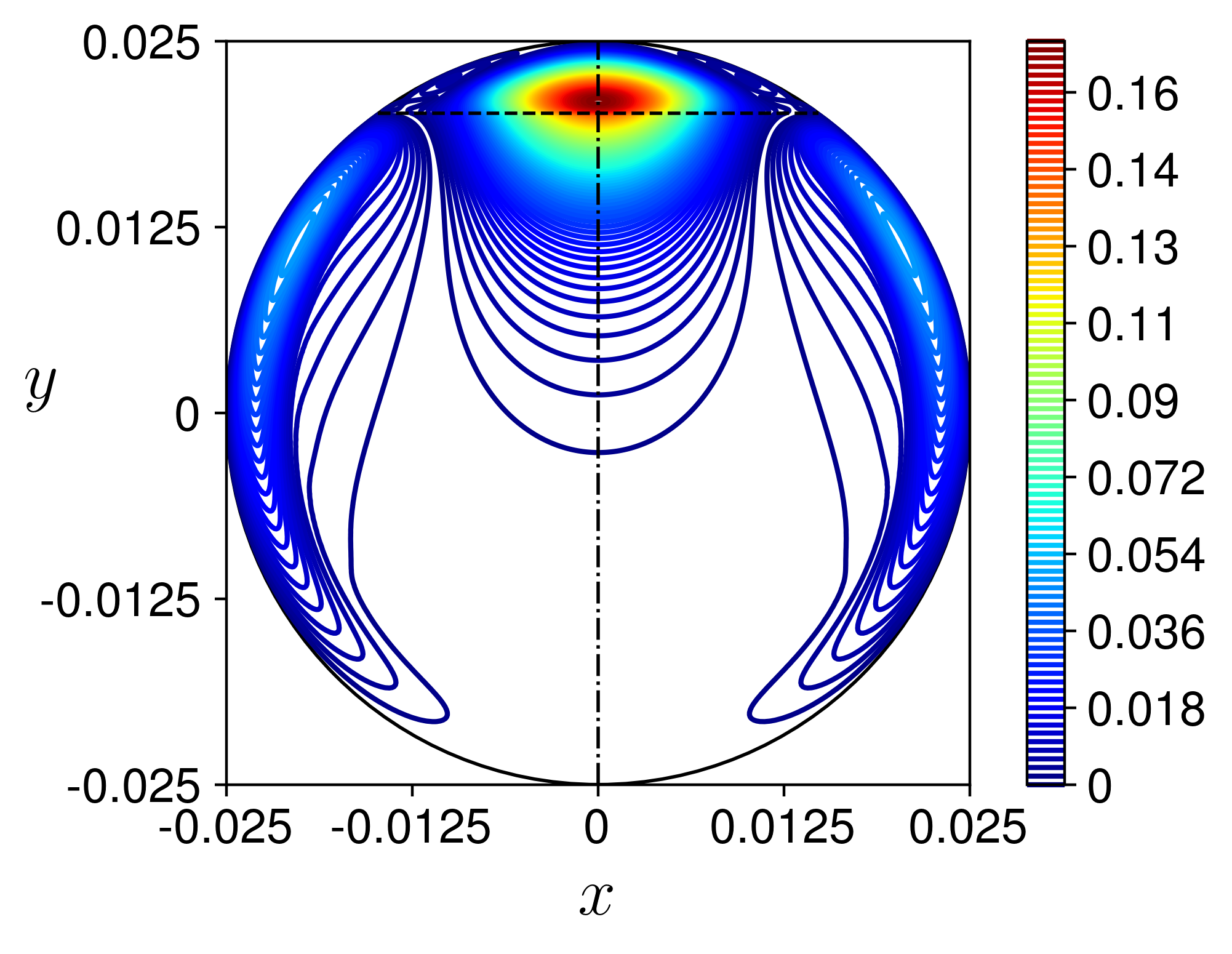}}
	\subfloat[$|u_z|/\max(|u_z|)$]{\includegraphics[width=0.33\textwidth,clip]{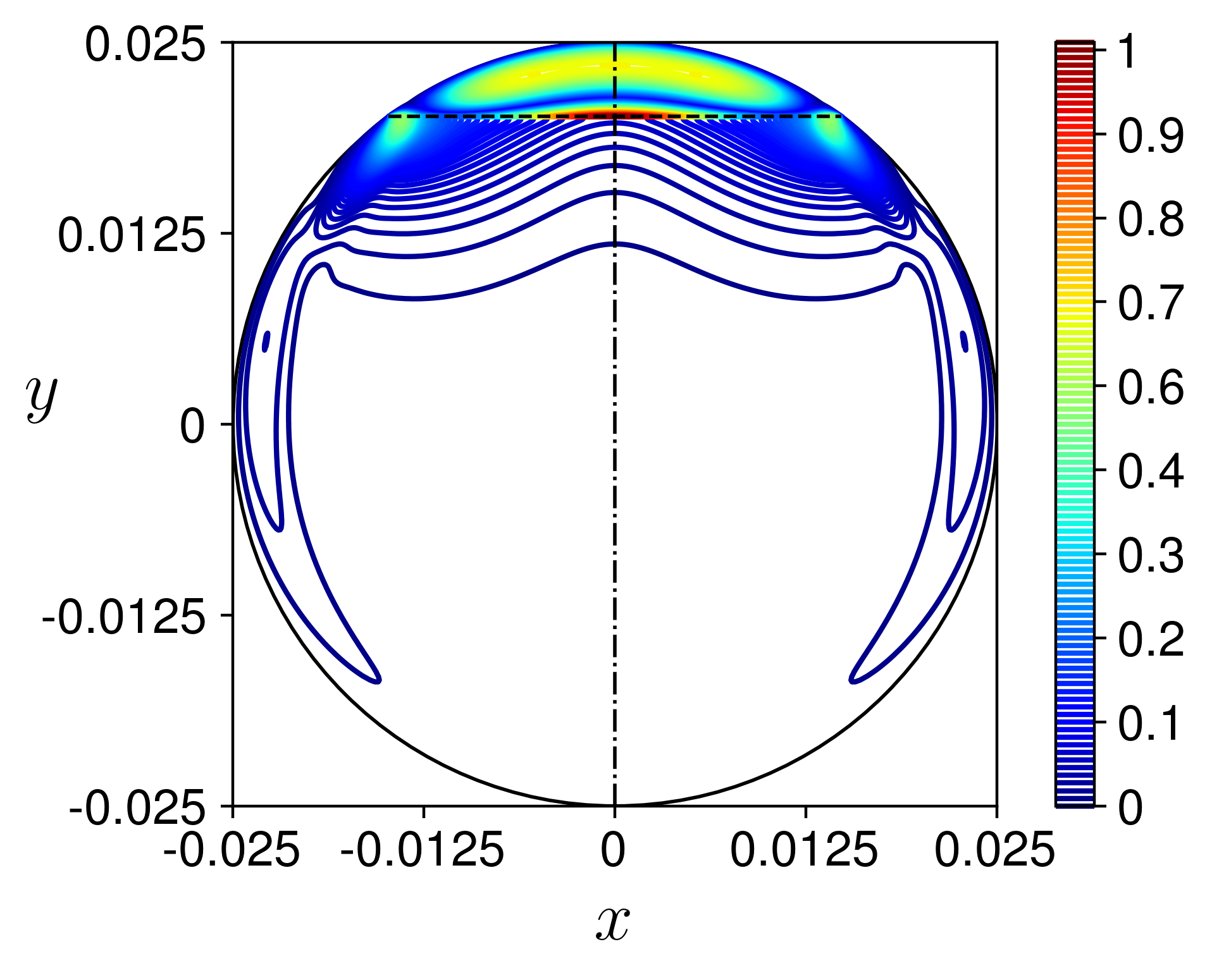}}
	\caption{\label{Fig: Holdup_0d52-0d95}Amplitude contours of the critical perturbation. Air--water flow in a circular pipe of diameter $D=0.05$m. (a) - (c) $h=0.52$. (d) - (f) $h=0.95$. (a), (d) Lateral component of velocity; (b), (e) vertical component of velocity; (c), (f) axial velocity. The unperturbed interface is denoted by a horizontal dashed black line, and the cross-section centerline - by a vertical dash-dot black line.}		
\end{figure}

\begin{figure}[h!]
	\centering	
	\subfloat[$|u_y|/\max(|u_y|)$]{\includegraphics[width=0.4\textwidth,clip]{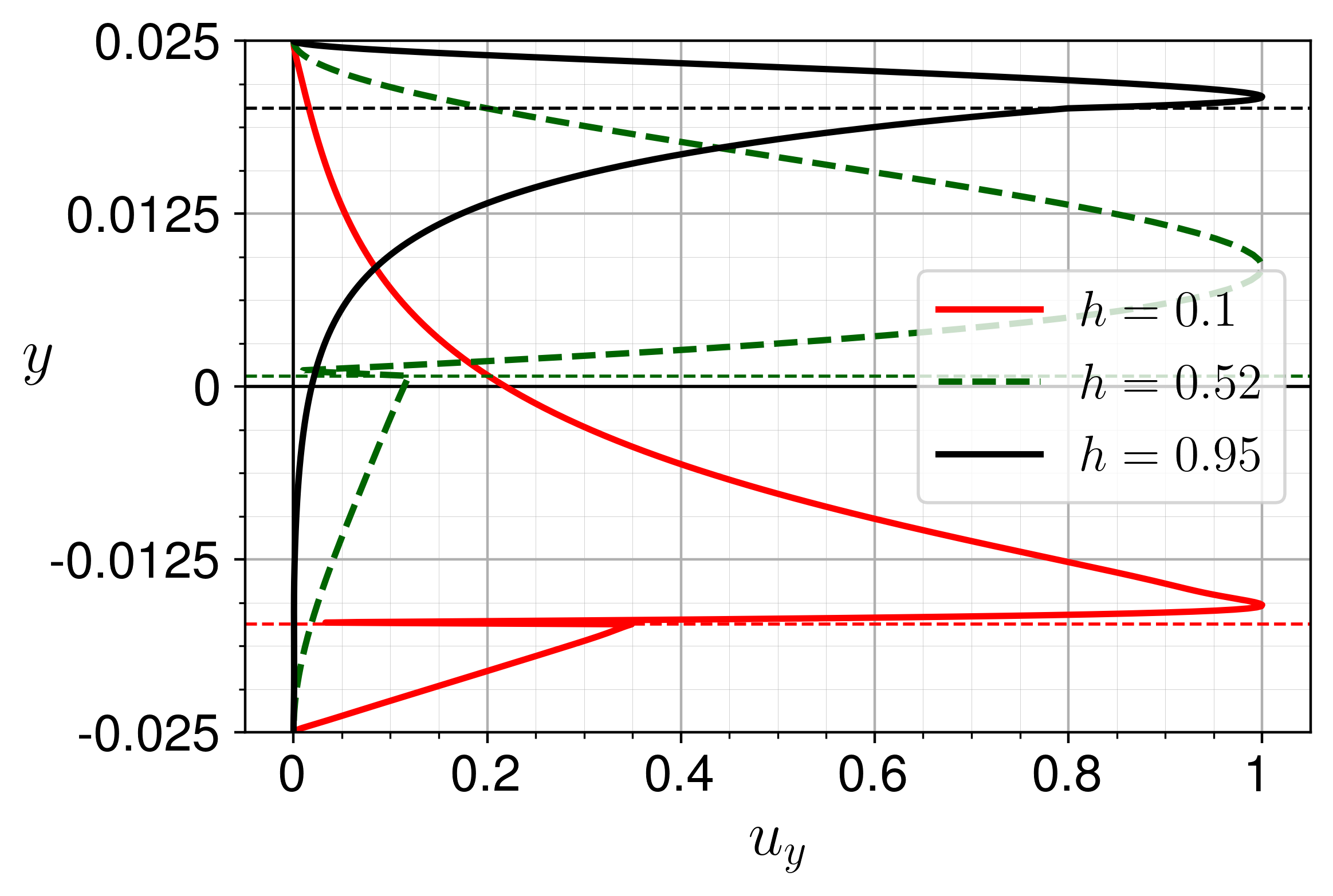}}
	\qquad
	\subfloat[$|u_z|/\max(|u_z|)$]{\includegraphics[width=0.4\textwidth,clip]{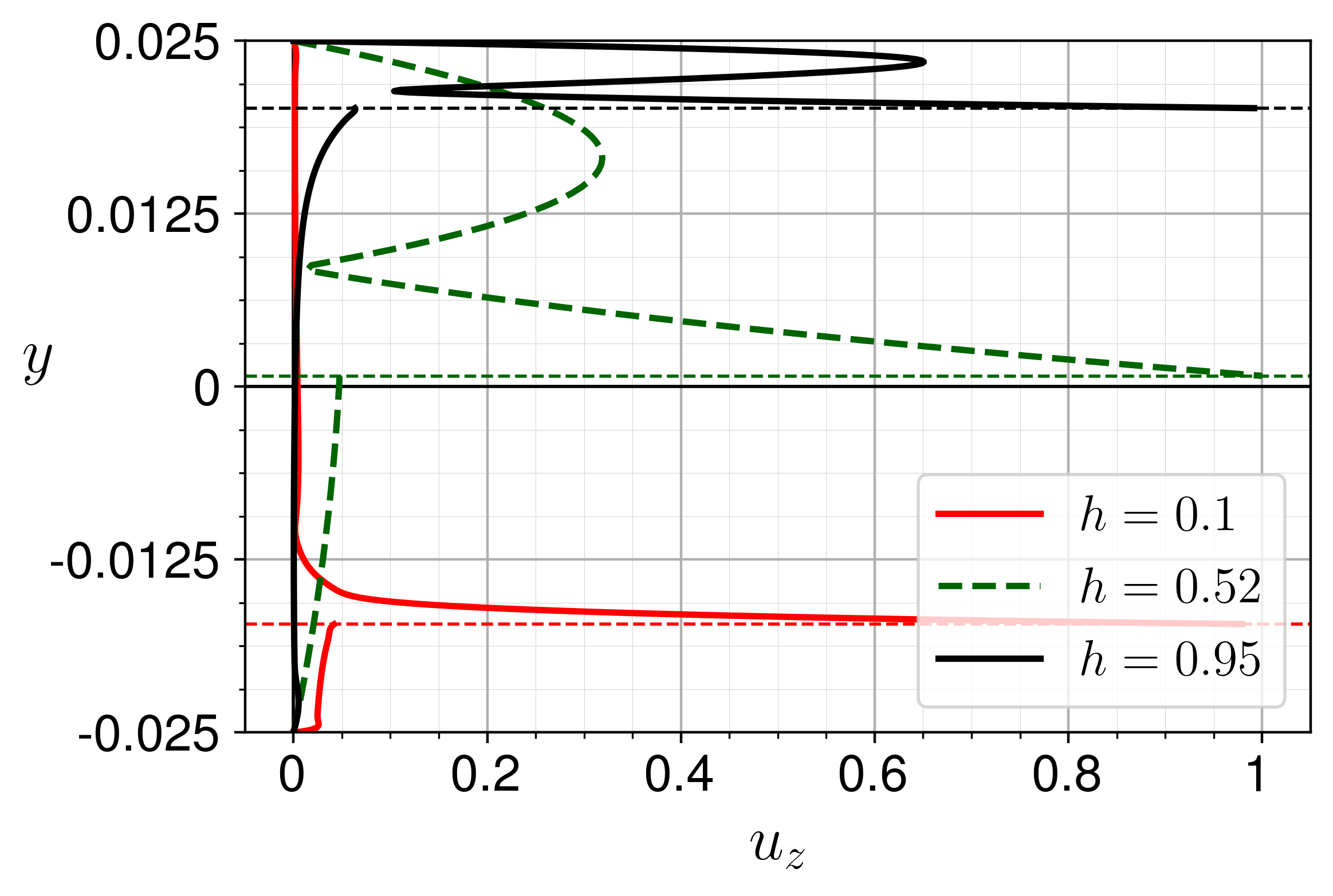}}
	\caption{\label{Fig: u_x_0}Velocity amplitude profiles of the critical perturbations (Fig.\ \ref{Fig: Holdup_0d1-0d51} and Fig.\ \ref{Fig: Holdup_0d52-0d95})  at $x=0$. Air--water pipe flow, $D=0.014$m. (a) Lateral component divided by its maximal value; (b) vertical component divided by its maximal value. The locations of the unperturbed interface is denoted by corresponding horizontal dashed lines.}	
\end{figure}

\begin{figure}[h!]
	\centering
	\subfloat[$|u_x|/\max(|u_x|)$]{\includegraphics[width=0.33\textwidth,clip]{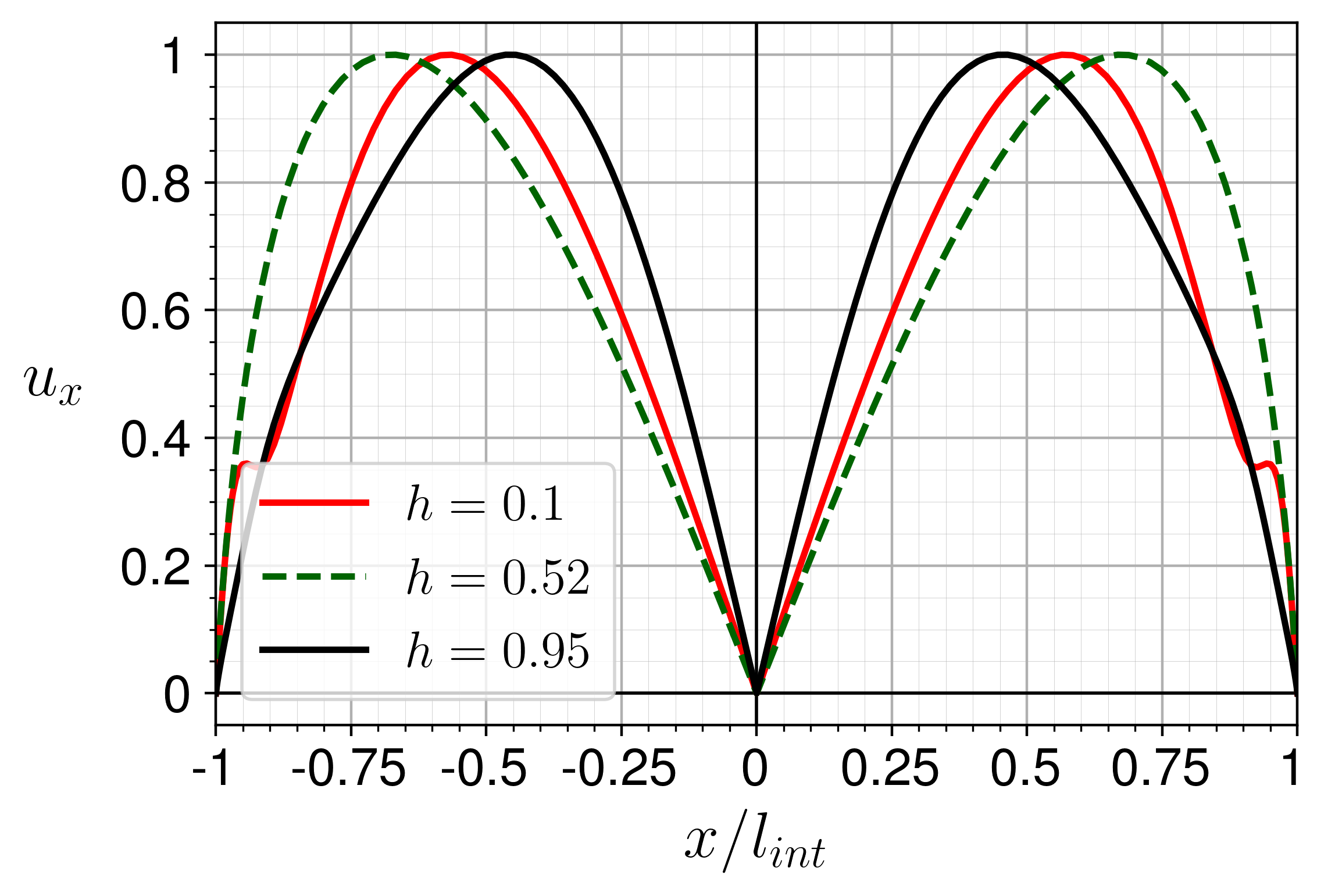}}
	\subfloat[$|u_y|/\max(|u_y|)$]{\includegraphics[width=0.33\textwidth,clip]{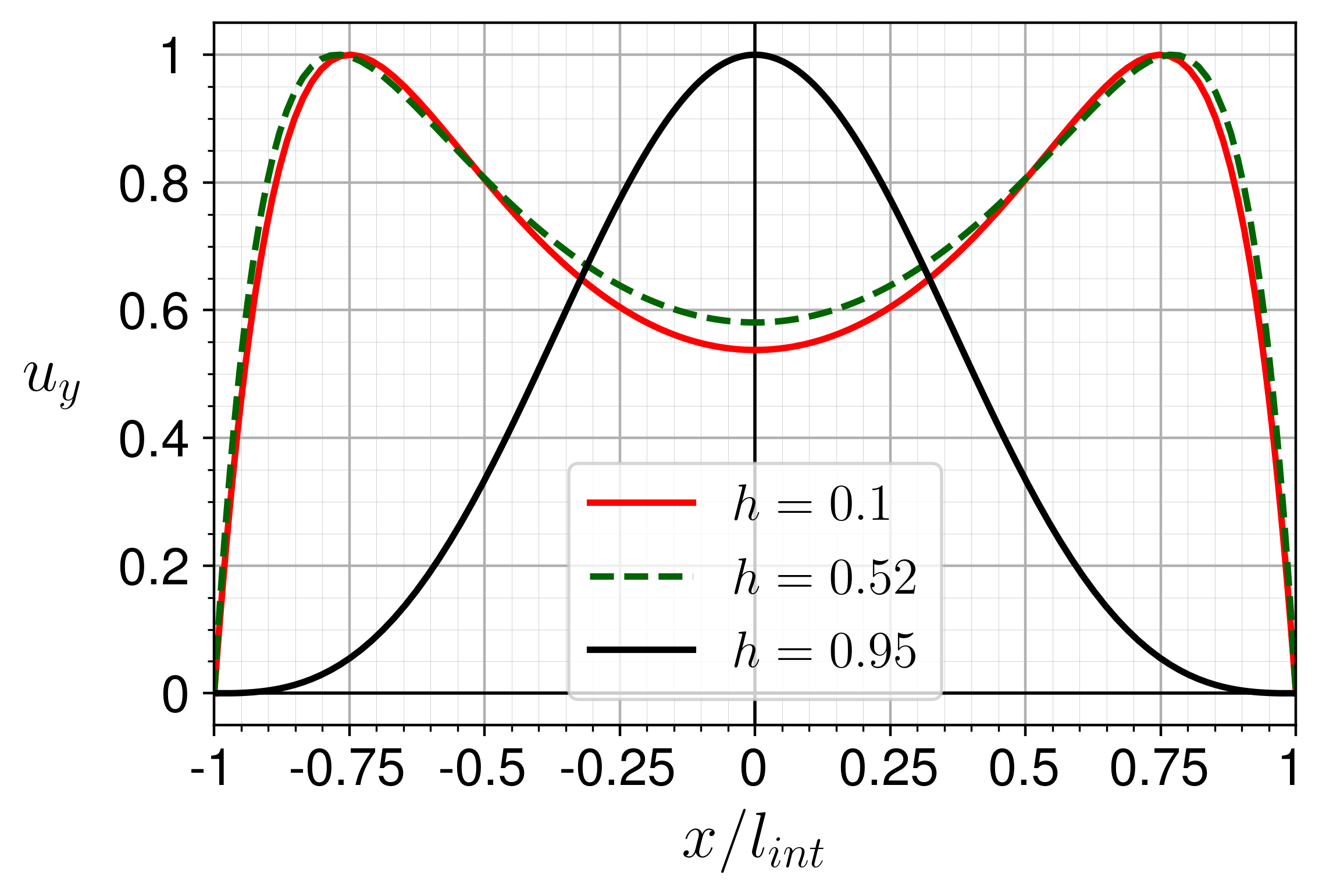}}
	\subfloat[$|\eta|/\max(|\eta|)$]{\includegraphics[width=0.33\textwidth,clip]{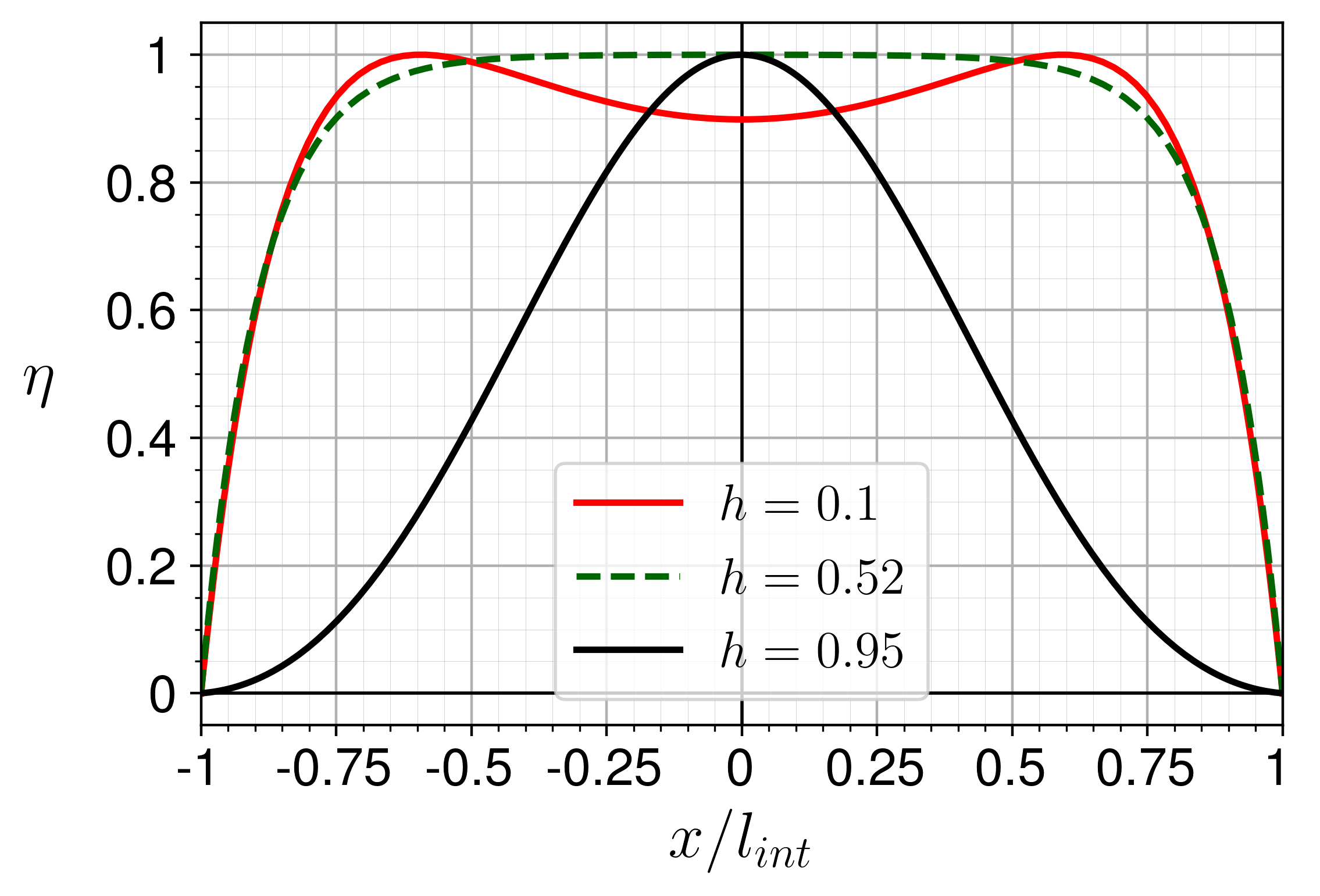}}
	\caption{\label{Fig: u_interface}Amplitude profiles of the critical perturbations at the interface (Fig.\ \ref{Fig: Holdup_0d1-0d51} and Fig.\ \ref{Fig: Holdup_0d52-0d95}). Air--water pipe flow, $D=0.014$m. (a) Lateral velocity component divided by its maximal value; (b) vertical velocity component divided by its maximal value; (c) interfacial displacement normalized by its maximum.}	
\end{figure}

The lateral, $u_x$, and vertical, $u_y$, velocity components are smaller than the axial one for all the critical perturbations obtained along the stability boundary for the considered pipe flow. For short-wave perturbations, they are an order of magnitude smaller, e.g., $\max(|u_x|)=0.18$ and $\max(|u_y|)=0.067$ for $h=0.1$, and either of these components can be the larger one (see Fig.\ \ref{Fig: Holdup_0d1-0d51} and Fig.\ \ref{Fig: Holdup_0d52-0d95}d-f). For long-wave perturbations, the axial component of velocity is larger by several orders of magnitude than the lateral and vertical components (compare color bar limits between Fig.\ \ref{Fig: Holdup_0d52-0d95}a,b and Fig.\ \ref{Fig: Holdup_0d52-0d95}c). The lateral and vertical velocities of long-wave perturbations are scaled by $\alpha$, i.e., $\displaystyle|u_x|/\alpha$ and $\displaystyle|u_y|/\alpha$ stay constant for $\displaystyle\alpha\le0.001$. Due to the flow symmetry, there is no flow across the vertical centerline, $x=0$, so that $u_x(x=0)=0$ (Fig.\ \ref{Fig: u_interface}a). The maximum of the perturbation amplitude $u_x$ lies either in the air phase in the vicinity of the interface (e.g., for small water holdups) or at the interface itself (e.g., for $h\ge0.52$), when there is a secondary maximum in the upper part of the pipe. The amplitude of the vertical velocity component, $u_y$, of the critical perturbation reaches its maximum in the air phase. For $h=0.1, 0.51$, the maximum is close to the triple points. Those regions of the pipe are also characterized by dense clustering of the perturbation amplitude contour lines, since significant interfacial and wall shear stresses act on the flow there. On the other hand, for larger holdups, i.e., $h=0.52, 0.95$, the maximum of $u_y$ is located at the centerline, $x=0$.

Fig.\ \ref{Fig: u_interface} shows the distribution along the interface of the two cross-sectional velocity components and the interfacial displacement of the critical perturbation. The latter is related to the vertical velocity through the kinematic boundary condition (Eq.\ \ref{Eq: BC_kinematic_lin}) that describes the interface as a transport barrier for each phase. Worth noting is that for low holdups, the maximal interface displacement is away from the pipe vertical centerline (Fig. \ref{Fig: u_interface}c). As mentioned above, in the considered air--water flow, the maximal values of the critical perturbation amplitude are reached at the interface, so that the interfacial instability can be expected. Thereby the interface will not remain smooth upon crossing the stability boundary. 

\subsection{Effect of pipe diameter}

In the previous section, we analyzed linear stability in a pipe of particular diameter of $D=0.05$m. In contrast to single-phase channel flow, stability characteristics of the two-phase stratified flow, in particular critical superficial velocities, are strongly affected by the channel size \cite[e.g.,][]{Barmak16a}. 

\begin{figure}[h!]
	\centering
	\subfloat[Stability map]{\includegraphics[width=0.33\textwidth,clip]{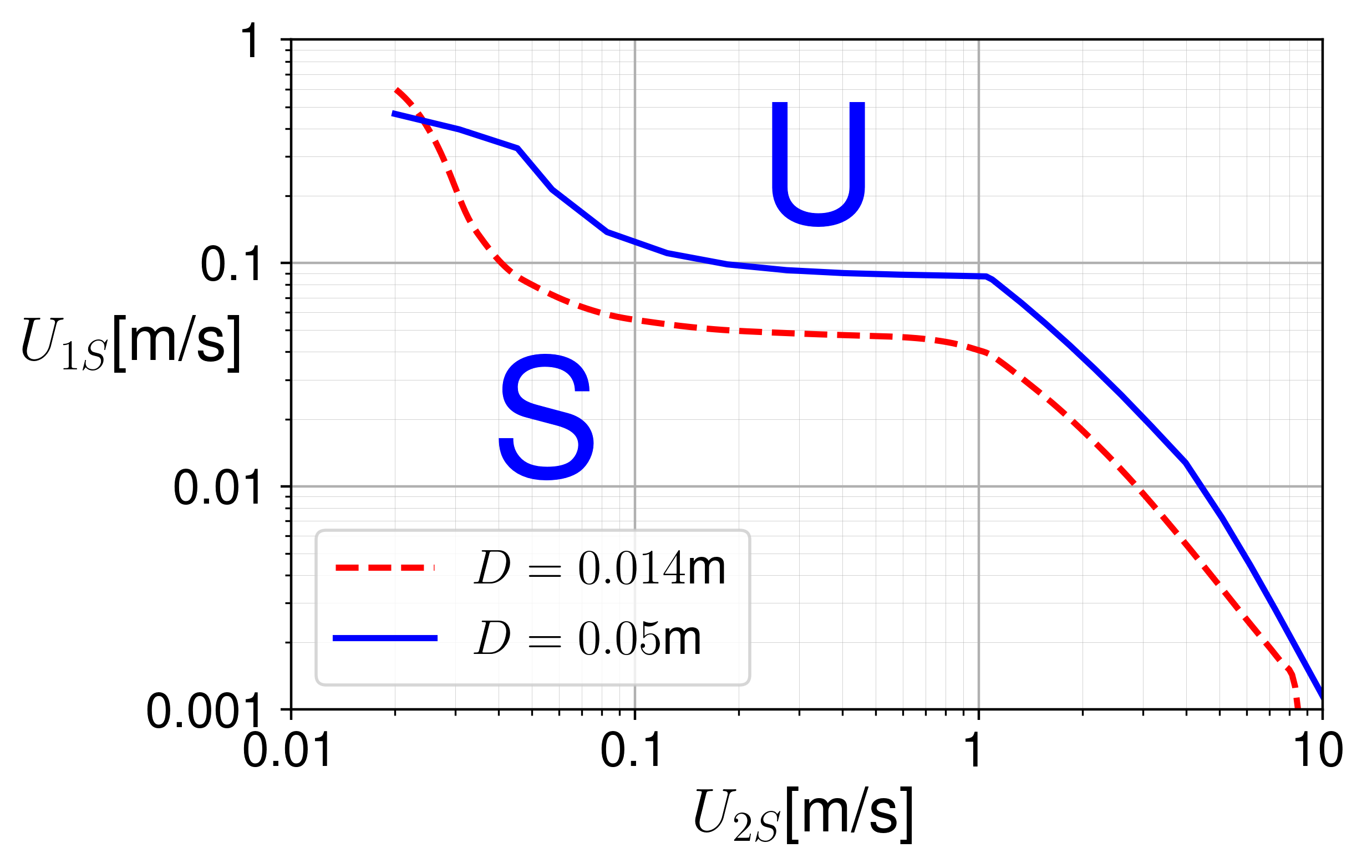}}
	\subfloat[]{\includegraphics[width=0.33\textwidth,clip]{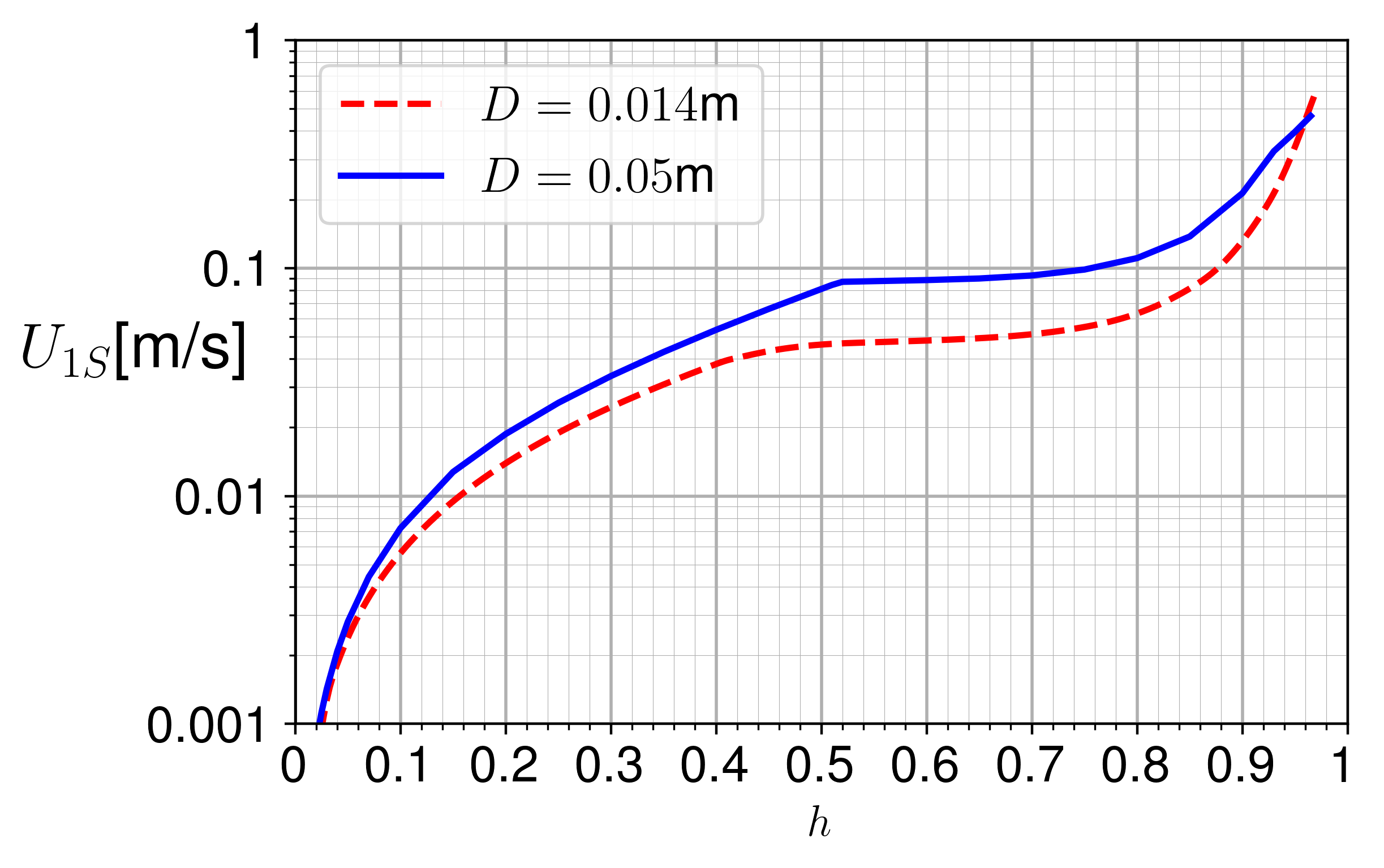}}
	\subfloat[]{\includegraphics[width=0.33\textwidth,clip]{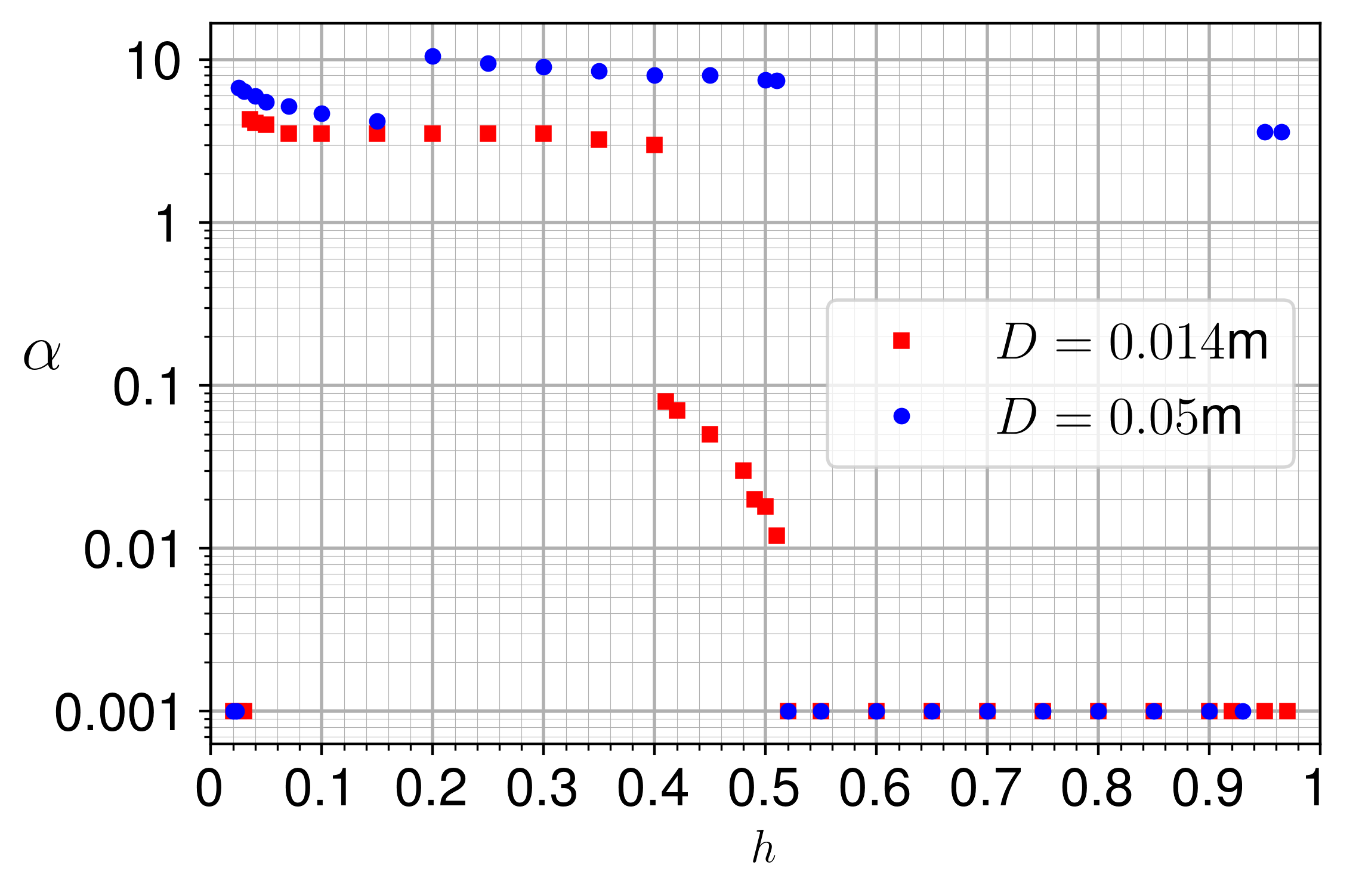}}
	\caption{\label{Fig: Stability_diameter_effect}Effect of the pipe diameter on stability boundary: (a) Stability boundaries on the flow pattern map; (b) critical water superficial velocity as a function of holdup.}	
\end{figure}

In order to analyze the effect of diameter, we consider a flow in a pipe of diameter $D=0.014$m. The stability boundary for the flow is obtained for the whole range of water holdups in the same manner as for the pipe of larger diameter. It is shown by dashed red line on the flow pattern map in Fig.\ \ref{Fig: Stability_diameter_effect}a. 

Comparing it with the stability boundary for $D=0.05$m (blue line), the decrease in diameter is found to result in a smaller stable region for the whole range of holdups. Note that the dimensionless base flow characteristics do not depend on the pipe diameters, therefore the same flow rate ratios in two systems correspond to the same holdups. The critical water superficial velocities for the same holdup (Fig.\ \ref{Fig: Stability_diameter_effect}b) are almost not affected by the pipe diameter for low values of the holdup ($h<0.05$). For high holdups ($h>0.9$), however, the critical perturbation in the pipe of smaller diameter is long wave, while there is a shift to short-wave perturbations for the larger diameter of $D=0.05$m (Fig.\ \ref{Fig: Stability_diameter_effect}c). For $D=0.014$m, there are three modes that can be distinguished for intermediate holdups. For a range of holdups less than $0.4$, the stability boundary is defined by the short-wave perturbations. The critcal wavenumber decreases gradually from $\alpha_\text{crit}\approx4.3$ (for $h\approx0.035$) to $3$ (for $h=0.4$). Then there is a mode change to the lower $\alpha_\text{crit}=0.08$ (higher wavelength). For higher holdups up to $0.52$, this critical mode evolves continuously into the long-wave perturbation with increase of holdup up to $0.52$, which then dominates the neutral stability boundary up to very high values of holdup.
{\renewcommand{\arraystretch}{1.5}
	\begin{table}[h!]
		\caption{\label{Tab: Effect_diameter}Critical parameters for air--water flow with low water holdup ($h=0.03$; $U_{1S}/U_{2S}=1.536\times10^{-4}$) in circular pipes of different diameters. $\We_{2S}= \rho_2 D U_{2S}^2/\sigma$. $\Rey_{2S} = \rho_2 D U_{2S}/\mu_2$.}				
		\centering		
		\begin{tabular}{|c|c|c|c|c|c|c|}
			\hline\hline
			$D$ & $U_{1S|\text{crit}}$ & $U_{2S|\text{crit}}$ & $\alpha_\text{crit}$ & $c_\text{crit}$ & $\We_{2S|\text{crit}}$ & $\Rey_{2S|\text{crit}}$
			\\
			$\big[m\big]$ & $\bigg[\dfrac{m}{s}\bigg]$ & $\bigg[\dfrac{m}{s}\bigg]$ & 
			& & &
			\\
			\hline
			0.002 & 0.00253 & 16.477 & $\to0$ & 0.0114 & 7.541 & 1830.8 \\
			\hline
			0.005 & 0.00175 & 11.416 & $\to0$ & 0.0114 & 9.051 & 3171.2 \\
			\hline
			0.01  & 0.00133 & 8.682  & $\to0$ & 0.0114 & 10.469 & 4823.4 \\
			\hline
			0.014 & 0.00128 & 8.303  & $\to0$ & 0.0113 & 13.405 & 6457.9 \\
			\hline
			0.02  & 0.00125 & 8.138  & 4.889  & 0.0236 & 18.398 & 9042.7 \\
			\hline
			0.05  & 0.00142 & 9.262  & 6.411  & 0.0243 & 59.571 & 25727.5 \\
			\hline
			0.1   & 0.00189 & 12.316 & 6.082  & 0.0237 & 210.668 & 68421.6 \\
			\hline
			0.2   & 0.00263 & 17.146 & 5.408  & 0.0234 & 816.663 & 190515.5 \\
			\hline\hline
		\end{tabular}
	\end{table}
}
\begin{figure}[h!]
	\centering
	\includegraphics[width=0.65\textwidth,clip]{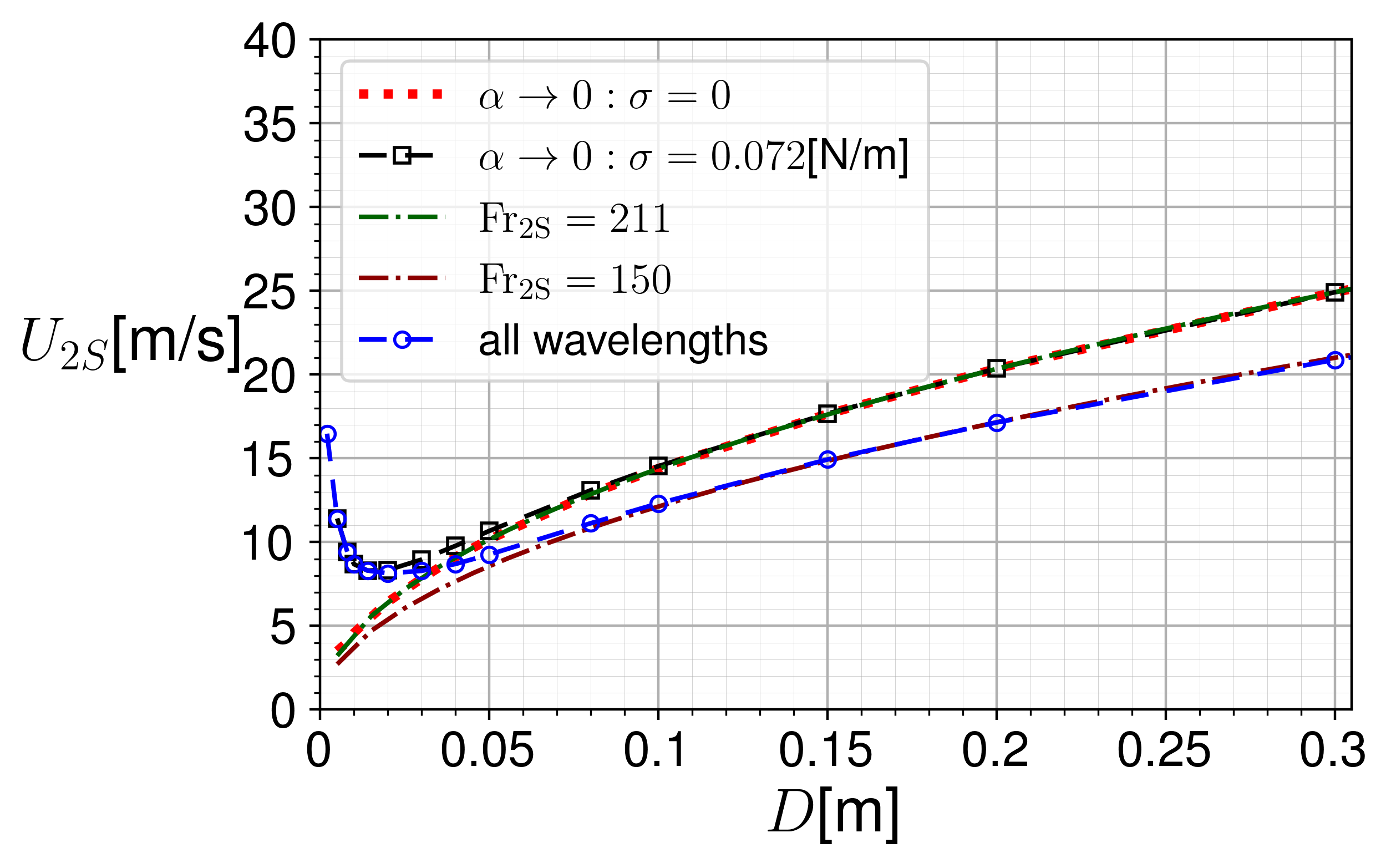}
	\caption{\label{Fig: Critical_U2s_effect_diameter}Effect of pipe diameter on the critical air velocity for low water holdup ($h=0.03$).}	
\end{figure}
\begin{figure}[h!]
	\centering
	\subfloat[$D = 0.005$m, $|u_z|/\max(|u_z|)$]{\includegraphics[width=0.33\textwidth,clip]{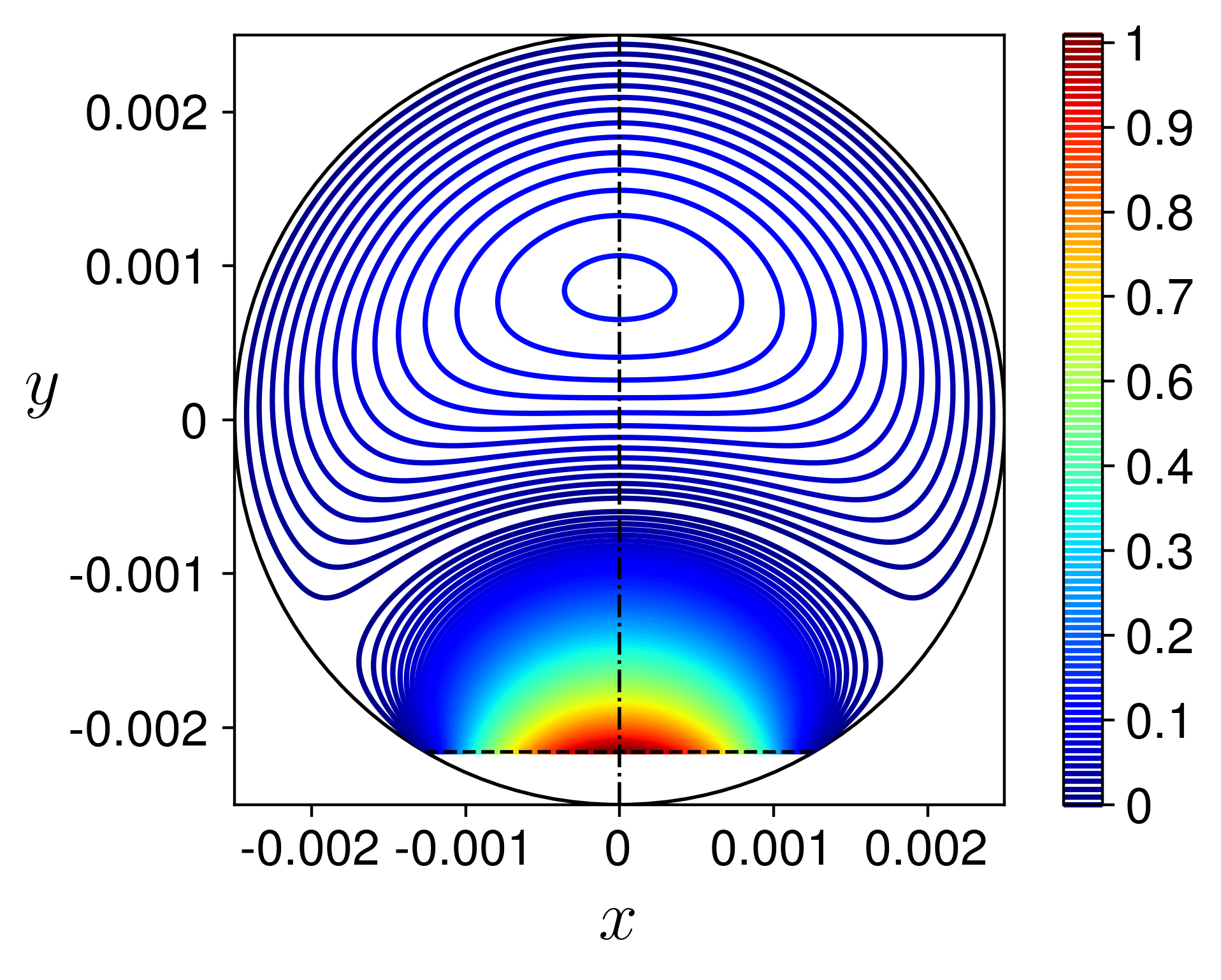}}
	\subfloat[$D = 0.2$m, $|u_z|/\max(|u_z|)$]{\includegraphics[width=0.33\textwidth,clip]{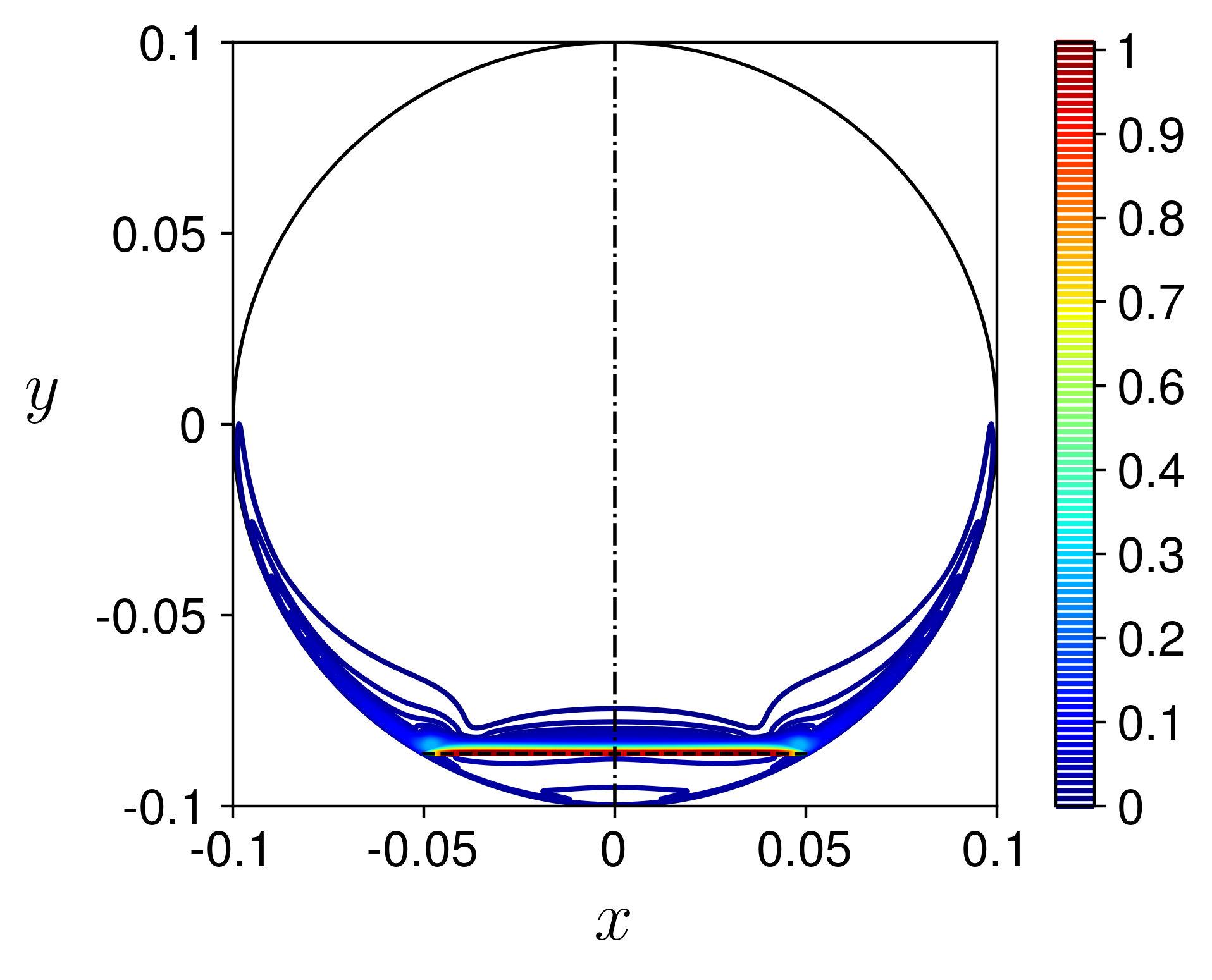}}
	\\
	\subfloat[$D = 0.2$m, $|u_x|/\max(|u_z|)$]{\includegraphics[width=0.33\textwidth,clip]{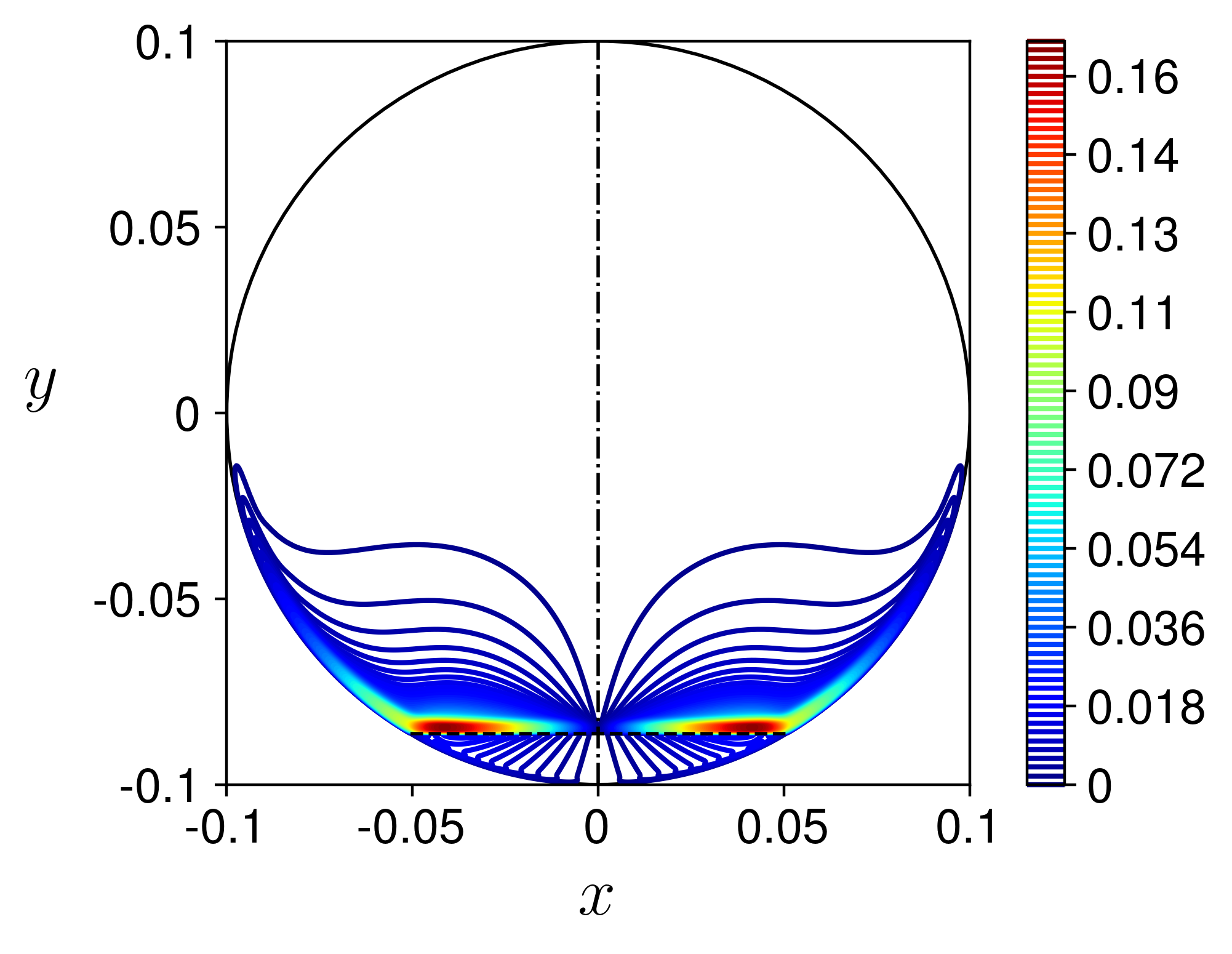}}
	\subfloat[$D = 0.2$m, $|u_y|/\max(|u_z|)$]{\includegraphics[width=0.33\textwidth,clip]{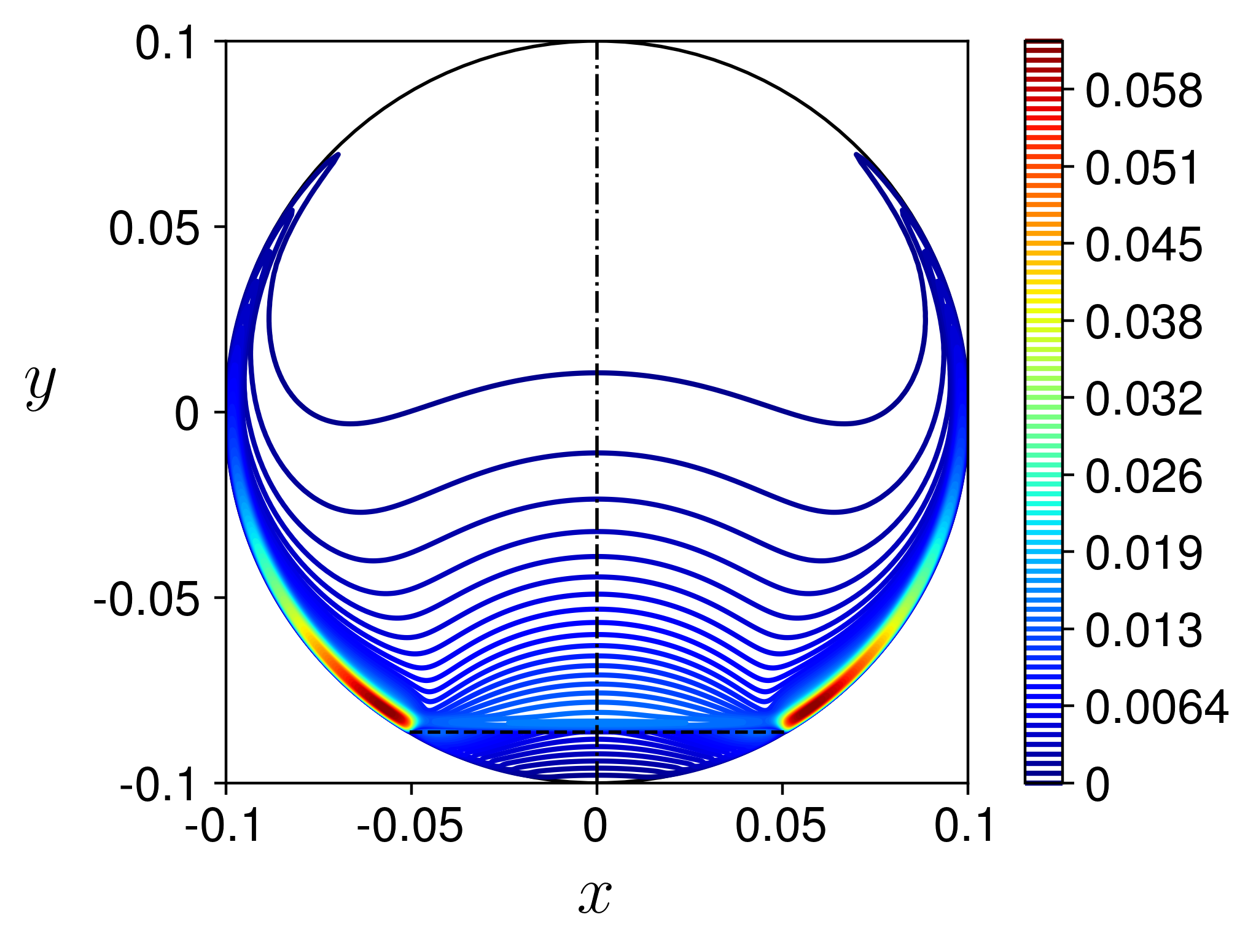}}
	\caption{\label{Fig: Effect_diameter_perturbations}Amplitude contours of the critical perturbation. Axial velocity in a circular pipe of diameter (a) $D=0.005$m (long-wave perturbation, $\alpha\to0$)  and (b) $D=0.2$m (short-wave perturbation, $\alpha=5.5$). (c) Lateral and (d) vertical components of the perturbation velocity for $D=0.2$m. The unperturbed interface is denoted by horizontal dashed black line, and the cross-section centerline -- by vertical dash-dot black line.}	
\end{figure}
\begin{figure}
	\centering
	\includegraphics[width=0.4\textwidth,clip]{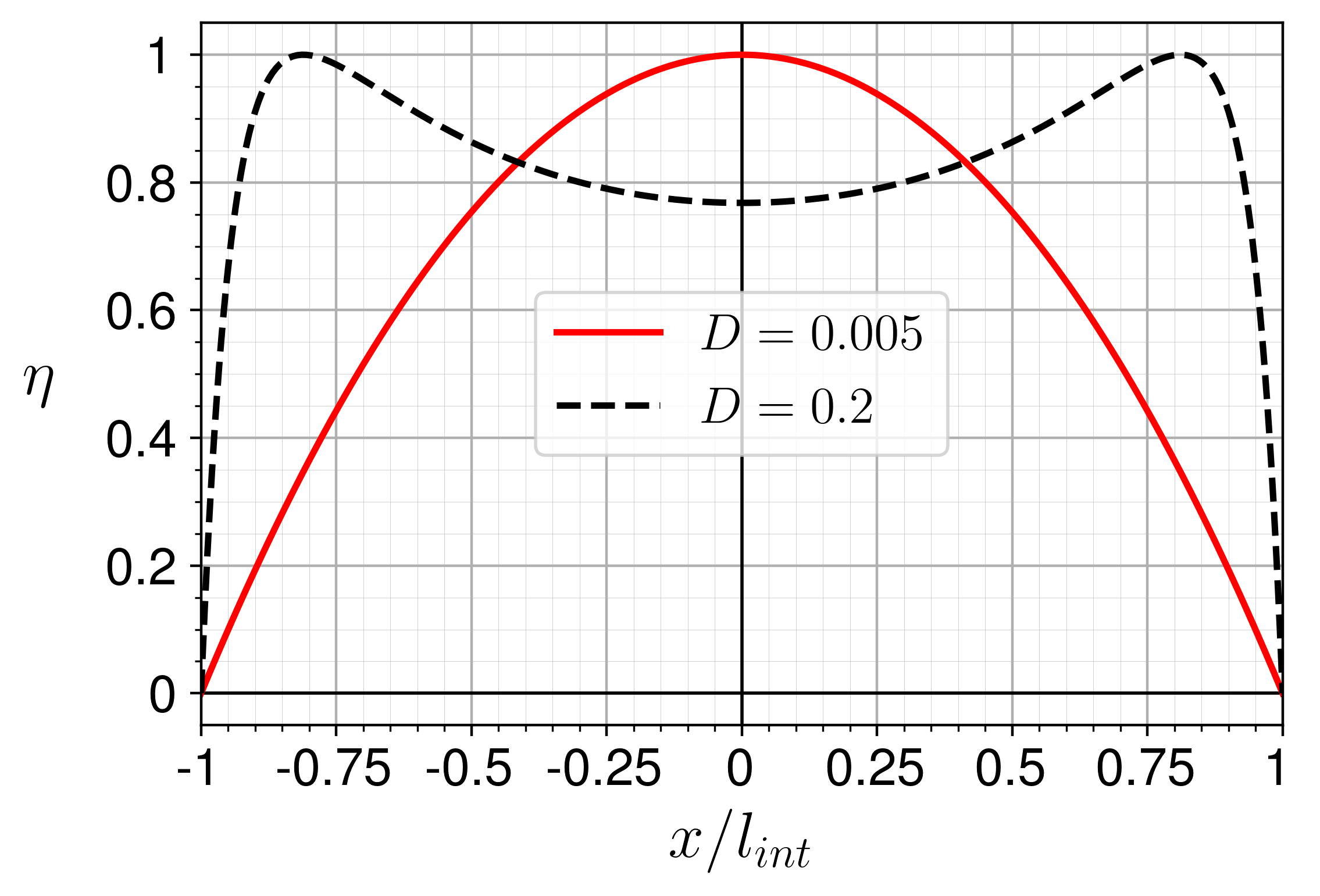}
	\caption{\label{Fig: Eta_effect_diameter}Amplitude profiles of the interfacial displacement (normalized by its maximum) of the long- and short-wave critical perturbations for $D=0.005$m and $D=0.2$m, respectively.}	
\end{figure}

The pipe diameter effect on the critical stability characteristics is shown in Table\ \ref{Tab: Effect_diameter} and Fig.\ \ref{Fig: Critical_U2s_effect_diameter} for low water holdup, where transition from the stratified-smooth to stratified-wavy flow patterns is expected (e.g., for $h=0.03$, which corresponds to $U_{1S}/U_{2S}=1.536\times10^{-4}$, in Fig.\ \ref{Fig: Critical_U2s_effect_diameter}). As shown, with zero surface tension and for long-wave perturbations, the growth in the critical air superficial velocity $U_{2S|\text{crit}}$ (dotted red line) is scaled with the superficial air Froude number ($\displaystyle\Fr_{2S|\text{crit}} = U_{2S|\text{crit}}^2/(gD)\approx211$, dash-dotted dark green line). Yet, for small diameters, the air--water surface tension has a strong stabilization effect also for long waves (due to the capillary force in the lateral direction) and $U_{2S|\text{crit}}$ decreases with $D$ (up to $D\approx0.02$m). However, the decrease of $U_{2S|\text{crit}}$ with $D$ is more moderate than that predicted by a constant Weber number (see Table\ \ref{Tab: Effect_diameter}). Once the effect of the surface tension becomes small, for pipe diameters of $D\ge0.1$m, $U_{2S|\text{crit}}$ increases according to the $\Fr_{2S}$-scaling. 

Considering only long-wave perturbation ($\alpha\to0$, dashed black line with square markers), the critical Froude number is the same as for the case with zero surface tension ($\displaystyle\Fr_{2S|\text{crit}}\approx211$) and the wave speed scaled by the average water base-flow velocity is $c_w=c_\text{crit} U_m h/ U_{1S}\approx2$ for $D\ge0.1$m. The axial-velocity pattern of the critical long-wave perturbation (e.g., for $D=0.005$m) is shown in Fig.\ref{Fig: Effect_diameter_perturbations}a. However, for $D\ge0.02$m, the short-wave perturbations become the critical one (dashed blue line with circle markers). Its wavenumber varies in the range of $\approx5-6.5$, and its wave speed is $c_w\approx4.5$. Another feature of these critical conditions is that $\Fr_{2S|\text{crit}}$ stays also approximately constant, $\displaystyle\Fr_{2S|\text{crit}}\approx150$ (dash-dotted dark red line in Fig.\ \ref{Fig: Critical_U2s_effect_diameter}). The patterns of the three velocity components of the critical short-wave perturbation are shown in Fig.\ \ref{Fig: Effect_diameter_perturbations}b-d and found to be similar to the critical (short-wave) perturbation for $D=0.05$m and the holdup of $0.1$ (Fig.\ \ref{Fig: Holdup_0d1-0d51}a-c). This perturbation can be classified as the interfacial mode of instability, although its pattern is different from that of the critical long-wave perturbation (Fig.\ \ref{Fig: Effect_diameter_perturbations}a). 

The two critical modes are also distinct on the interface (Fig.\ \ref{Fig: Eta_effect_diameter}): whereas the maximum of its displacement is at the center for long waves, the interface deforms stronger near the pipe wall. This can be explained by the strong wall shear, the result of which are also seen in the velocity patterns of the short-wave critical perturbation (Fig.\ \ref{Fig: Effect_diameter_perturbations}b-d). For all the considered pipe diameters, the velocity pattern of the long-wave perturbation looks the same as in Fig.\ \ref{Fig: Effect_diameter_perturbations}a, even when it ceases to be the most unstable perturbation for $D\ge0.02$m.

\subsection{Effect of channel geometry}

As mentioned in the introduction, most stability studies considered two-phase stratified flows in the simplified two-plate geometry. Such analysis completely disregards the presence of lateral walls in the channel cross-section, thereby assuming it to be one dimensional. The only geometrical parameter characterizing such a system is the distance between two plates. If this distance is equal to the pipe diameter $D=0.014$m, then one can compare stability characteristics obtained in the simplified geometry with those obtained in channels with the realistic square duct and circular pipe cross sections. The details of stability analysis in the TP geometry can be found in \cite{Barmak16a}.

\begin{figure}[h!]
	\centering
	\subfloat[Stability map]{\includegraphics[width=0.33\textwidth,clip]{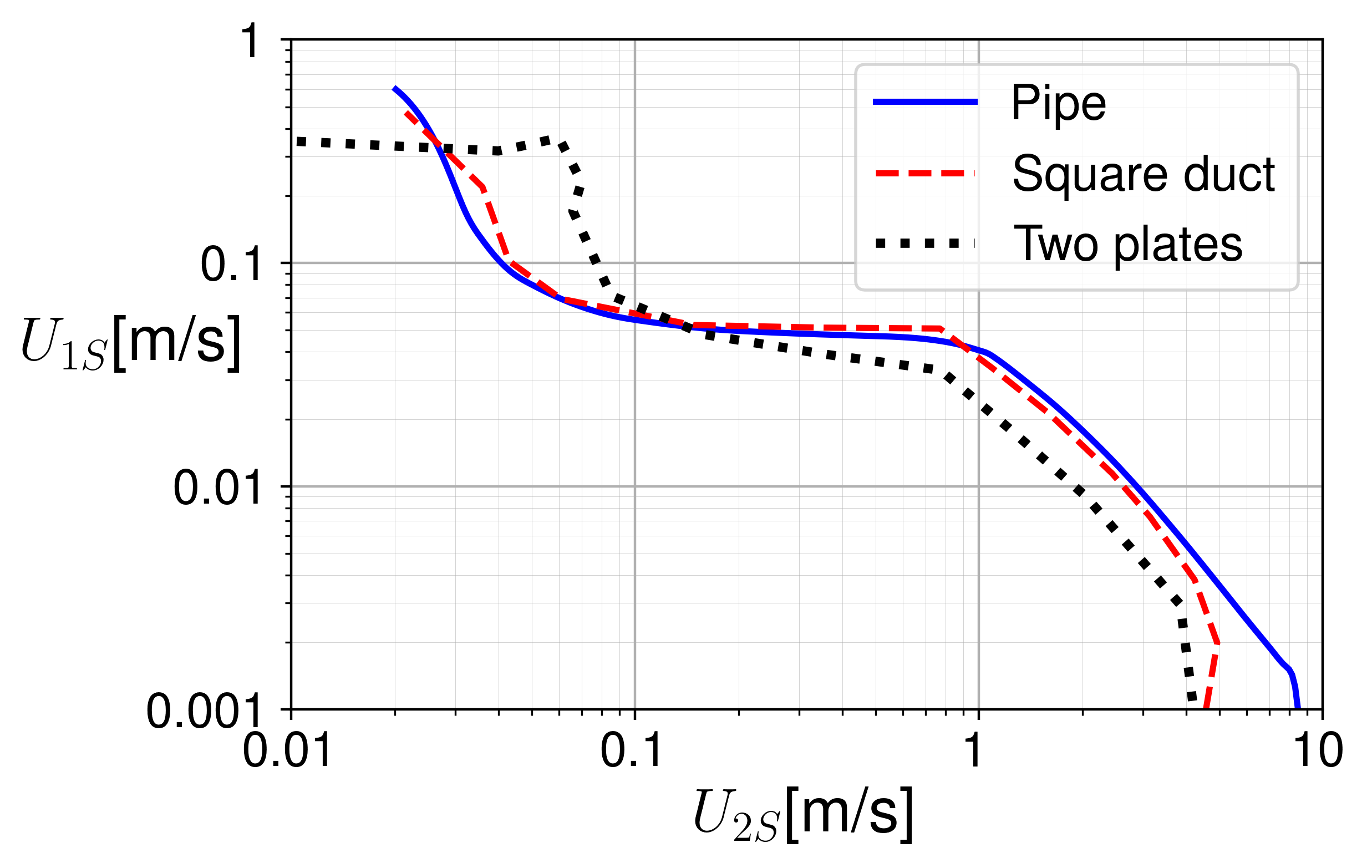}}
	\subfloat[]{\includegraphics[width=0.33\textwidth,clip]{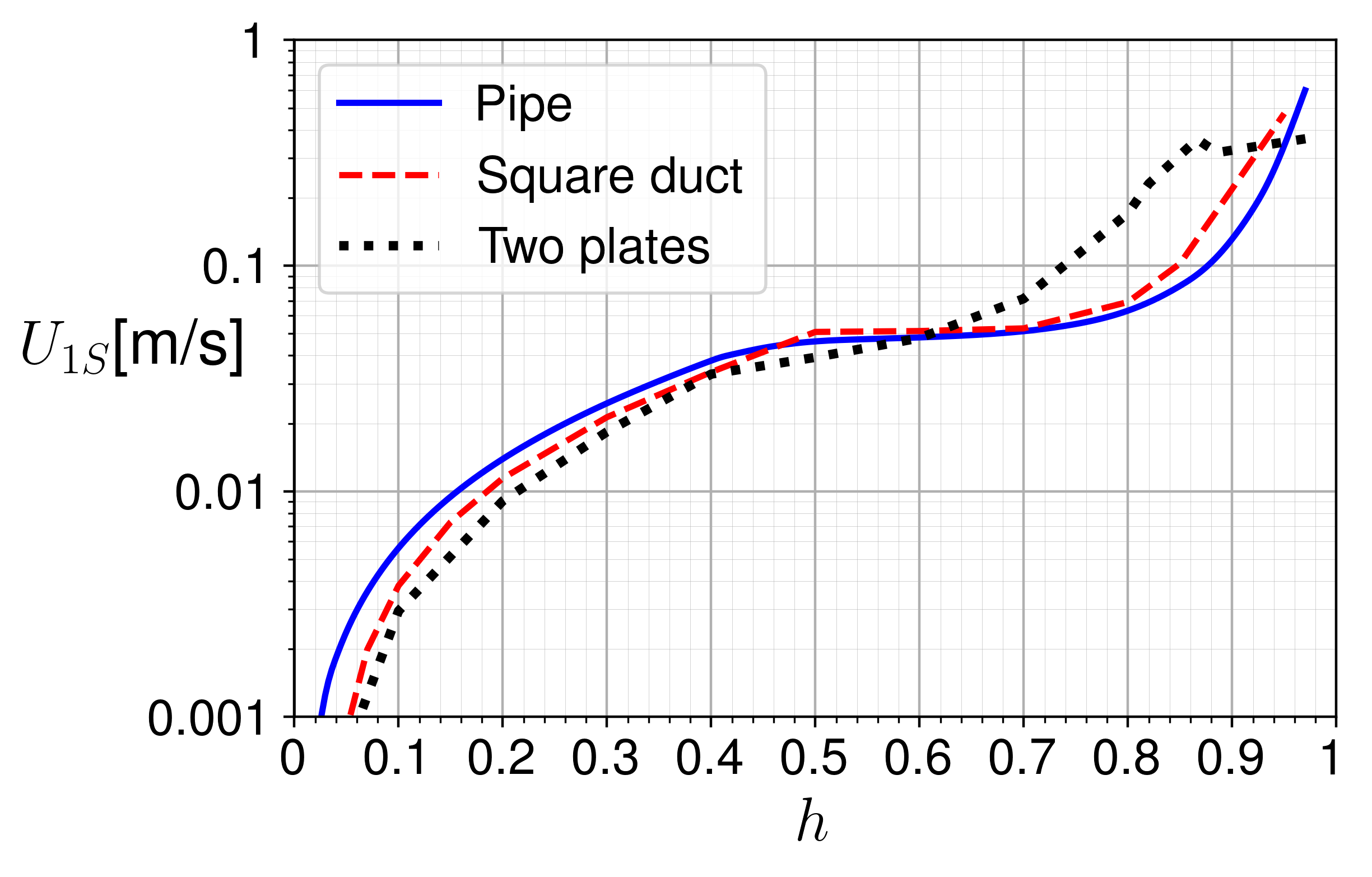}}
	\subfloat[]{\includegraphics[width=0.33\textwidth,clip]{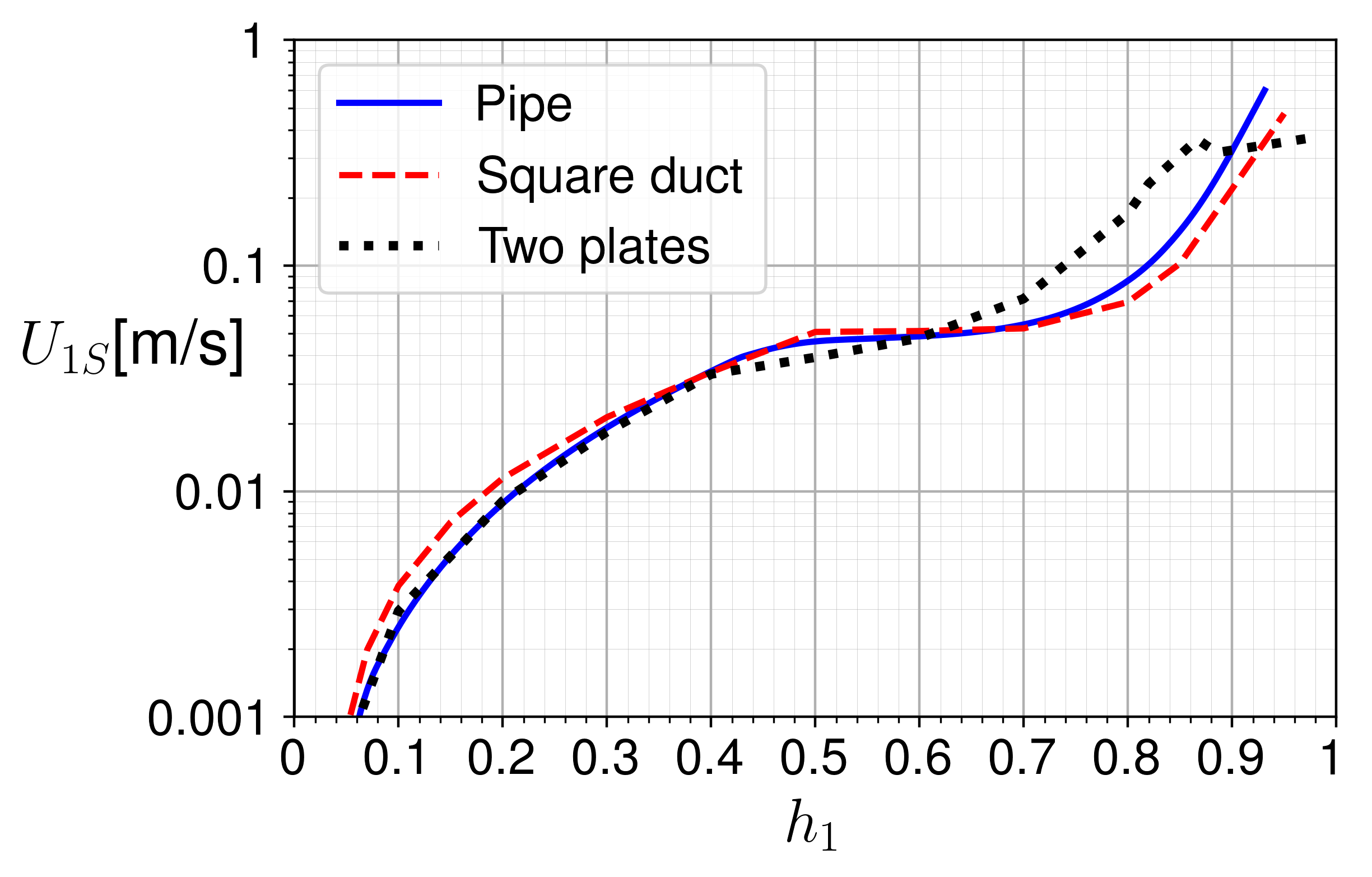}}
	\caption{\label{Fig: Stability_map_geometry_effect}Effect of geometry on stability boundaries of air-water ($\displaystyle\mu_{1 2} = 55$, $\displaystyle\rho_{1 2} = 1000$, $\sigma = 0.072$ N/m) flow in channels with characteristic size $D=0.014$m: (a) Stability map; (b) critical water superficial velocity as a function of holdup; (c) critical water superficial velocity as a function of the (dimensionless) interface height.}	
\end{figure}

Stability boundary for air--water flow between two plates is shown by black dots in Fig.\ \ref{Fig: Stability_map_geometry_effect}. Comparing it with the stability boundary for the circular pipe geometry, which is shown in blue and discussed in detail above (shown in blue, Fig.\ \ref{Fig: Stability_map_geometry_effect}a), one can see the apparent difference between results for the whole range of holdups. For low holdups up to $h\approx0.6$, the critical superficial velocities are lower in the TP geometry than in the pipe, while for higher holdups, they are higher. The holdup in the TP geometry is, in fact, the same as the (dimensionless) thickness of the lower water layer, $h_1$ (Sec.\ \ref{Sec: Formulation}). In circular pipe, on the other hand, the holdup $h$, defined as the relative area of the cross section occupied by the water layer, is always different from $h_1$ except for $h=0.5$ (i.e., for $\phi_0=\pi/2$, compare $(\phi_0-0.5\sin2\phi_0)/\pi$ with $0.5(1-\cos\phi_0)$, Sec.\ \ref{Sec: Formulation}), when water occupies exactly the half of the pipe. The difference between $h_1$ and $h$ is larger the further the holdup is from $0.5$. Comparing the critical water superficial velocity obtained in the pipe and TP geometries (Fig.\ \ref{Fig: Stability_map_geometry_effect}b), it is seen that the difference between them is also larger when the holdup is smaller and further away from $0.5$. This points at possible relevance of comparison of the critical parameters for the same water level, i.e., the same $h_1$. Indeed, once the critical $U_{1S}$ is redrawn in terms of $h_1$ (Fig.\ \ref{Fig: Stability_map_geometry_effect}c), the stability curve for the pipe almost overlaps with that for the TP geometry for the range of $\displaystyle h<0.4$. 

Although the comparison of different geometries has been demonstrated for particular channel size, the difference between the critical velocities in the pipe and two-plate geometries (when compared for the same  $h_1$) remains relatively small for small water layer heights also for other values of $D\in[0.01,0.1]$m (typically used in lab-scale experiments). However, for high water holdups, the stability boundary for the TP geometry diverges from both those of the pipe and square duct, which yield similar stability characteristics as a result of the critical long-wave perturbation in both geometries. In the absence of the side walls in the TP geometry, short-wavelength perturbations bound the stable region, which is otherwise unbounded for low air superficial velocities with respect to long waves \cite{Barmak16a}.

\begin{figure}[h!]
	\centering
	\subfloat[Pipe]{\includegraphics[width=0.35\textwidth,clip]{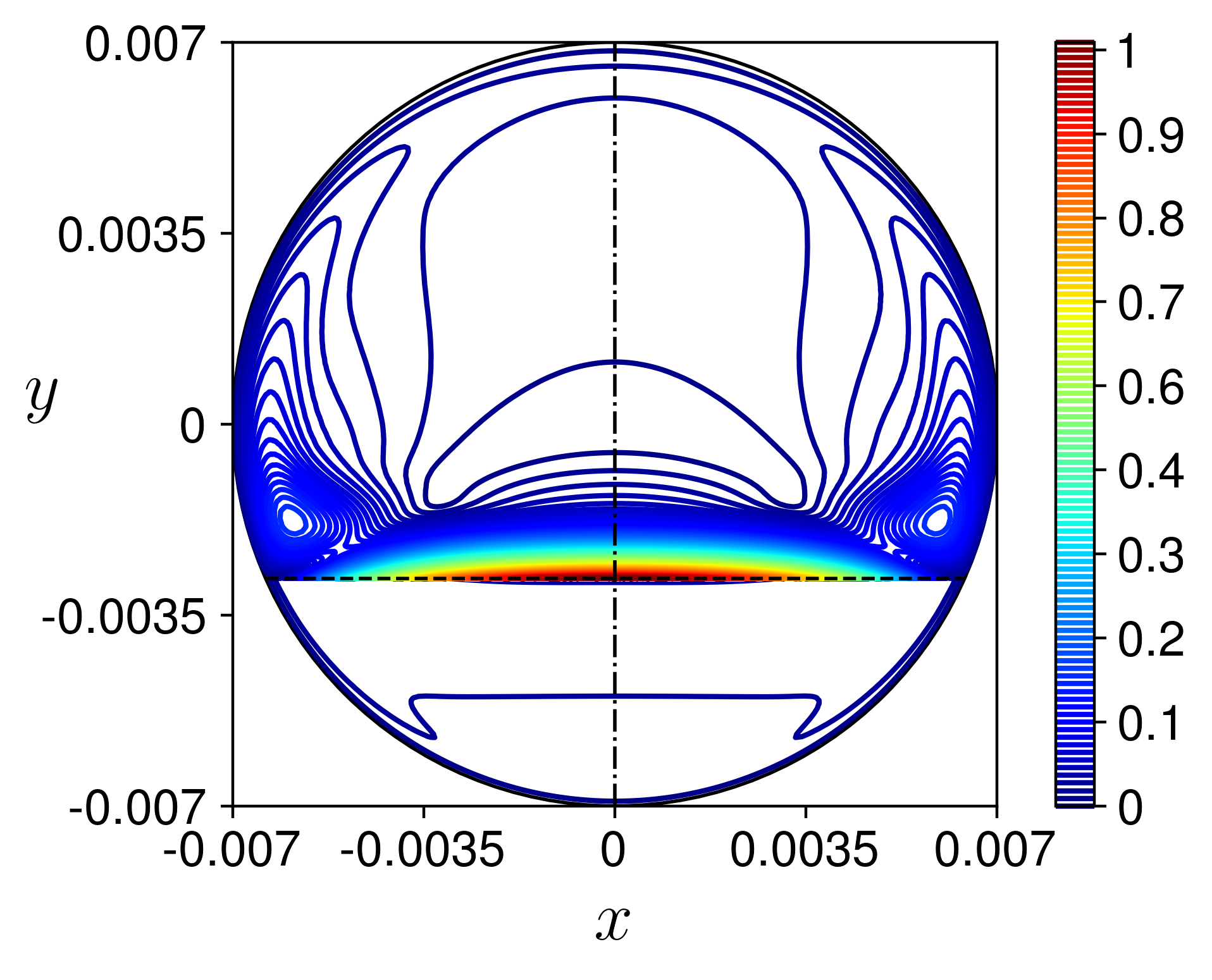}}
	\subfloat[Square duct]{\includegraphics[width=0.35\textwidth,clip]{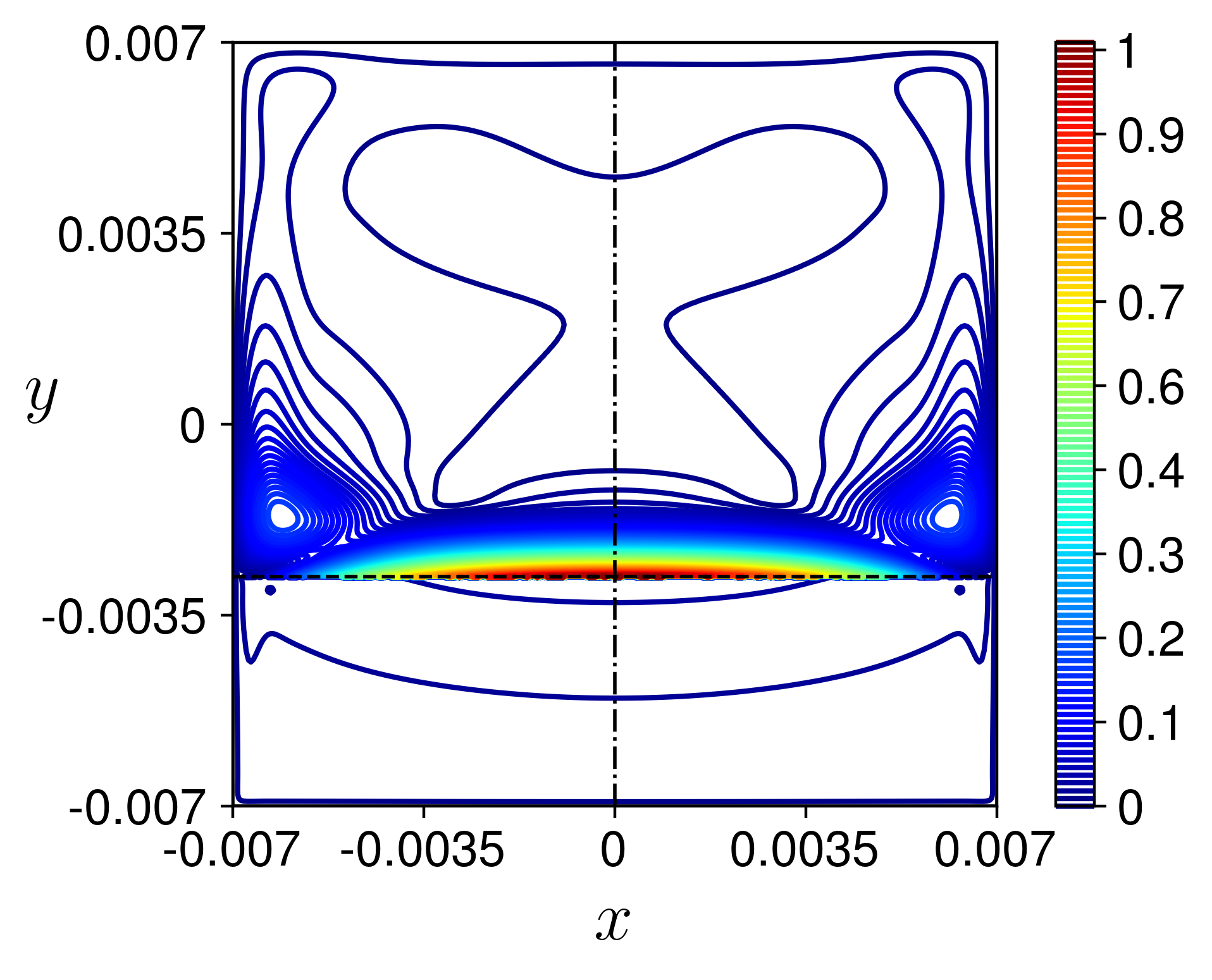}}
	\caption{\label{Fig: u_z_geometry_effect}Effect of geometry on the axial velocity of the critical perturbation of air-water flow in channels with characteristic size $D=0.014$m: (a) Circular pipe, $h=0.252$ ($h_1 = 0.3$); (b) square duct, $h=0.3$.}	
\end{figure}

\begin{figure}[h!]
	\centering
	\subfloat[$|u_z|/\max(|u_z|)$]{\includegraphics[width=0.33\textwidth,clip]{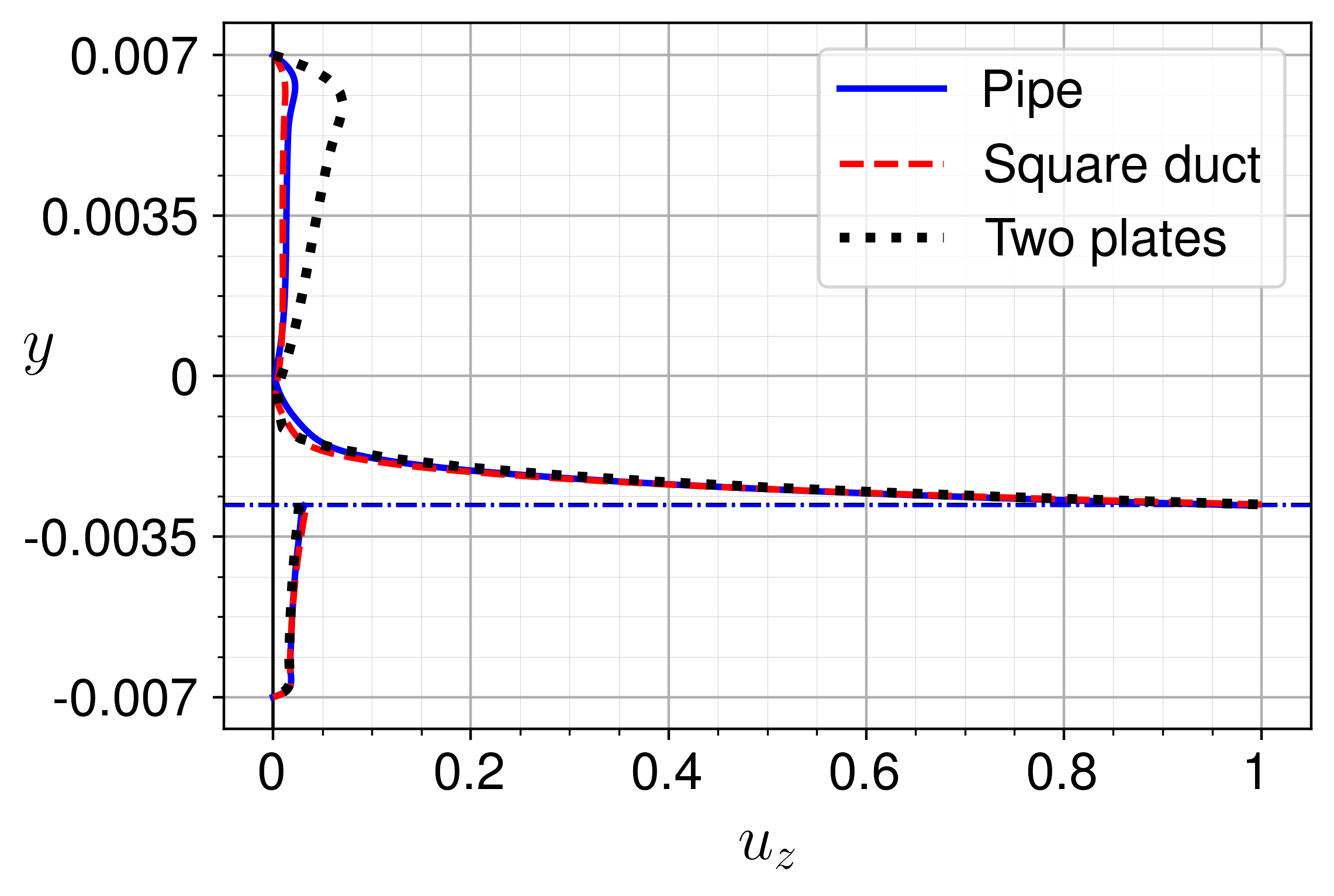}}
	\subfloat[$|u_{x,y}|/\max(|u_z|)$]{\includegraphics[width=0.33\textwidth,clip]{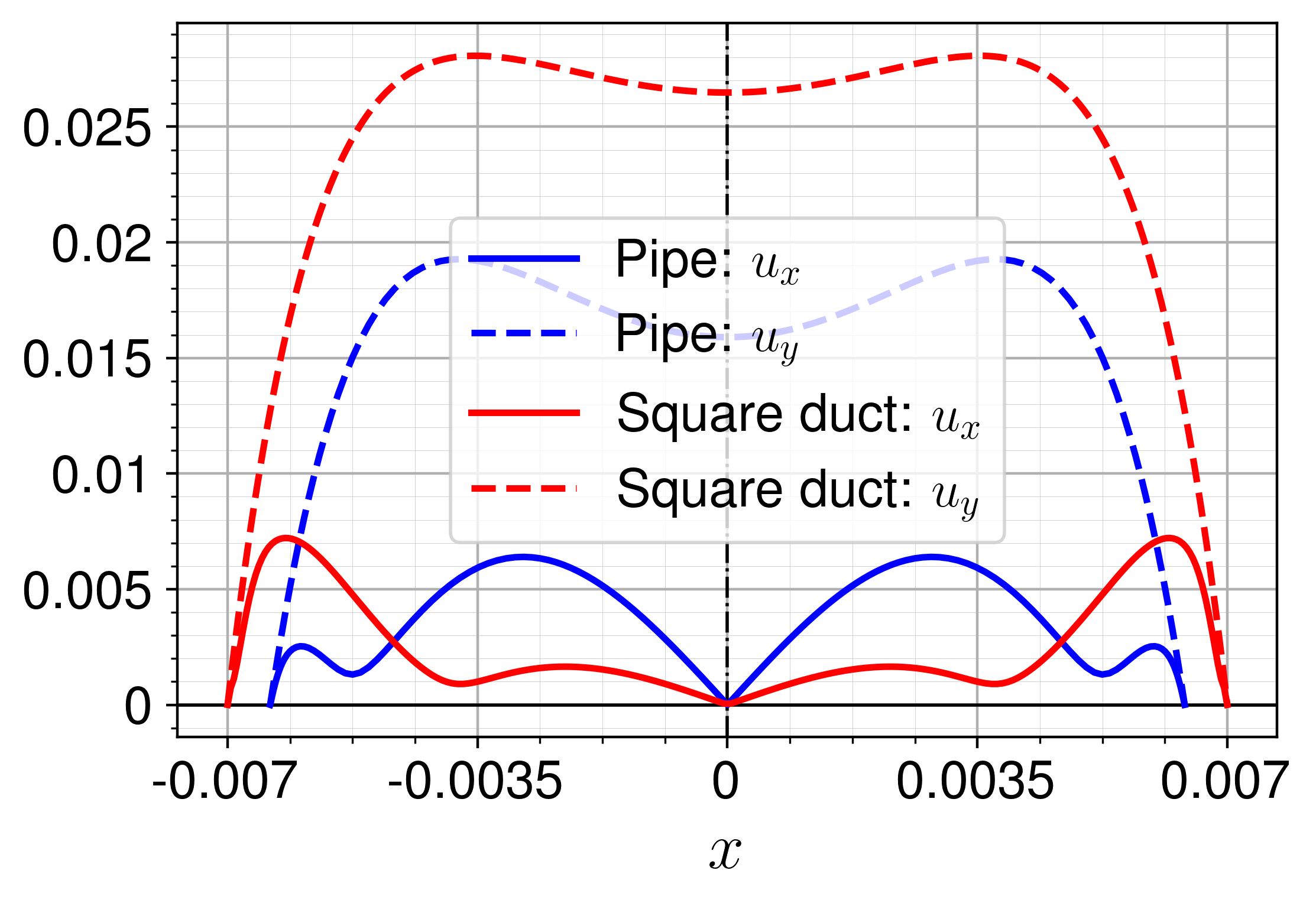}}
	\subfloat[$|\eta|/\max(|\eta|)$]{\includegraphics[width=0.33\textwidth,clip]{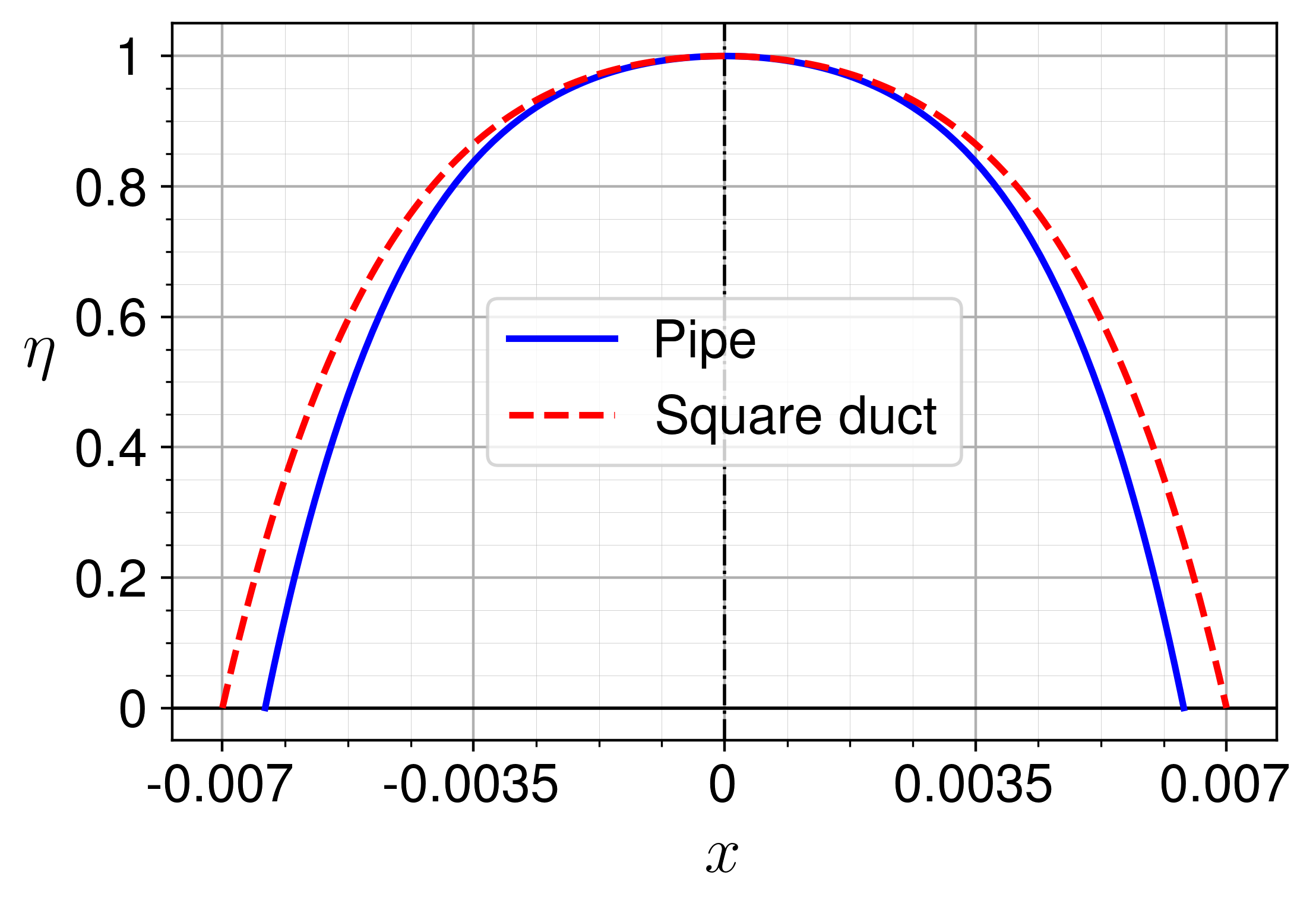}}
	\caption{\label{Fig: Perturbation_geometry_effect}Effect of geometry on the critical perturbation profile of air-water flow in channels with characteristic size $D=0.014$m: (a) Axial velocity on the vertical centerline; (b) lateral and vertical velocities at the interface; (c) interface displacement. $h=0.252$ ($h_1 = 0.3$) for pipe flow and $h=0.3$ for square duct and TP geometries.}	
\end{figure}

\begin{figure}[h!]
	\centering
	\subfloat[Pipe]{\includegraphics[width=0.3\textwidth,clip]{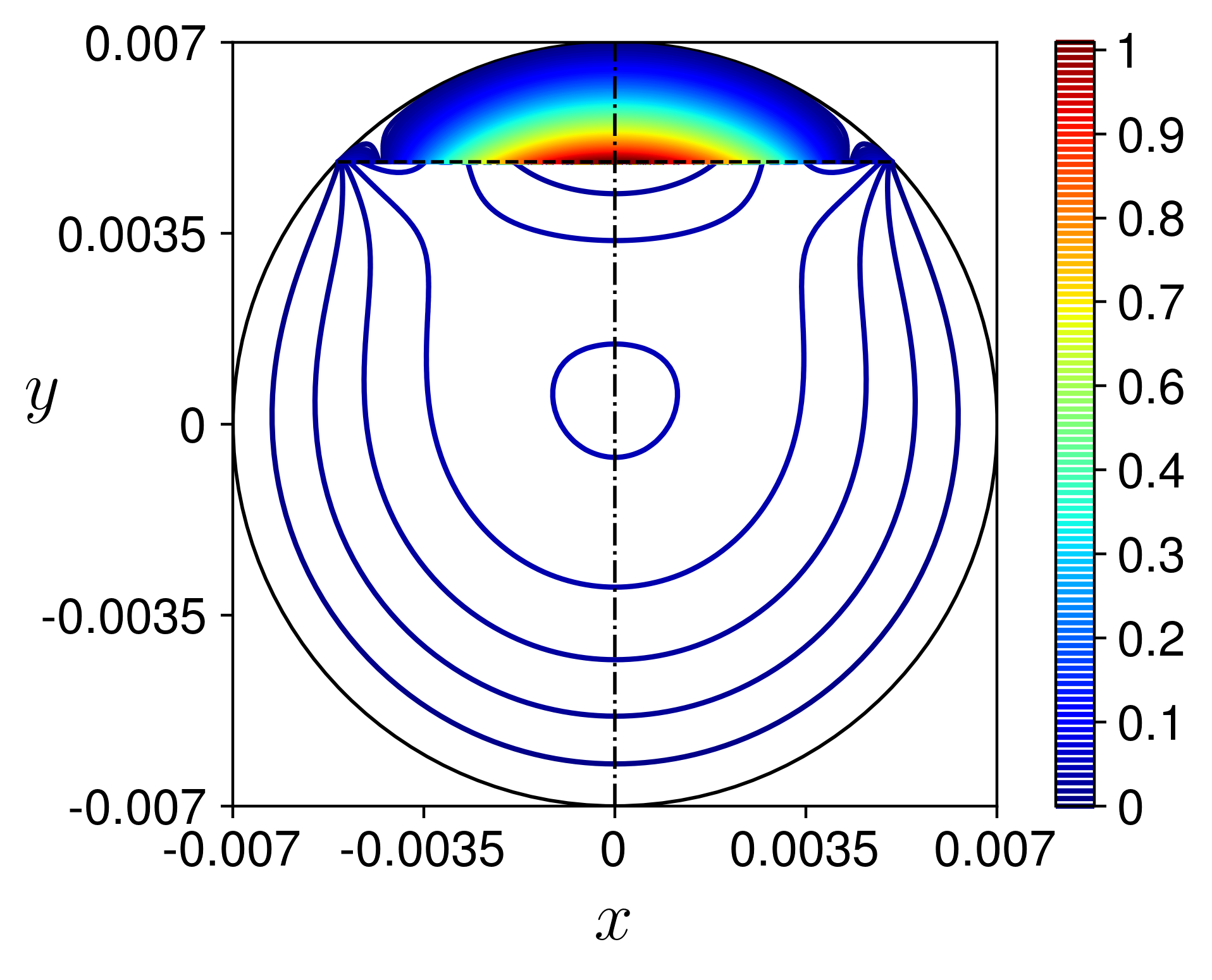}}
	\subfloat[Square duct]{\includegraphics[width=0.3\textwidth,clip]{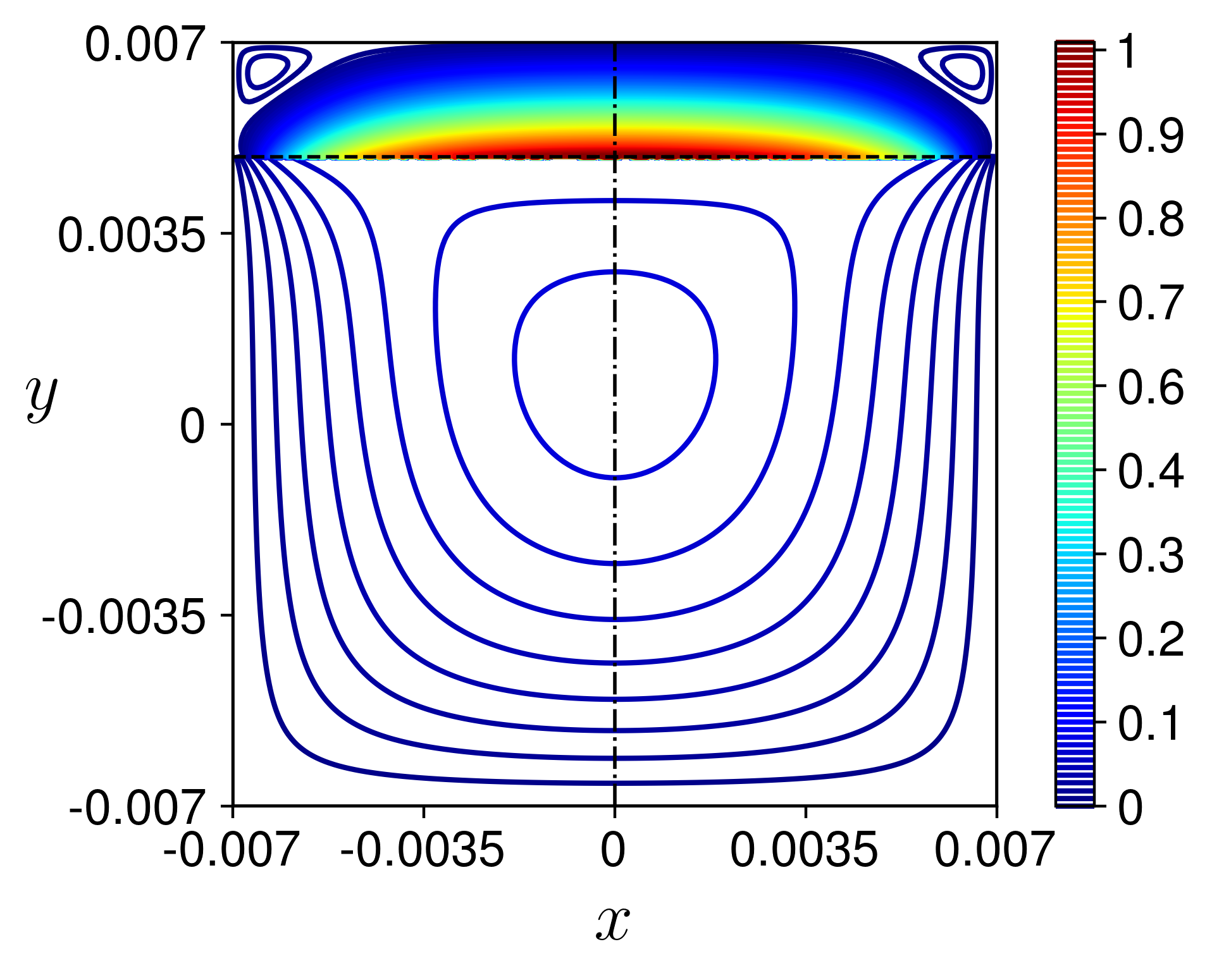}}
	\subfloat[$|u_z|/\max(|u_z|)$]{\includegraphics[width=0.37\textwidth,clip]{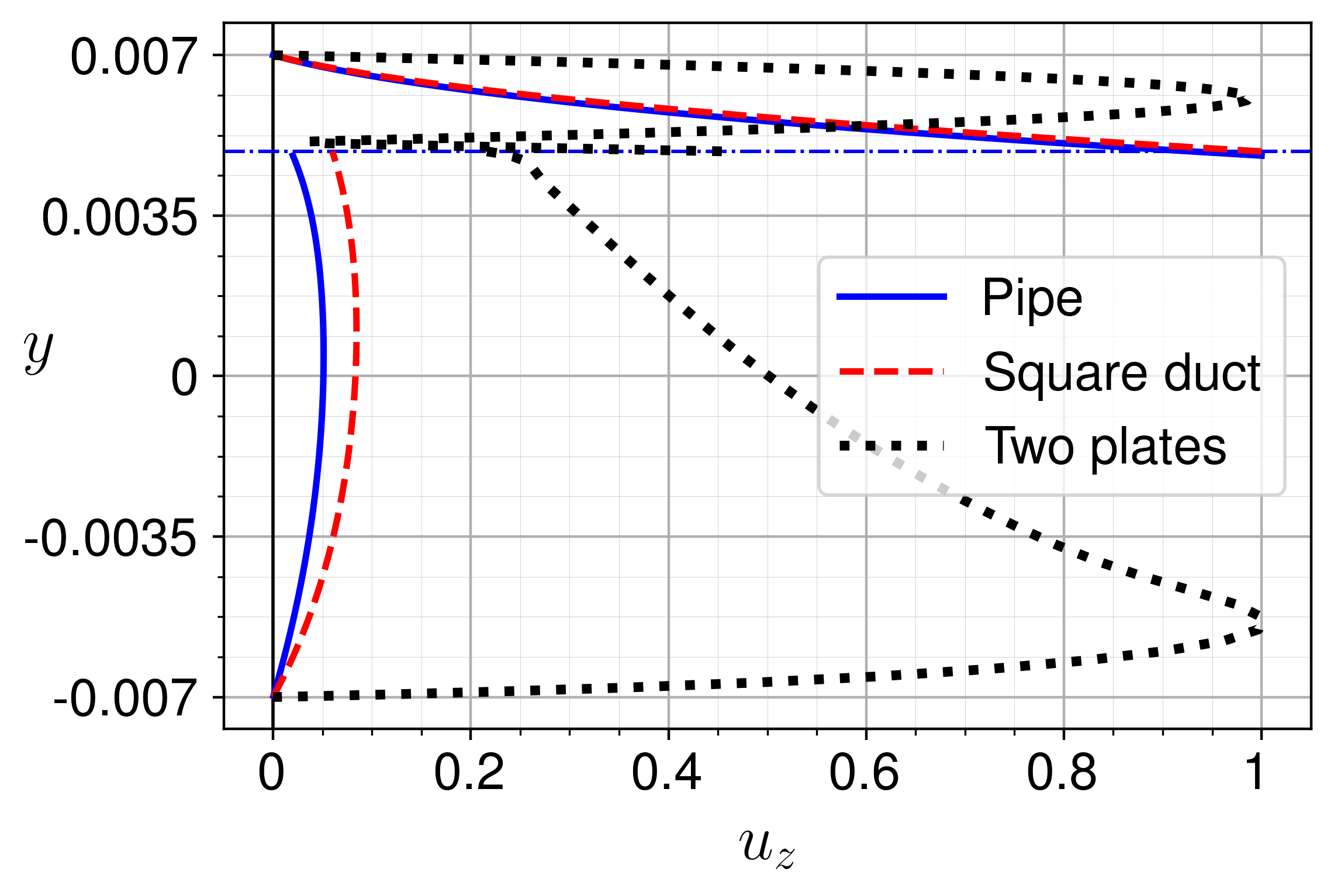}}
	\caption{\label{Fig: Perturbation_geometry_effect_h1_0d85}Effect of geometry on the critical perturbation profile of air-water flow in channels with characteristic size $D=0.014$m: (a) Circular pipe, $h=0.905$ ($h_1 = 0.85$); (b) square duct, $h=h_1=0.85$; (c) axial velocity on the vertical centerline.}	
\end{figure}

In order to compare the critical perturbations for the different geometries, we consider the same $h_1 = 0.3$, which in case of the circular pipe corresponds to $h \approx 0.252$ ($\phi_0 \approx 1.159$). The critical perturbation wavenumbers are found to be similar: $\alpha_\text{crit} = 3.5$ -- for pipe, $\alpha_\text{crit} = 4.038$ -- for square duct, $\alpha_\text{crit} = 3.286$ -- for TP. Two-dimensional contours of the amplitude of the axial velocity are shown in Fig.\ \ref{Fig: u_z_geometry_effect}a and b for circular and square cross sections, respectively. Similar behavior is observed far from the side walls in the center region of the cross section. When the results for the vertical centerline are compared, they show similar critical perturbation profiles for the three considered geometries (Fig.\ \ref{Fig: Perturbation_geometry_effect}a). The only difference is the presence of a secondary maximum of the axial velocity in the air layer, close to the upper wall, in the TP geometry. The velocity fields in the circular and square cross sections are  different as a result of the different shape of the walls. In the pipe, the contours of the axial velocity are stretched along the curved wall by the wall shear stresses (Fig.\ \ref{Fig: u_z_geometry_effect}a). At the interface, the behavior of the lateral and vertical velocities of the critical perturbation is somewhat different depending on geometry (Fig.\ \ref{Fig: Perturbation_geometry_effect}b), however their amplitudes are two orders of magnitude smaller than $u_z$. The interface displacement $\eta$ is also not the same in pipe and square duct due to different contact lengths between air and water (Fig.\ref{Fig: Perturbation_geometry_effect}c). 

For higher holdups, from $h\approx0.6$ up to $\approx0.9$, higher critical superficial velocities are predicted in the TP model than those obtained both in the pipe and duct. Moreover, the velocity pattern of the critical perturbation the pipe and square duct geometries is different that that obtained in the TP geometry (see Fig.\ \ref{Fig: Perturbation_geometry_effect_h1_0d85}). In the latter, the maximum of the axial velocity component of the critical (short-wave, $\alpha_\text{crit}=1.333$) perturbation is in the bulk of the air and water layers (see dotted black line in Fig.\ \ref{Fig: Perturbation_geometry_effect_h1_0d85}c), indicating shear mode of instability. On the other hand, in the pipe and square duct, the critical perturbation is long-wave and corresponds to interfacial mode of instability.

In spite of similarities between the stability characteristics for thin water layers in the pipe and TP geometries, the effect of the channel size in these two geometries is different when a wider range of sizes is considered. In the TP geometry and for thin water layers, linear instability was found to be triggered by long-wave perturbation and the critical air velocity is scaled by the air Froude number up to a channel size of about $0.025$m \cite{Barmak16a}. In larger channels, the critical air velocity in the TP geometry was found to be due to short waves, and the critical air velocity is scaled by the air Reynolds number (i.e., $U_{2S|\text{crit}}$ decreases with the channel size). On the other hand, in circular pipes, due to the lateral confinement, the surface tension has a strong stabilization effect not only for short waves, but also for long-wave perturbations. As a result, the diameter effect on the critical air velocity for low liquid holdups is not scaled by the air Froude number for $D<0.1$m, even when instability is set due to long waves. As shown in Fig. \ref{Fig: Critical_U2s_effect_diameter}, for $D\ge0.02$m short-wave disturbances become most unstable, Nevertheless, for $D\ge0.1$m, where the effect of the surface tension due to the lateral confinement becomes small also for low water holdups, the diameter effect on critical air velocity is found to be scaled by the air Froude number, whereby the critical air superficial velocity increases proportionally to $\sqrt{D}$. The value of critical Froude number is however smaller than the critical Froude number that corresponds to long-wave instability.

The different scaling (with channel size) of the critical air velocity obtained in large channels and pipes of large diameters may be attributed to an inherent difference between the stability limit of the Reynolds number in single-phase (e.g, air) flow in these two geometries. In the TP geometry, instability in single phase flow is triggered by a short wave due to the shear mode of instability, which is associated with $\Rey_\text{crit}=5772$ \cite{Orszag71}, However, in pipe geometry, the linear stability analysis does not yield critical $\Rey$ for single-phase flow. Apparently, the shear mode of instability in the gas phase is not triggered even in the presence of a thin water layer, and the predicted stability boundary corresponds to the interfacial instability. The latter is associated with a critical Froude number.

\section{Conclusions}

Linear stability analysis of two-phase stratified air–water flow in horizontal pipes has been performed. The stability problem is formulated and solved for all possible infinitesimal three-dimensional disturbances, and taking into account possible deformation of the air–water interface. The numerical approach is based on the problem formulation in the bipolar coordinates, with the consequent discretization by the finite volume method. The whole computational approach, including calculation of the base flow and solution of the linear stability eigenvalue problem, is described and validated in \cite{Barmak23}. Solving the problem in bipolar coordinates allows us not only to reproduce numerically the analytical solution of \cite{Goldstein15} for the base flow, but also to predict a correct exponential growth of the interfacial shear stresses near the triple points, as reported in \cite{Goldstein21a}.

The whole problem is described by six dimensionless parameters that makes the corresponding parametric studies not feasible. To focus on a certain two-phase system, we consider air-water flow, which has been subject to extensive experimental research. The regions of stable and unstable superficial velocities of each phase are presented as stability diagrams to allow direct comparison with the experimental flow-pattern data. As a case study we consider a flow in a pipe of 50 mm diameter, which was used in well-known experiments of \cite{Mandhane74} and \cite{Barnea82}. The present numerical results are found to agree well with the experimental findings.

The wavenumbers and wave speeds of the critical perturbations, as well as patterns of their three velocity components are reported along with the stability boundaries. Moreover, several modes of the critical perturbations are revealed. Long waves are found to be the critical perturbation over part of the stability boundary. Perturbations of the axial velocity are found to be larger than the two other components for all the critical perturbations obtained along the stability boundary. The vertical velocity, however, defines the behavior of the interface displacement via the kinematic boundary condition. For the considered air–water flows, the maximum of the amplitude of the axial velocity component of the critical perturbation is always reached at the interface, so that the interfacial instability is always observed.

Due to the confinement effect in the lateral direction, the surface tension has a strong stabilization effect not only for short, but also for long-wave disturbances, in particular for low-water holdups. As a result, the critical air velocity that triggers interfacial instability, and thereby transition to stratified-wavy flow of thin water layers, is found to be scaled by the air Froude number only for large pipe diameters ($D\ge0.1$m). The critical disturbances are found to be short waves, and the corresponding critical air Froude number is $\Fr_{2S|\text{crit}}\approx150$. Long-wave instability overpredicts the critical air flow rate and is triggered at $\Fr_{2S|\text{crit}}\approx211$.

\begin{acknowledgments}
	This research was supported by Israel Science Foundation (ISF) grant No 1363/23.
\end{acknowledgments}

\bibliography{Barmak_instability_pipe}

\end{document}

%% file: figures/Fig1.pdf_tex
\begingroup%
  \makeatletter%
  \providecommand\color[2][]{%
    \errmessage{(Inkscape) Color is used for the text in Inkscape, but the package 'color.sty' is not loaded}%
    \renewcommand\color[2][]{}%
  }%
  \providecommand\transparent[1]{%
    \errmessage{(Inkscape) Transparency is used (non-zero) for the text in Inkscape, but the package 'transparent.sty' is not loaded}%
    \renewcommand\transparent[1]{}%
  }%
  \providecommand\rotatebox[2]{#2}%
  \newcommand*\fsize{\dimexpr\f@size pt\relax}%
  \newcommand*\lineheight[1]{\fontsize{\fsize}{#1\fsize}\selectfont}%
  \ifx\svgwidth\undefined%
    \setlength{\unitlength}{464.12417915bp}%
    \ifx\svgscale\undefined%
      \relax%
    \else%
      \setlength{\unitlength}{\unitlength * \real{\svgscale}}%
    \fi%
  \else%
    \setlength{\unitlength}{\svgwidth}%
  \fi%
  \global\let\svgwidth\undefined%
  \global\let\svgscale\undefined%
  \makeatother%
  \begin{picture}(1,0.38455043)%
    \lineheight{1}%
    \setlength\tabcolsep{0pt}%
    \put(0,0){\includegraphics[width=\unitlength,page=1]{Fig1.pdf}}%
    \put(0.94828761,0.17816357){\color[rgb]{0,0,0}\makebox(0,0)[lt]{\lineheight{1.25}\smash{\begin{tabular}[t]{l}$z$\end{tabular}}}}%
    \put(0.89552185,0.24927359){\color[rgb]{0,0,0}\makebox(0,0)[lt]{\lineheight{1.25}\smash{\begin{tabular}[t]{l}$x$\end{tabular}}}}%
    \put(0.81001663,0.3616326){\color[rgb]{0,0,0}\makebox(0,0)[lt]{\lineheight{1.25}\smash{\begin{tabular}[t]{l}$y$\end{tabular}}}}%
    \put(0,0){\includegraphics[width=\unitlength,page=2]{Fig1.pdf}}%
    \put(0.0160396,0.14447676){\color[rgb]{0,0,0}\makebox(0,0)[lt]{\lineheight{1.25}\smash{\begin{tabular}[t]{l}$D$\end{tabular}}}}%
    \put(0,0){\includegraphics[width=\unitlength,page=3]{Fig1.pdf}}%
    \put(0.33363041,0.05398896){\color[rgb]{0,0,0}\makebox(0,0)[lt]{\lineheight{1.25}\smash{\begin{tabular}[t]{l}$Q_1$\end{tabular}}}}%
    \put(0,0){\includegraphics[width=\unitlength,page=4]{Fig1.pdf}}%
    \put(0.33363041,0.22250918){\color[rgb]{0,0,0}\makebox(0,0)[lt]{\lineheight{1.25}\smash{\begin{tabular}[t]{l}$Q_2$\end{tabular}}}}%
    \put(0.51400309,0.05032972){\color[rgb]{0,0,0}\makebox(0,0)[lt]{\lineheight{1.25}\smash{\begin{tabular}[t]{l}$\rho_1, \mu_1$\end{tabular}}}}%
    \put(0.48543697,0.18418246){\color[rgb]{0,0,0}\makebox(0,0)[lt]{\lineheight{1.25}\smash{\begin{tabular}[t]{l}$\rho_2, \mu_2$\end{tabular}}}}%
  \end{picture}%
\endgroup%

%% file: figures/Fig2a.pdf_tex
\begingroup%
  \makeatletter%
  \providecommand\color[2][]{%
    \errmessage{(Inkscape) Color is used for the text in Inkscape, but the package 'color.sty' is not loaded}%
    \renewcommand\color[2][]{}%
  }%
  \providecommand\transparent[1]{%
    \errmessage{(Inkscape) Transparency is used (non-zero) for the text in Inkscape, but the package 'transparent.sty' is not loaded}%
    \renewcommand\transparent[1]{}%
  }%
  \providecommand\rotatebox[2]{#2}%
  \newcommand*\fsize{\dimexpr\f@size pt\relax}%
  \newcommand*\lineheight[1]{\fontsize{\fsize}{#1\fsize}\selectfont}%
  \ifx\svgwidth\undefined%
    \setlength{\unitlength}{151.33764312bp}%
    \ifx\svgscale\undefined%
      \relax%
    \else%
      \setlength{\unitlength}{\unitlength * \real{\svgscale}}%
    \fi%
  \else%
    \setlength{\unitlength}{\svgwidth}%
  \fi%
  \global\let\svgwidth\undefined%
  \global\let\svgscale\undefined%
  \makeatother%
  \begin{picture}(1,0.89030788)%
    \lineheight{1}%
    \setlength\tabcolsep{0pt}%
    \put(0,0){\includegraphics[width=\unitlength,page=1]{Fig2a.pdf}}%
    \put(0.51677805,0.58367591){\color[rgb]{0,0,0}\makebox(0,0)[lt]{\lineheight{1.25}\smash{\begin{tabular}[t]{l}$\phi_0$\end{tabular}}}}%
    \put(0.53978718,0.37536352){\color[rgb]{0,0,0}\makebox(0,0)[lt]{\lineheight{1.25}\smash{\begin{tabular}[t]{l}$A_2$\end{tabular}}}}%
    \put(0.53978718,0.09406718){\color[rgb]{0,0,0}\makebox(0,0)[lt]{\lineheight{1.25}\smash{\begin{tabular}[t]{l}$A_1$\end{tabular}}}}%
    \put(0,0){\includegraphics[width=\unitlength,page=2]{Fig2a.pdf}}%
    \put(0.06209708,0.20912325){\color[rgb]{0,0,0}\makebox(0,0)[lt]{\lineheight{1.25}\smash{\begin{tabular}[t]{l}F$_1$\end{tabular}}}}%
    \put(0.25822076,0.43389051){\color[rgb]{0,0,0}\makebox(0,0)[lt]{\lineheight{1.25}\smash{\begin{tabular}[t]{l}$\rho_2, \mu_2$\end{tabular}}}}%
    \put(0.25821647,0.16776893){\color[rgb]{0,0,0}\makebox(0,0)[lt]{\lineheight{1.25}\smash{\begin{tabular}[t]{l}$\rho_1, \mu_1$\end{tabular}}}}%
    \put(-0.22192885,0.94245326){\color[rgb]{0,0,0}\makebox(0,0)[lt]{\begin{minipage}{1.36638646\unitlength}\raggedright \end{minipage}}}%
    \put(0,0){\includegraphics[width=\unitlength,page=3]{Fig2a.pdf}}%
    \put(0.05447606,0.82002319){\color[rgb]{0,0,0}\makebox(0,0)[lt]{\lineheight{1.25}\smash{\begin{tabular}[t]{l}$y$\end{tabular}}}}%
    \put(0.18191375,0.70369043){\color[rgb]{0,0,0}\makebox(0,0)[lt]{\lineheight{1.25}\smash{\begin{tabular}[t]{l}$x$\end{tabular}}}}%
    \put(-0.12743499,0.96461227){\color[rgb]{0,0,0}\makebox(0,0)[lt]{\begin{minipage}{1.31859841\unitlength}\raggedright \end{minipage}}}%
    \put(0.55283896,0.78958731){\color[rgb]{0,0,0}\makebox(0,0)[lt]{\lineheight{1.25}\smash{\begin{tabular}[t]{l}P\end{tabular}}}}%
    \put(0.94712797,0.20912325){\color[rgb]{0,0,0}\makebox(0,0)[lt]{\lineheight{1.25}\smash{\begin{tabular}[t]{l}F$_2$\end{tabular}}}}%
    \put(0,0){\includegraphics[width=\unitlength,page=4]{Fig2a.pdf}}%
  \end{picture}%
\endgroup%

%% file: figures/Fig2b.pdf_tex
\begingroup%
  \makeatletter%
  \providecommand\color[2][]{%
    \errmessage{(Inkscape) Color is used for the text in Inkscape, but the package 'color.sty' is not loaded}%
    \renewcommand\color[2][]{}%
  }%
  \providecommand\transparent[1]{%
    \errmessage{(Inkscape) Transparency is used (non-zero) for the text in Inkscape, but the package 'transparent.sty' is not loaded}%
    \renewcommand\transparent[1]{}%
  }%
  \providecommand\rotatebox[2]{#2}%
  \newcommand*\fsize{\dimexpr\f@size pt\relax}%
  \newcommand*\lineheight[1]{\fontsize{\fsize}{#1\fsize}\selectfont}%
  \ifx\svgwidth\undefined%
    \setlength{\unitlength}{333.55932521bp}%
    \ifx\svgscale\undefined%
      \relax%
    \else%
      \setlength{\unitlength}{\unitlength * \real{\svgscale}}%
    \fi%
  \else%
    \setlength{\unitlength}{\svgwidth}%
  \fi%
  \global\let\svgwidth\undefined%
  \global\let\svgscale\undefined%
  \makeatother%
  \begin{picture}(1,0.53637734)%
    \lineheight{1}%
    \setlength\tabcolsep{0pt}%
    \put(0,0){\includegraphics[width=\unitlength,page=1]{Fig2b.pdf}}%
    \put(0.16526412,0.14017867){\color[rgb]{0,0,0}\makebox(0,0)[lt]{\lineheight{1.25}\smash{\begin{tabular}[t]{l}$\rho_2, \mu_2$\end{tabular}}}}%
    \put(0,0){\includegraphics[width=\unitlength,page=2]{Fig2b.pdf}}%
    \put(0.48561106,0.44101254){\color[rgb]{0,0,0}\makebox(0,0)[lt]{\lineheight{1.25}\smash{\begin{tabular}[t]{l}$\phi_0+\pi$\end{tabular}}}}%
    \put(0.05111886,0.27759109){\color[rgb]{0,0,0}\makebox(0,0)[lt]{\lineheight{1.25}\smash{\begin{tabular}[t]{l}F$_1$\end{tabular}}}}%
    \put(0.46714138,0.00989857){\color[rgb]{0,0,0}\makebox(0,0)[lt]{\lineheight{1.25}\smash{\begin{tabular}[t]{l}$\phi_0$\end{tabular}}}}%
    \put(0.47195959,0.27938979){\color[rgb]{0,0,0}\makebox(0,0)[lt]{\lineheight{1.25}\smash{\begin{tabular}[t]{l}$\pi$\\\end{tabular}}}}%
    \put(0.45760621,0.5044888){\color[rgb]{0,0,0}\makebox(0,0)[lt]{\lineheight{1.25}\smash{\begin{tabular}[t]{l}$\phi$\\\end{tabular}}}}%
    \put(0.89943636,0.27977569){\color[rgb]{0,0,0}\makebox(0,0)[lt]{\lineheight{1.25}\smash{\begin{tabular}[t]{l}$\xi$\\\end{tabular}}}}%
    \put(0.16526412,0.34677565){\color[rgb]{0,0,0}\makebox(0,0)[lt]{\lineheight{1.25}\smash{\begin{tabular}[t]{l}$\rho_1, \mu_1$\end{tabular}}}}%
    \put(0.77063154,0.27759109){\color[rgb]{0,0,0}\makebox(0,0)[lt]{\lineheight{1.25}\smash{\begin{tabular}[t]{l}F$_2$\end{tabular}}}}%
    \put(0.48027671,0.07985715){\color[rgb]{0,0,0}\makebox(0,0)[lt]{\lineheight{1.25}\smash{\begin{tabular}[t]{l}P\end{tabular}}}}%
  \end{picture}%
\endgroup%